\documentclass[prb,twocolumn,showpacs,amsmath,amssymb,floatfix,superscriptaddress]{revtex4-1}

\usepackage{stringstrings} 

\usepackage{graphicx}
\usepackage{xspace}
\usepackage{commath}
\usepackage{bm}
\usepackage{float}
\usepackage{amsmath}
\usepackage{fancyhdr}
\usepackage{afterpage}
\usepackage{url}
\usepackage{bm}
\usepackage{epstopdf}
\usepackage{multirow}
\usepackage{caption}
\usepackage{tikz}
\usepackage[]{threeparttable}
\usepackage{adjustbox}
\usepackage{soul}
\usepackage{setspace}
\usepackage[normalem]{ulem}
\usepackage{enumitem}
\usepackage{pythonhighlight}
\usepackage{chemformula}
\usepackage{xurl}
\usepackage{aligned-overset}

\usepackage{color}
\usepackage{hyperref}
\usepackage{listings}
\usepackage{geometry}

\graphicspath{{./figures/}}

\newcommand{\shift}{\texttt{shift}\xspace}
\newcommand{\stretchmorph}{\texttt{stretch}\xspace} 
\newcommand{\squeeze}{\texttt{squeeze}\xspace}
\newcommand{\scale}{\texttt{scale}\xspace}
\newcommand{\smear}{\texttt{smear}\xspace}
\newcommand{\smearpdf}{\texttt{smear-pdf}\xspace}
\newcommand{\shape}{\texttt{shape}\xspace}
\newcommand{\funcx}{\texttt{funcx}\xspace}
\newcommand{\funcy}{\texttt{funcy}\xspace}
\newcommand{\funcxy}{\texttt{funcxy}\xspace}

\definecolor{tagcolor}{HTML}{21a889}

\newcounter{saveenumi}
\newcommand{\seti}{\setcounter{saveenumi}{\value{enumi}}}
\newcommand{\conti}{\setcounter{enumi}{\value{saveenumi}}}

\newcommand{\qmax}{\ensuremath{Q_{\mathrm{max}}}\xspace}

\newcommand{\qmin}{\ensuremath{Q_{\mathrm{min}}}\xspace}

\newcommand{\rw}{\ensuremath{R_w}\xspace}
\newcommand{\rws}{\ensuremath{R_w}s\xspace}

\newcommand{\delr}{\ensuremath{dr}\xspace}

\newcommand{\delq}{\ensuremath{dQ}\xspace}

\newcommand{\iq}{\ensuremath{I(Q)}\xspace}
\newcommand{\iqs}{\ensuremath{I(Q)}s\xspace}
\newcommand{\sq}{\ensuremath{S(Q)}\xspace}
\newcommand{\fq}{\ensuremath{F(Q)}\xspace}
\newcommand{\gr}{\ensuremath{G(r)}\xspace}
\newcommand{\rr}{\ensuremath{R(r)}\xspace}
\newcommand{\q}{\ensuremath{Q}\xspace}
\newcommand{\ir}{\ensuremath{r}\xspace}
\newcommand{\dd}{\ensuremath{\mathrm{d}}}

\renewcommand{\vec}[1]{\mathbf{#1}}

\newcommand{\sjba}[1]{}

\newcommand{\mta}[1]{}

\newcommand{\eq}[1]{Eq.~\ref{eq:#1}}
\newcommand{\eqs}[1]{Eqs.~\ref{eq:#1}}
\newcommand{\fig}[1]{Fig.~\ref{fig:#1}}

\newcommand{\sect}[1]{Section~\ref{sec:#1}}

\newcommand{\tabl}[1]{Table~\ref{table:#1}}

\newcommand{\insitu}{\textit{in situ}\xspace}

\definecolor{dgreen}{HTML}{008000}

\definecolor{codegreen}{rgb}{0,0.6,0}
\definecolor{codegray}{rgb}{0.5,0.5,0.5}
\definecolor{codepurple}{rgb}{0.58,0,0.82}
\definecolor{backcolour}{rgb}{0.95,0.95,0.92}
\lstdefinestyle{mystyle}{
    backgroundcolor=\color{backcolour},
    commentstyle=\color{codegreen},
    keywordstyle=\color{magenta},
    numberstyle=\tiny\color{codegray},
    stringstyle=\color{codepurple},
    basicstyle=\ttfamily\footnotesize,
    breakatwhitespace=false,
    breaklines=true,
    captionpos=b,
    keepspaces=true,
    numbers=left,
    numbersep=5pt,
    showspaces=false,
    showstringspaces=false,
    showtabs=false,
    tabsize=2
}
% ...
\newcommand{\be}{\begin{enumerate}}
\newcommand{\ee}{\end{enumerate}}
\newcommand{\bes}{\begin{enumerate}[wide, labelwidth=!, labelindent=0pt, label=\textbf{\textcolor{blue}{\arabic*}.}]}
\newcommand{\ees}{\end{enumerate}}

\newcommand{\pdfgui}{\textsc{PDFgui}\xspace}
\newcommand{\diffpy}{\textsc{diffpy}\xspace}
\newcommand{\pdfgetxthree}{\textsc{PDFgetX3}\xspace}
\newcommand{\morph}{\textsc{diffpy.morph}\xspace}
\newcommand{\utils}{\textsc{diffpy.utils}\xspace}

\newcommand{\pyfai}{\textsc{pyFAI}\xspace}

\begin{document}

\title{\textsc{diffpy.morph}: Python tools for model independent comparisons between sets of 1D functions}

\author{Andrew Yang}
\altaffiliation{Current address: Department of Applied Physics and Materials Science, California Institute of Technology, Pasadena, CA 91125, USA}
\affiliation{Department of Applied Physics and Applied Mathematics, Columbia University, New York, NY 10027, USA}
\author{Christopher L. Farrow}
\affiliation{Department of Applied Physics and Applied Mathematics, Columbia University, New York, NY 10027, USA}
\author{Pavol Juh\'as}
\affiliation{Google Research, Google, Mountain View, CA 94043, USA}
\author{Luis Kitsu Iglesias}
\affiliation{Electrochemical Energy Systems Laboratory, Department of Mechanical and Process Engineering, ETH Zurich, 8092 Zurich, Switzerland}
\affiliation{Department of Chemical and Biological Engineering, University of Colorado Boulder, Boulder, CO 80309, USA}
\author{Chia-Hao Liu}
\affiliation{Department of Applied Physics and Applied Mathematics, Columbia University, New York, NY 10027, USA}
\author{Samuel D. Marks}
\affiliation{Advanced Photon Source, Argonne National Laboratory, Lemont, IL 60439, USA}
\affiliation{Materials Science and Engineering Program, University of Colorado Boulder, CO 80309, USA}
\author{Vivian R. K. Wall}
\affiliation{Department of Chemistry, University of California, Berkeley, CA 94720, USA}
\author{Joshua Safin}
\affiliation{Department of Materials Science and Engineering, The University of Tennessee, Knoxville, TN 37996, USA}
\author{Sean M. Drewry}
\affiliation{Department of Materials Science and Engineering, The University of Tennessee, Knoxville, TN 37996, USA}
\author{Caden Myers}
\affiliation{Department of Applied Physics and Applied Mathematics, Columbia 
University, New York, NY 10027, USA}
\author{Dillon F. Hanlon}
\affiliation{Department of Physics, Arizona State University, Tempe, AZ 85287, USA}
\author{Nicholas Leonard}
\affiliation{Department of Physics, Arizona State University, Tempe, AZ 85287, USA}
\author{Cedomir Petrovic}
\altaffiliation{Current address: Shanghai Key Laboratory of Material Frontiers Research in Extreme Environments (MFree), Shanghai Advanced Research in Physical Sciences (SHARPS), Shanghai 201203, P. R. China}
\affiliation{Brookhaven National Laboratory, Upton, NY 11973, USA}
\author{Ahhyun Jeong}
\affiliation{Department of Chemistry and James Franck Institute, University of Chicago, Chicago, IL 60637, USA}
\author{Dmitri V. Talapin}
\affiliation{Department of Chemistry and James Franck Institute, University of Chicago, Chicago, IL 60637, USA}
\author{Linda F. Nazar}
\affiliation{Department of Chemistry, University of Waterloo, Waterloo, ON N2L 3G1, Canada}
\author{Haidong Zhou}
\affiliation{Department of Physics and Astronomy, The University of Tennessee, Knoxville, TN 37996, USA}
\author{Samuel W. Teitelbaum}
\affiliation{Department of Physics, Arizona State University, Tempe, AZ 85287, USA}
\author{Tim B. van Driel}
\affiliation{SLAC National Accelerator Laboratory, Stanford University, Menlo Park, CA 94025, USA}
\author{Soham Banerjee}
\affiliation{Department of Applied Physics and Applied Mathematics, Columbia University, New York, NY 10027, USA}
\author{Emil S.  Bozin}
\affiliation{Center for Solid State Physics and New Materials, Institute of Physics Belgrade, 11080 Belgrade, Serbia}
\author{Michael F. Toney}
\affiliation{Department of Chemical and Biological Engineering, University of Colorado Boulder, Boulder, CO 80309, USA}
\affiliation{Materials Science and Engineering Program, University of Colorado Boulder, CO 80309, USA}
\affiliation{Renewable and Sustainable Energy Institute,  University of Colorado Boulder, CO 80309, USA}
\author{Katharine Page}
\affiliation{Department of Materials Science and Engineering, The University of Tennessee, Knoxville, TN 37996, USA}
\affiliation{Neutron Scattering Division, Oak Ridge National Laboratory, Oak Ridge, TN 37831, USA}
\author{Naomi S. Ginsberg}
\affiliation{Department of Chemistry, University of California, Berkeley, CA 94720, USA}
\affiliation{Department of Physics, University of California, Berkeley,  CA 94720, USA}
\affiliation{Molecular Biophysics and Integrated Bioimaging Division and Materials Sciences and Chemical Sciences Divisions, Lawrence Berkeley National Laboratory, Berkeley, CA 94720, USA}
\affiliation{Kavli Energy NanoSciences Institute, University of California, Berkeley,  CA 94720, USA}
\affiliation{STROBE, NSF Science \& Technology Center, University of California, Berkeley,  CA 94720, USA}
\author{Simon J.  L.  Billinge}
\email{sbillinge@ucsb.edu}
\altaffiliation{Current address: Materials Department, University of California, Santa Barbara, CA 93106, USA}
\affiliation{Department of Applied Physics and Applied Mathematics, Columbia University, New York, NY 10027, USA}

\date{\today}

\begin{abstract} 
\morph addresses a need to gain scientific insights from 1D scientific spectra in model independent ways.  
A powerful approach for this is to take differences between pairs of spectra and look for meaningful changes that might indicate underlying chemical, structural, or other modifications.
The challenge is that the difference curve may contain uninteresting differences such as experimental inconsistencies and benign physical changes such as the effects of thermal expansion.
\morph allows researchers to apply simple transformations, or ``morphs", to one of the datasets to remove the unwanted differences revealing, when they are present, non-trivial differences.
\morph is an open-source Python package available on the Python Package Index and conda-forge.  Here, we describe its functionality and apply it to solve a range of experimental challenges on diffraction and PDF data from x-rays and neutrons, though we note that it may be applied to any 1D function  in principle.
\end{abstract}

\maketitle

\section{Introduction}
    
Neutron and X-ray powder diffraction (NXPD) and atomic pair distribution function (PDF) analysis are powerful tools for probing the local and average structure of materials~\cite{dinnebier01GeneralIntroPowderDiffraction2008, egami;b;utbp12}.
They can provide many insights; for example, one can compare NXPD patterns and PDFs of samples measured at different temperatures to search for structural changes~\cite{billi;b;apdfap24}. 
This may be done in a model independent way by comparing two patterns and looking for meaningful changes, for example, by plotting the difference curve between the patterns or computing similarity metrics such as the profile-weighted agreement factor (\rw~\cite{dinnebier01GeneralIntroPowderDiffraction2008,  egami;b;utbp12}) or the Pearson correlation coefficient (PCC~\cite{pearsonVIINoteRegression1895}).
This task is made more difficult in the presence of less interesting effects such as expansion/contraction due to temperature changes, thermal smearing due to increased thermal motion, and certain experimental artifacts.
These are effects that may not represent ``interesting" structural changes, but can produce large contributions to the residual curves and \rws, hiding more scientifically interesting structural changes.  
A challenge is to remove the uninteresting changes to reveal scientifically interesting ones when they are present.

Here we present the Python program \morph that provides a solution to this problem by first applying simple transformations, or ``morphs", to spectra to partially or fully account for these non-interesting effects, allowing any interesting changes that can't be removed by the morph to become apparent.

We note that this approach is completely general and may be applied beyond diffraction and atomic pair distribution function~\cite{egami;b;utbp12} data to any kinds of 1D spectra.  
The approach is fast and model independent and is a powerful tool for interrogating data in real-time as it is being collected.

Here we describe the approach and test it on a number of different real-world examples from our own studies.
We show how it can be used to reveal when a structural phase transition occurs, and when it does not, in temperature series PDF data.  
We then present an example where a thermodynamic parameter $\beta$, the (linear) thermal expansion coefficient, is extracted directly from temperature-dependent data without modeling,
and in contrast, where the temperature of a sample can be extracted given a known $\beta$, again in a model independent way.
We present an example where experimental inconsistencies could be removed between datasets allowing them to be compared despite the inconsistencies.
Finally, we also present some more advanced use-cases, where we use \morph to find optimal parameters in a data reduction workflow by morphing a low \qmax high-throughput data to a reference high \qmax target dataset.

\morph is easy to use and flexible. It is open source (BSD license)~\cite{bsd3clause88} and available on PyPi~\cite{morphPyPiRelease25} and conda-forge~\cite{morphCondaForgeRelease25}.  It has a command-line interface and API allowing it to be integrated into users' Python scripts.  The program can read most column-style data files (e.g. `.xy', `.dat', `.chi', `.gr'), with and without header information.
It also automatically handles comparison of two functions plotted on different grids by using linear interpolation.

Though this software was originally designed in the context of PDFs, \morph and its morphs can be easily applied to any general function.
For example, a neutron scattering experimentalist could subtract the hydrogen contribution to a density of states (DOS) by morphing a known hydrogen DOS to their measured DOS and taking the difference curve, or a computer vision scientist comparing two pictures taken at different brightnesses can give the (flattened) 2D images to \morph to correct for intensity differences.

\section{Motivating Examples}
\label{sec:motivating}

Here, we introduce some examples that motivated our development of \morph but which also illustrate how it may be applied in different situations.



\subsection{Identifying a Structural Phase Transition}
\label{sec:temp-morphs}

\subsubsection{Identifying whether or not a phase transition has occurred by using morphs}
\label{sec:identify-phase-transition}

\begin{figure}[tbp]
    \centering
\includegraphics[width=1\linewidth]{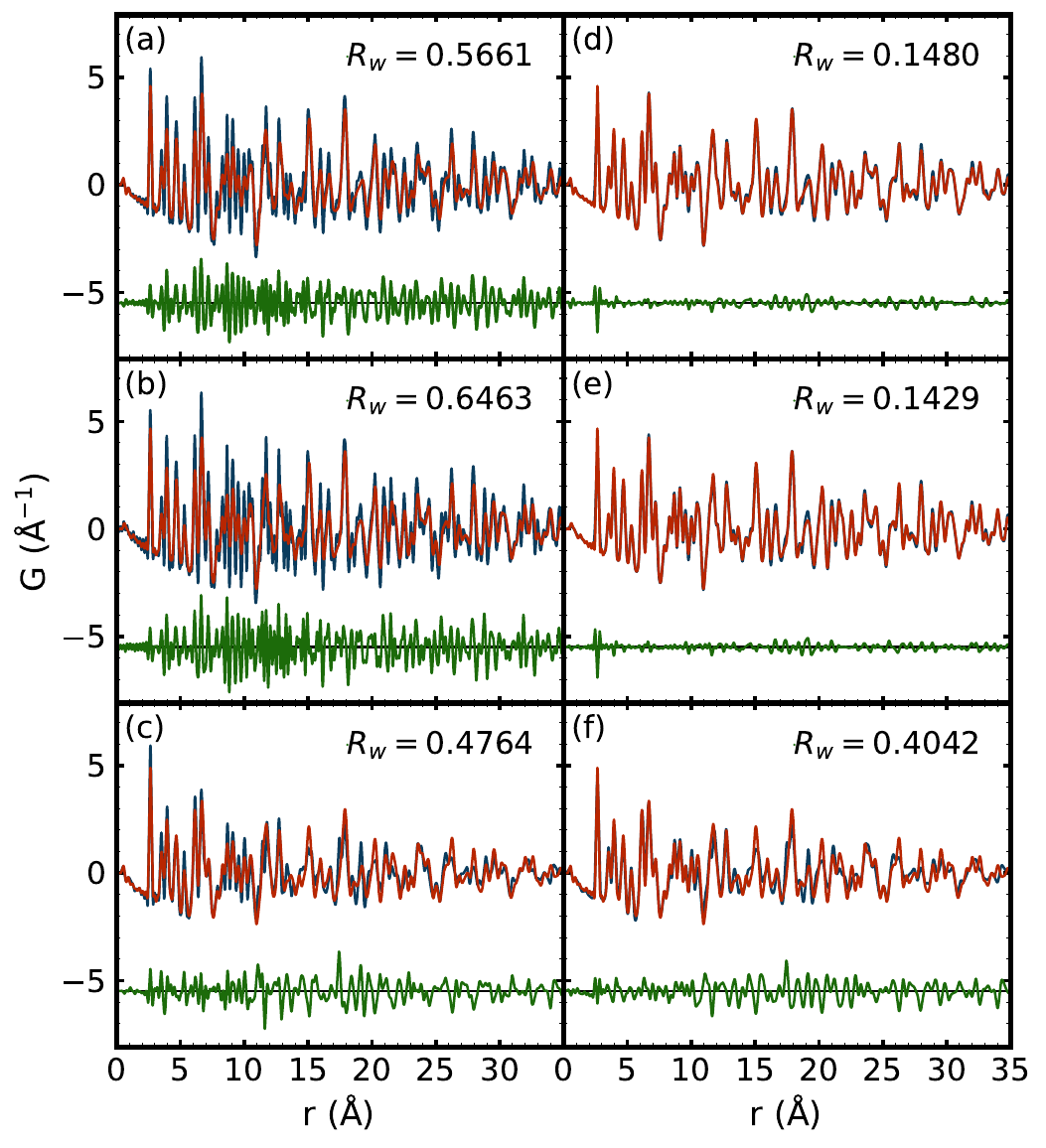}
    \caption{PDFs of IrTe$_2$ samples with various forms of supstitution measured at $10$~K (blue) and $300$~K (red).  The difference curves (green) are plotted below.  All \rw values are computed on the grid $r \in [0~\text{\AA}, 35~\text{\AA}]$ with spacing $\delr = 0.01$.  (a) Samples are Rhodium-substituted Ir$_{\text{0.8}}$Rh$_{\text{0.2}}$Te$_2$.  (b) Platinum-substituted Ir$_{\text{0.95}}$Pt$_{\text{0.05}}$Te$_2$.  (c) Pure IrTe$_2$.  Plots (d), (e), and (f) have the same high-temperature PDF as (a), (b), and (c) respectively, but a ``stretching", ``scaling", and ``smearing" morph has been applied to the low-temperature PDF to best fit the high-temperature target PDF.}
    \label{fig:10-300-Morph}
\end{figure}

\fig{10-300-Morph}(a-c) shows pairs of PDFs that were collected at 10~K (blue) and 300~K (red) with the difference curve plotted below.
We would like to answer the questions:  (1) ``Do any of these show a structural phase transition in this range?" and (2) ``If so, which?".
We might be able to do this by comparing the two PDFs and considering various difference measures.
A large difference measure should indicate that there may be a phase transition, and a smaller measure that there is not.
We see the largest difference measures for (b) ($\rw = 0.6463$) then (a) ($\rw = 0.5661$), with the smallest for (c) ($\rw = 0.4764$), which might suggest that it is more likely to find a structural phase transition in (a) and (b).
However, this is a wide temperature range, and there are significant thermal effects.
\fig{10-300-Morph}(d-f) show the corresponding pairs of PDFs after applying ``stretching", ``scaling", and ``smearing" morphs, described further below, to the lower-temperature data.
In this case, the difference measures become very small for (d) and (e) but remain somewhat large with significant unresolved signal in the green difference curve for (f).

The sample in (c,f) is an IrTe$_2$ compound, and in (a,d) and (b,e), the Iridium is substituted with Rhodium and Platinum respectively.  Single crystals of IrTe$_2$ where grown with a self-flux method~\cite{pretrovicIrTe2SamplePreparation14}, while Rhodium-substituted~\cite{petrovicIrRhTe2SamplePreparation18,yangIrRhTe2SamplePreparation12} and Platinum-substituted~\cite{pyonIrPtTe2SamplePreparation12} polycrystals were synthesized using a solid-state reaction method.

After morphing, it is clear that pure IrTe$_2$ has a phase transition in this temperature range, but substituting Pt, even at the five percent level, suppresses the phase transition.
This was not apparent in the unmorphed data but is confirmed by structural modeling in the original study~\cite{banerjeeBozinLocalFlucctuatingDimers18}.

\begin{figure}[tbp]
    \centering
    \includegraphics[width=1\linewidth]{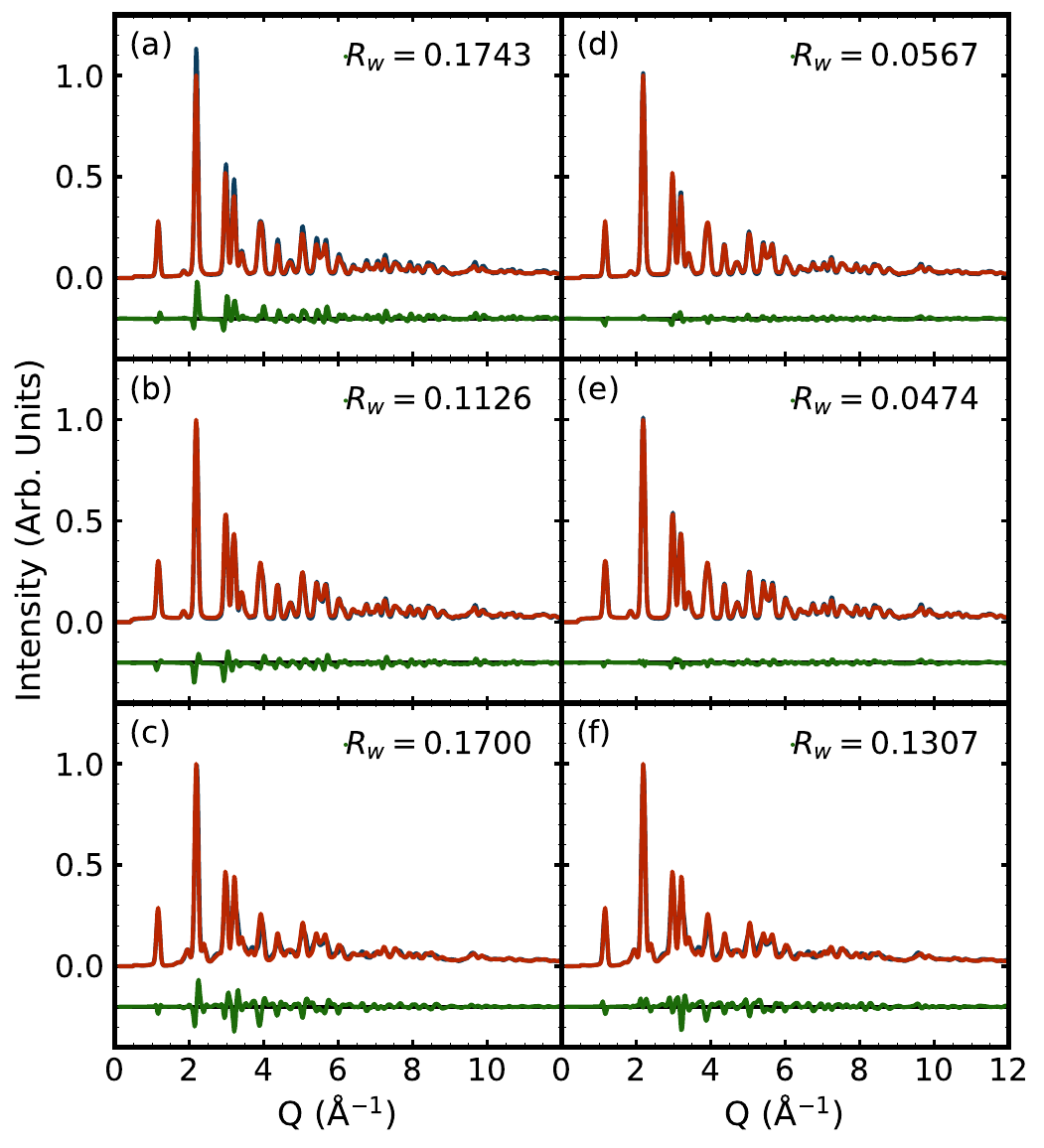}
    \caption{\iqs of IrTe$_2$ samples with various forms of substitution measured at $10$~K (blue) and $300$~K (red).  The difference curves (green) are plotted below.  All \rw values are computed on the grid $Q \in [0~\text{\AA}^{-1}, 12~\text{\AA}^{-1}]$ with spacing $\delq = 0.02$.  (a) Rhodium-substituted Ir$_{\text{0.8}}$Rh$_{\text{0.2}}$Te$_2$.  (b) Platinum-substituted Ir$_{\text{0.95}}$Pt$_{\text{0.05}}$Te$_2$.  (c) Pure IrTe$_2$.  The \iqs plotted in (a-c) were used to generate the PDFs in \fig{10-300-Morph}(a-c).  Plots (d), (e), and (f) have the same high-temperature \iq as (a), (b), and (c) respectively, but a \stretchmorph and \scale morph has been applied to the low-temperature \iq to best fit the high-temperature target \iq.}
    \label{fig:10-300-iq-Morph}
\end{figure}

Similar problems arise when comparing 1D diffraction intensities, \iq, prior to computing a PDF.  
\fig{10-300-iq-Morph}(a-c) shows the same analysis as carried out in \fig{10-300-Morph} but this time on the \iq data instead of the PDF.   
%

As before, prior to morphing, it is much less clear from the difference curve which of these cases have structural phase transitions and which do not, but this is very apparent after morphing.

\begin{figure}[tbp] \includegraphics[width=1\columnwidth]{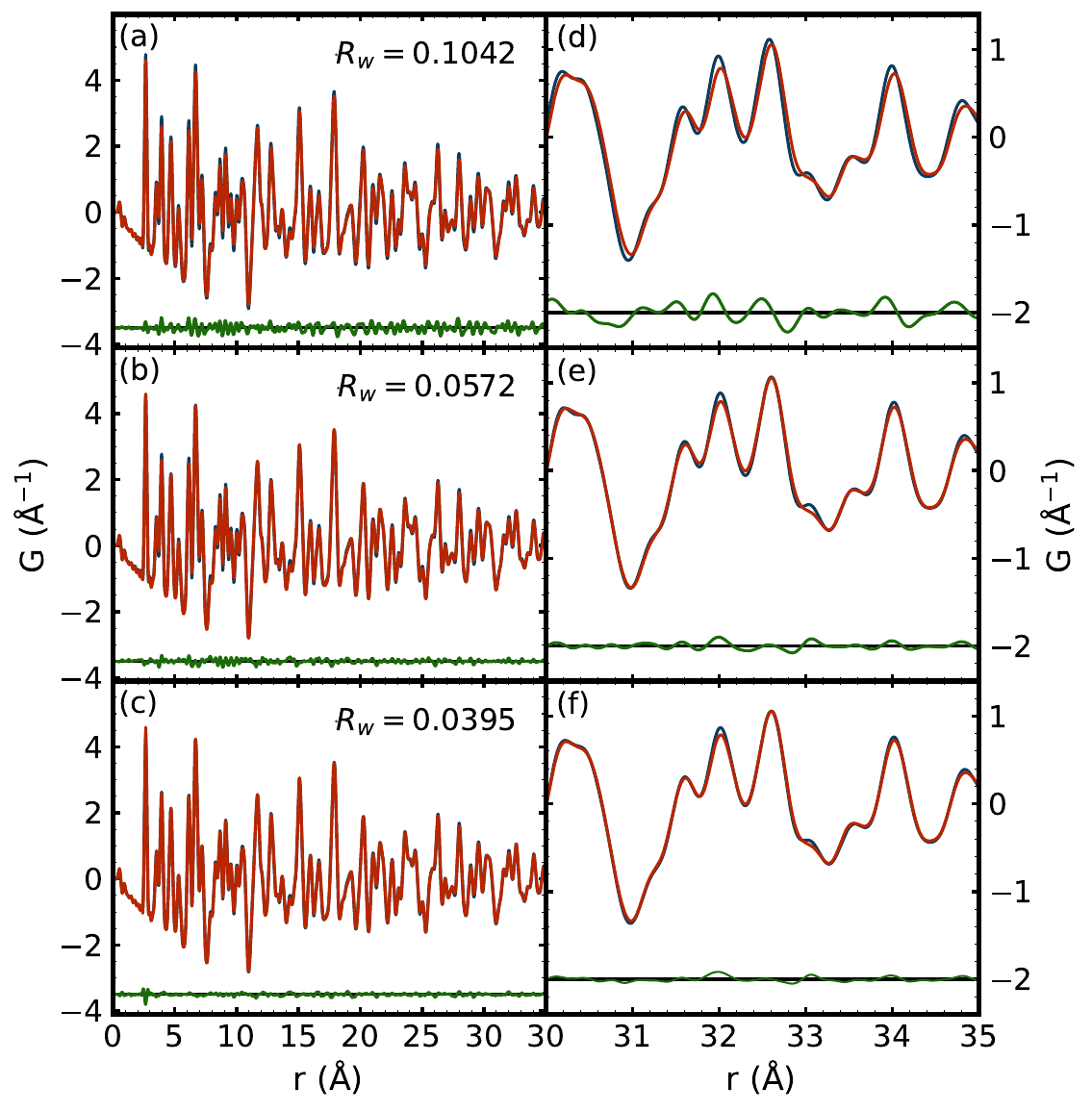}
    \caption{ 
    Morphed PDFs of Ir$_{\text{0.8}}$Rh$_{\text{0.2}}$Te$_2$ samples measured at $250$~K (blue) plotted against unmorphed PDFs measured at $300$~K (red).  The difference curves (green) are plotted below. All \rw values are computed on the grid $\ir \in [0~\text{\AA}, 35~\text{\AA}]$ with spacing $\delr = 0.01$. (a) No morph is applied.  (b) A \stretchmorph and \scale morph is applied to the lower-temperature PDF.  (c) A \stretchmorph, \scale, and \smear morph is applied to the lower-temperature PDF.  Plots (d), (e), and (f) are the same as (a), (b), and (c) respectively zoomed into the region $30~\text{\AA} \leq r \leq 35~\text{\AA}$.}
\label{fig:250-300-Rh-Morph}
\end{figure}

We now investigate in more detail how these morphing transformations actually affected the data.
In \fig{250-300-Rh-Morph}(a) we compare PDFs from Ir$_{\text{0.8}}$Rh$_{\text{0.2}}$Te$_2$ at 250~K (blue) and 300~K (red).
The agreement factor $\rw = 0.1042$, and there are some ripples in the green difference curve plotted below.
By close inspection, we can observe that the peaks in the warmer sample are broader and shifted to higher-\ir.  This is most evident in the region $30~\text{\AA} \leq r \leq 35~\text{\AA}$, plotted on an expanded scale in \fig{250-300-Rh-Morph}(d).  
More evidence of peak shifting across the entire \ir-grid is provided in Supplemental Material, Fig.~\ref{sup-fig:grad-diff-250K-300K}.

The peak shifts could come from thermal expansion and the peak broadening from increased thermal motion.
\morph allows us to investigate whether a uniform ``stretch" of the lattice of the lower-temperature sample can account for the peak shifts across the whole range of the PDF without having to know and model the structure of the material.

To accomplish this \morph first interpolates the lower-temperature PDF onto a slightly stretched \ir-grid $r \mapsto r/(1 + \eta)$.  This we call the \stretchmorph morph with a stretch parameter $\eta$.
Then, \morph  multiplies the PDF by a constant scale factor~$s$.
This is the ``\scale\unskip" morph with scale factor $s$.
The result of applying these two transformations is shown in \fig{250-300-Rh-Morph}(b).  The \rw is significantly reduced, as are ripples in the difference curve.

From a practical perspective, it is tedious to find the exact \stretchmorph and \scale morphs to best explain the difference, so we wrap the transformation in a least-squares regression method to find the $\eta$ and $s$ values that minimize the \rw between the PDF of the low-temperature signal after morphing (which we denote the ``morphed" signal $G_m(r)$) and the untouched high-temperature PDF (the ``target" signal $G_t(r)$).

Although \rw is reduced by almost a factor of two by the \stretchmorph and \scale morphs, differences remain.
We would like to see the extent to which they can be removed by a uniform broadening, or smearing, of the radial distribution function (RDF) of the lower-temperature sample. 
We accomplish this by computing the RDF from the morphed PDF, convolving the RDF with a normalized Gaussian function with variance $\sigma^2$, and convert the smeared RDF back to a PDF. 
We then vary the $\sigma^2$ of that convolution function until we get the best agreement between the morphed PDF and the target PDF.
This is the ``\smear\unskip" morph with smearing parameter $\sigma$.  
The result of turning on the \smear morph transformation on top of the \stretchmorph and \scale morphs is shown in \fig{250-300-Rh-Morph}(c). 
This results in a further reduction in \rw of 31\% and an almost flat difference curve.
For more details on how the morphs are implemented, see \sect{morphing-transforms}.

In the absence of a structural phase transition, the one place in the difference curve where significant fluctuations remain after morphing is in the vicinity of the first peak at $r=2.7$~\AA.  This is evident in the difference curve in \fig{10-300-Morph}(d,e) and
\fig{250-300-Rh-Morph}(c).
This phenomenon is well-known and is due to changes in the correlated atomic motion~\cite{jeongCorrelatedAtomicMotion03}.  This could be removed with a smear-morph applied over just the range of the PDF incorporating this first peak.  It would allow the temperature dependence of the broadening of the first peak to be extracted separately from the higher-\ir peaks and extract information about the degree of motional correlations, though we don't do that here.
It also shows how careful morphing can remove some features and not others, revealing key insights.

\subsubsection{Finding the temperature of a phase transition by analyzing multiple morphs}
\label{sec:phase-transition}

We have seen how \morph is helpful in identifying whether or not a phase transition has occurred, but it can also help identify the temperature range in which the transition occurs.

\begin{figure}[tbp]
    \centering
    \includegraphics[width=1\linewidth]{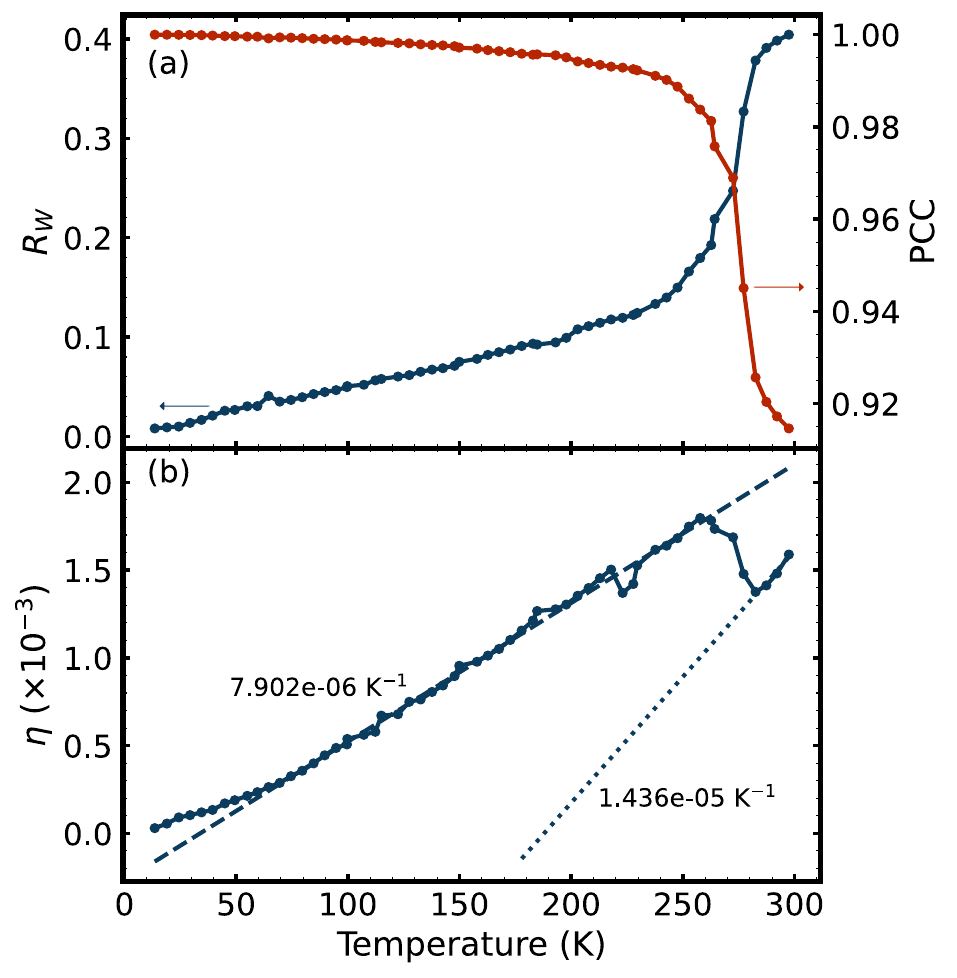}
    \caption{(a) \rw (blue) and Pearson correlation coefficient (red) computed between a morphed and target PDF from  IrTe$_2$ PDFs measured at temperatures ranging from $9$~K to $297$~K.  (b) The \stretchmorph parameter $\eta$ found in each morph indicating thermal expansion. This parameter sharply drops above 263~K. The drop indicates contraction at the phase transition. Dashed lines are linear fits to the low and high temperature regions.}
    \label{fig:IrTe2-All}
\end{figure}

On a dataset of IrTe$_2$ PDFs measured from $10$~K to $300$~K, we have used \morph's capability to create multiple morphs (\stretchmorph\unskip, \scale\unskip, \smear\unskip) of the lowest-temperature PDF with each higher-temperature PDF in the set as a target.
In \fig{IrTe2-All}(a), we plot the resulting (post-morph) \rws.
The \rws vary slowly over the entire temperature range until $263$~K where there is a rapid change in slope up till $283$~K, where it again slows down.
The same can be seen in the PCC curve shown in red.
A phase transition (from low-temperature triclinic to high-temperature trigonal) has been observed near $280$~K~\cite{banerjeeBozinDSTransition18}.
The \morph analysis shows the onset of this phase transition on cooling beginning at around $283$~K with the distortion growing smoothly on cooling until around $263$~K but using a completely model-independent analysis and a simple command in a terminal.

\subsubsection{Comparison of \morph with model refinement techniques}
\label{sec:pdfmorph-pdfgui-comparison}

\begin{figure}[tbp]
    \centering
    \includegraphics[width=1\linewidth]{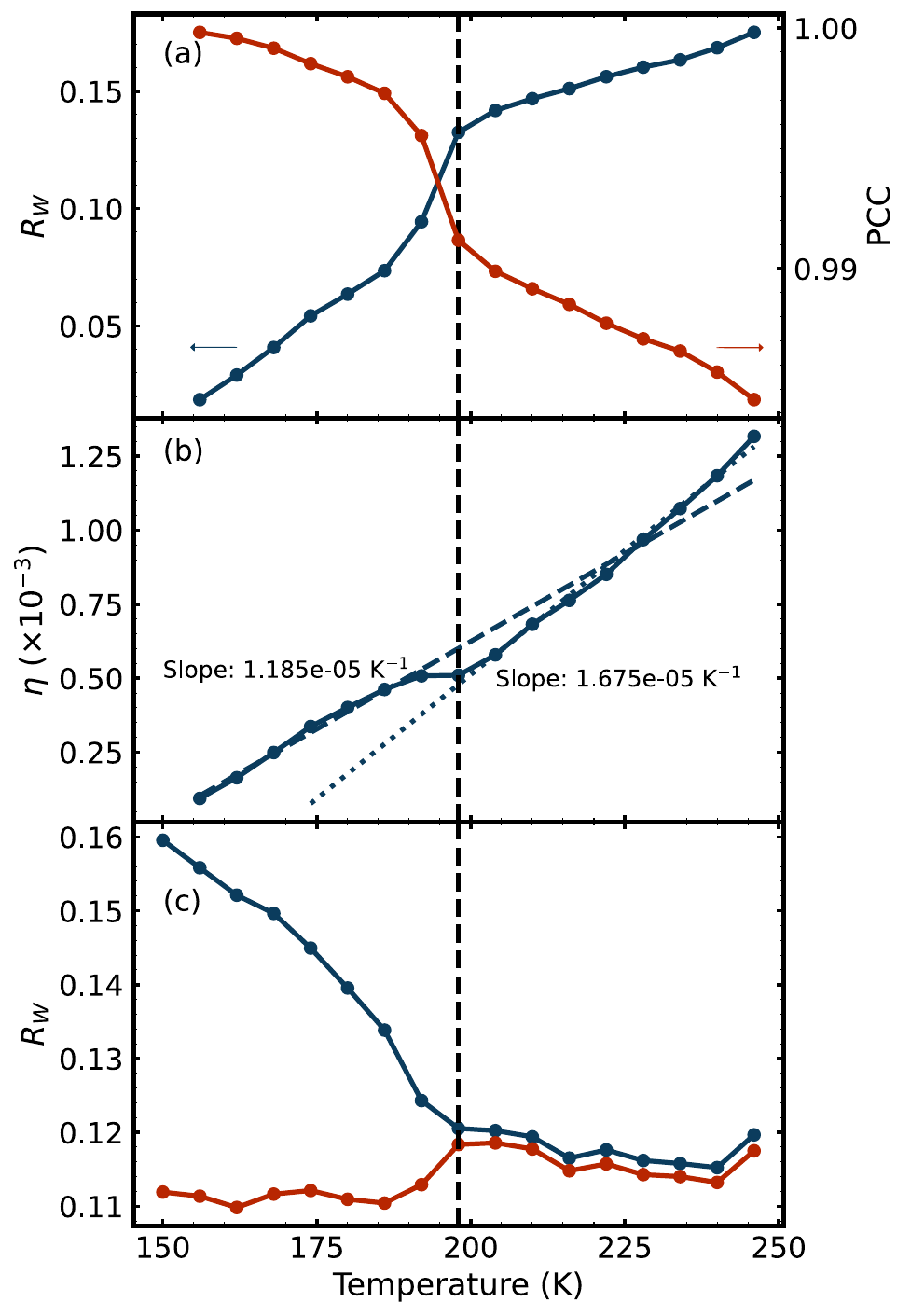}
    \caption{(a) \rw (blue) and PCC (red) after morphing the PDF from SrFe$_2$As$_2$ measured at $T=150$~K on to the PDF measured at target at temperatures up to $246$~K.
    (b) The \stretchmorph parameter $\eta$ obtained from each morph. The dashed and dotted lines show best-fit linear curves in the temperature ranges $156$~K to $192$~K and $198$~K to $246$~K, respectively. The slope gives the thermal expansion coefficient in the respective temperature ranges.
    (c) The \rws obtained by fitting a tetragonal (blue) and orthorhombic (red) structure to each PDF on the $r$-range $[0.5~\text{\AA},\;100~\text{\AA}]$.
The black vertical dashed line is a guide to the eye at the first measured point above the structural phase transition.}
    \label{fig:SrFe2As2-All}
\end{figure}

This method of identifying phase transition temperatures using \morph produces results that agree with structural model fitting techniques such as \pdfgui~\cite{farrowPDFfit2PDFgui}.
There is a known structural change in SrFe$_2$As$_2$  between $192$~K and $198$~K~\cite{billi;b;apdfap24}.
PDFs of SrFe$_2$As$_2$ measured from $150$~K to $246$~K were sourced from data published alongside the textbook ``Atomic Pair Distribution Function Analysis: A Primer"~\cite{billi;b;apdfap24,pdfttpdata23}.
In \fig{SrFe2As2-All}(a), we have plotted the \rw and PCC values resulting from morphing the PDF of SrFe$_2$As$_2$ measured at $150$~K to PDFs of the same sample measured at the various higher temperatures.
We observe a steep descent in the \rw and ascent in the PCC within the temperature range of the phase transition.  We have also used \morph to plot the \stretchmorph parameter used in the morphs across temperature in \fig{SrFe2As2-All}(b).  While this parameter increases linearly from $156$~K to $192$~K and from $198$~K to $245$~K, indicating some isotropic expansion in those ranges, from $192$~K to $198$~K, the \stretchmorph parameter does not change, further confirming that some more interesting structural change is occuring at those temperatures.  Structural modeling refinement done by \pdfgui~\cite{farrowPDFfit2PDFgui} shows that the tetragonal model only fits well past $198$~K, indicating some structural phase change occurs past this temperature.  This result was obtained in a model-independent \morph analysis.



\subsection{Extracting Material Properties using Morph Comparisons}

\subsubsection{Thermal expansion and Debye temperature using stretch morph on temperature dependent x-ray data}
\label{sec:mat-prop-thermal}

Using a series of PDFs measured at different temperatures, one can estimate the linear thermal expansion coefficient of a material using the \stretchmorph morph. For nearly harmonic materials, one can further compute the Debye temperature by looking at the derivative of the stretch parameter with respect to temperature.

In \fig{IrTe2-All}(b) the value of the \stretchmorph morph parameter $\eta(T)$ computed on the grid $r \in [0~\text{\AA}, 35~\text{\AA}]$ is plotted against temperature for the IrTe$_2$ sample whose x-ray PDF was measured vs. temperature.
We see this grow smoothly, albeit noisily.  By finding the gradient of $\eta(T)$ we can extract the thermal expansion coefficient as we discuss below.

The \stretchmorph parameter obtained from morphing between a temperature $T_a$ and $T_b$ is an average isotropic strain $\epsilon(T_a, T_b)$ of the material induced by the change in temperature (shown in Supplemental Material, \sect{sup-temp}).  We can compute a linear thermal expansion coefficient $\beta(T_a)$ using
\begin{align}
    \beta(T_a) &= \frac{\eta(T_b) - \eta(T_a)}{(T_b - T_a)(1 + \eta(T_a))}\\
    &\approx \frac{\eta(T_b) - \eta(T_a)}{T_b - T_a},\label{eq:beta-from-morph}
\end{align}
where the approximation holds for $\eta(T_a) << 1$. 
Each $\beta(T_a)$ was computed from discrete temperature data using finite differences of $\eta(T)$.
From \fig{IrTe2-All}(b), we compute an average thermal expansion coefficient of $7.902\cdot10^{-6}$~K$^{-1}$ for the triclinic phase of IrTe$_2$ (in range $T \in [107~\text{K}, 263~\text{K}]$) and $1.436\cdot10^{-5}$~K$^{-1}$ for the trigonal phase (in range $T \in [283~\text{K}, 297~\text{K}]$). Note that the \stretchmorph yields an average linear expansion coefficient that more accurately reflects the actual expansion the more isotropic the material and
if the material expansion is highly anisotropic, the stretch morph will not remove the effects of the thermal expansion.

\begin{figure}[tbp]
    \centering
\includegraphics[width=1\linewidth]{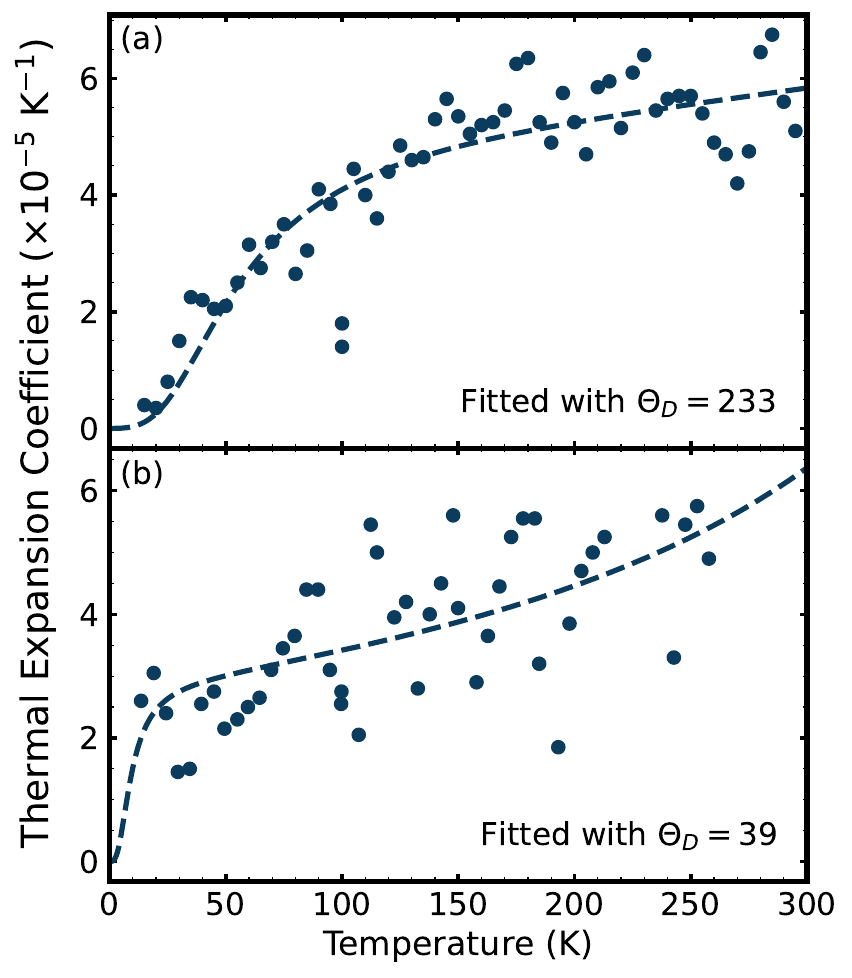}
    \caption{Plots of thermal expansion coefficients (extracted using the approximation in \eq{beta-diff-approx}) at various temperatures.  The samples involved are (a) Ir$_{\text{0.95}}$Pt$_{\text{0.05}}$Te$_2$ and (b) IrTe$_2$.  Data is fitted to \eq{beta-qh-fit} with parameters $A$, $B$, and $\Theta_D$ (see text).  The fits have $R^2$ values of (a) $0.7937$ and (b) $0.4143$.}
    \label{fig:IrTe2-Quasiharmonic}
\end{figure}

We can also get a sense of the anharmonicity of the material by examining the thermal expansion coefficient as a function of temperature. This $\beta(T)$ can be modeled by the temperature derivative of the stretch parameter $\eta(T)$
\begin{align}
    \beta(T) &\approx \frac{\partial \eta(T)}{\partial T}\label{eq:beta-diff-approx}
\end{align}
per \eq{beta-from-morph}.
In \fig{IrTe2-Quasiharmonic}, we plot the thermal expansion coefficient, extracted from the morphed data (using central differences to compute the derivative of $\eta(T)$), over a large temperature range for, (a), Pt-substituted and, (b), pure IrTe$_2$.
We then fit the data to the curve for the (linear) thermal expansion $\beta_{G,A}$,
using a quasiharmonic model given by, 
\begin{align}
    \beta_{G,A}(T;A,B,\Theta_D) &\approx A\frac{\left(\frac{T}{\Theta_D}\right)^3\int_0^{\Theta_D/T}\frac{x^4e^x}{(e^x-1)^2}dx}{1 - BTe^{-\Theta_D/(2T)}}\label{eq:beta-qh-fit},
\end{align}
with fitting parameters $A$, $B$, and $\theta_D$ (Debye temperature).
The expression is obtained from the Gr\"uneisen equation of state~\cite{gruneisenTheoryMonatomic1912,changBulkModulusTemperatureDependence1967}, where the forms of the isochoric heat capacity (numerator) and adiabatic bulk modulus (denominator) come from the Debye model~\cite{schroederIntroduction21}, and Anderson approximation~\cite{andersonDerivationWachtman1966,changBulkModulusTemperatureDependence1967}, respectively.  For this model to work, we must assume a constant heat-capacity ratio $C_V/C_P$~\cite{changBulkModulusTemperatureDependence1967,schroederIntroduction21} to ensure that the isothermal and adiabatic bulk moduli are proportional, assume the change in volume is small, and assume a constant Gr\"uneisen parameter~\cite{gruneisenTheoryMonatomic1912,andersonDerivationWachtman1966,changBulkModulusTemperatureDependence1967}.
Anharmonic (non-harmonic and non-quasi-harmonic) materials are likely to violate these conditions.
We see that while the Platinum-substituted sample has a nice agreement and sensible Debye temperature of $233$~K~\cite{petrovicIrRhTe2SamplePreparation18,pyonIrPtTe2SamplePreparation12}, the opposite is true for the pure sample with an anomalous upturn in $\beta$ at high-$T$ (\fig{IrTe2-Quasiharmonic}(b)) and an abnormally low Debye temperature $39$~K, suggesting significant anharmonicity.  The anharmonicity of pure IrTe$_2$ has been explored previously using Raman spectrometry~\cite{lazarevicProbingIrTePRC14}.

\subsubsection{Using \morph for \insitu sample thermometry}
\label{sec:temp-from-neutron}

The converse of finding the thermal expansion coefficient from temperature series data is using a known thermal expansion coefficient to estimate sample temperature in a model independent way.

There are few available techniques for precisely measuring the temperature of diffraction experiment samples heated to high temperatures.  One such technique is pyrometry, but thermal gradients caused by uneven heating of the material and ablation in the scattering region can greatly limit the accuracy of the measurement~\cite{ushakovYSZHighTemperatureThermalExpansion15,aldebertPyrometryLimitationsHighTemperature84}.  Further, pyrometry measures only the surface temperature, which makes it unreliable for transmission geometry scattering experiments.

This motivates the development and use of alternative methods for obtaining the bulk average temperature of a sample under \insitu conditions.  
In \sect{phase-transition}, we demonstrated how \morph can be used to identify whether a phase transition has occurred between two PDFs.  This can be matched with a known phase transition temperature to identify the temperatures of these PDFs (this works also for \iqs).  On the other hand, if a sample does not undergo a phase transition, the following example shows how one can use the known thermal expansion coefficient of the sample material to obtain the average temperatures of the sample at the moment the \iqs and PDFs were acquired\cite{imteyazRapidThermalProcessingChamber15}.

\begin{figure}[tbp]
    \centering
    \includegraphics[width=1\linewidth]{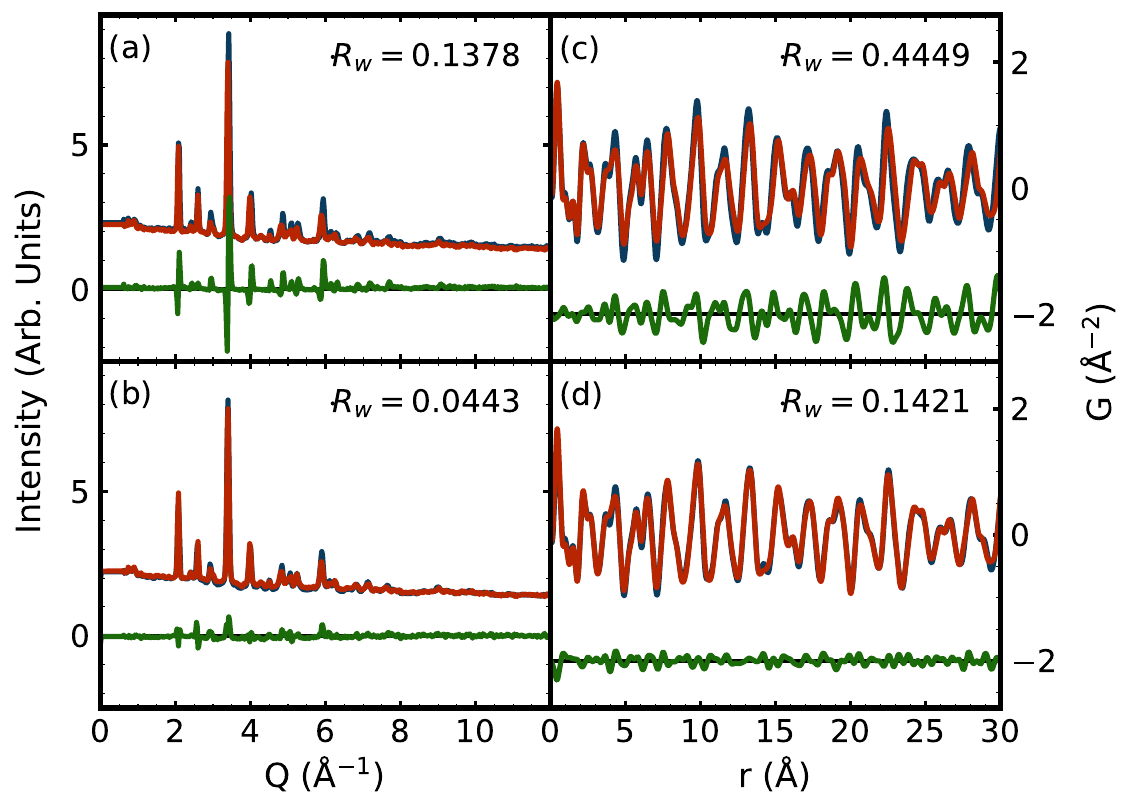}
    \caption{\iqs ((a) and (b)), and PDFs ((c) and (d)) of YSZ samples heated with a $30$~W (plotted in blue) and $50$~W (plotted in red) laser.  Difference curves are plotted offset below in green.
    No morphs are applied in (a) and (c). \stretchmorph and \scale morphs are applied to the lower-power (blue) signal in (a) to obtain the blue \iq in (b). \stretchmorph, \scale, and \smear morphs are applied to the lower-power signal in (c) to obtain the blue PDF in (d).
    The \rw is computed from the entire $Q$, or $r$, range plotted. 
    }
    \label{fig:neutron-morph}
\end{figure}

A commercially-obtained sample of yttria-stabilized zirconia (YSZ) was laser heated from room temperature with laser powers ranging from $0$~W to $50$~W. 
At each power level, a neutron powder diffraction \iq was measured, and the corresponding PDF, \gr, computed. 
In \fig{neutron-morph}(a), we show the \iq functions computed from laser-heated YSZ samples at powers $30$~W and $50$~W.
The difference curve shows a significant signal, due to thermal expansion from the additional $20$~W of power.  Similarly, the difference curve between the corresponding PDFs (shown in \fig{neutron-morph}(c)) also shows a significant signal. 
However, in (b) and (d), after applying a \stretchmorph and \scale (as well as a \smear morph for the PDF data) on the $30$~W data with the $50$~W data as a target, the majority of the differences are removed.

The yittria content of the YSZ samples used was estimated to be $5.32$~mol\% using calculated lattice parameters~\cite{witzPhaseDiagram07}, with a corresponding thermal expansion coefficient of $\beta = 11.29\cdot10^{-6}$~1/K.  This thermal expansion coefficient is in agreement with existing literature~\cite{hayashiYSZThermalExpansionCoefficient05,krogstadTetragonalYSZHighTemperatureThermalExpansion15,ushakovYSZHighTemperatureThermalExpansion15}. 
With this thermal expansion, we can use the \stretchmorph morph to estimate the temperature of the laser-heated YSZ samples directly.

\begin{figure}[tbp]
    \centering
    \includegraphics[width=1\linewidth]{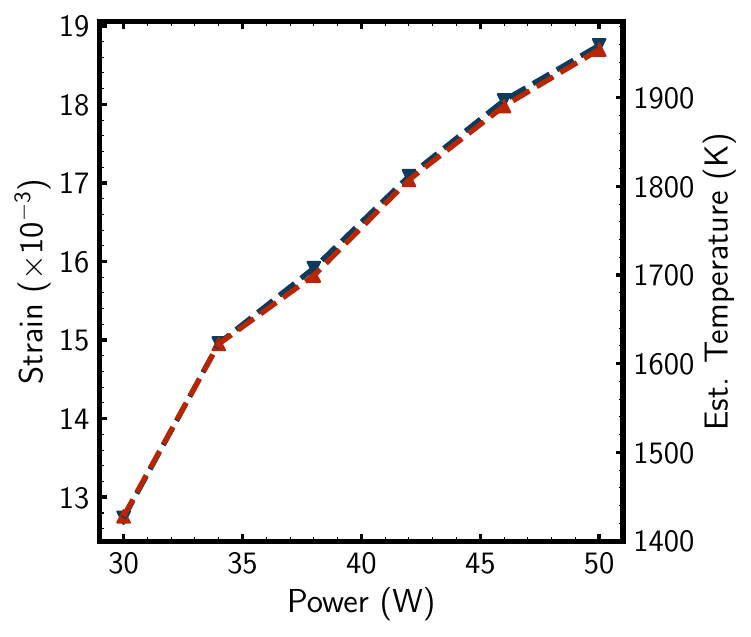}
    \caption{Strains computed by morphing $0$~W YSZ PDF data (blue) and \iq data (red) to higher-power YSZ data. The corresponding temperature is given by the right axis.}
    \label{fig:neutron-est-temp}
\end{figure}

For small $\Delta T = T_b - T_a$, we assume $\eta(T_a) \approx \eta(T_b)$.
Rearranging \eq{beta-from-morph}, we get
\begin{align}
    \Delta T &\approx \frac{\eta_{gr}}{\beta},
\end{align}
where $\eta_{gr}$ is the PDF \stretchmorph parameter.
Using the \stretchmorph parameter and $\beta$, we can obtain $\Delta T$ between the room temperature ($298$~K) sample and each laser-powered sample.  The actual temperature is then
\begin{align}
    T &= 298~\text{K} + \Delta T.
\end{align}
The strain, equivalent to $\eta_{gr}$ (see Supplemental Material \sect{sup-temp}), and the corresponding estimated temperatures (right axis label) are plotted in blue in \fig{neutron-est-temp}.

We also carried out the same analysis directly on the \iq data rather than the PDF.  (A similar approach has been used to determine sample temperatures on x-ray \iq data~\cite{imteyazRapidThermalProcessingChamber15}.)
In this case the \stretchmorph parameters, $\eta_{iq}$, correspond to a strain $\epsilon$ of
\begin{align}
    \epsilon &= (1 + \eta_{iq})^{-1} - 1
\end{align}
(see Supplemental Material, \sect{sup-temp-iq}) and thus a temperature change of
\begin{align}
    \Delta T &\approx \frac{(1 + \eta_{iq})^{-1} - 1}{\beta}.
\end{align}
The strains and corresponding temperatures extracted from the \iq morphs correspond to the red curve in \fig{neutron-est-temp}. These are in excellent agreement with each other.
The largest discrepancy between the two curves is a difference of $<0.3\%$ at $38$~W.

This example not only shows how temperatures can be estimated using \morph, but its applicability to not just x-ray diffraction signals, but also neutron data.

\subsubsection{Nanoparticle shape information from shape morphs}
\label{sec:mat-prop-shape}

The shape and size of a nanoparticle can greatly affect its (electronic, optical, etc.) properties~\cite{goyalSizeProperties20,tolbertCdSeSizeProperties94}.
If one has PDF measurements of a bulk and a nanoparticle of the same material, it is possible to use a nanoparticle ``\shape\unskip" morph to determine the morphological parameters without a structural model.
By way of example, we show how it is possible to determine the radius of a spherical nanoparticle using a \shape morph.

This morph can be very useful if you have a PDF calculated from a bulk model but your data is from a nanocrystalline sample.
\morph will then multiply the bulk PDF by the chosen nanoparticle characteristic function~\cite{farrowBillingeSasPdf09,egami;b;utbp12}, $\gamma_0(r)$, automatically adjusting the parameters of the model to find the best agreement to the nanoparticle PDF. 

If a good fit is obtained, the model choice and refined shape parameters can be considered useful, though other models should be tested to search for alternative possible candidates~\cite{farrowBillingeSasPdf09,banerjeeQuantitativeStructuralCharacterization2019c,banerjeeClusterminingApproachDetermining2020a,taoRapidModellingLocalStructure24}.
Significant difference curve signals or \rws indicate large deviations from the desired shape and new shape models may be tried.  

This approach has previously been used to estimate diameters of spherical CdSe nanoparticles consistent with those obtained from transmission electron microscopy, ultraviolet-visible spectroscopy, and photoluminescence measurements~\cite{masadehCdSeNano07}.

\begin{figure}[tbp]
    \centering
    \includegraphics[width=1\linewidth]{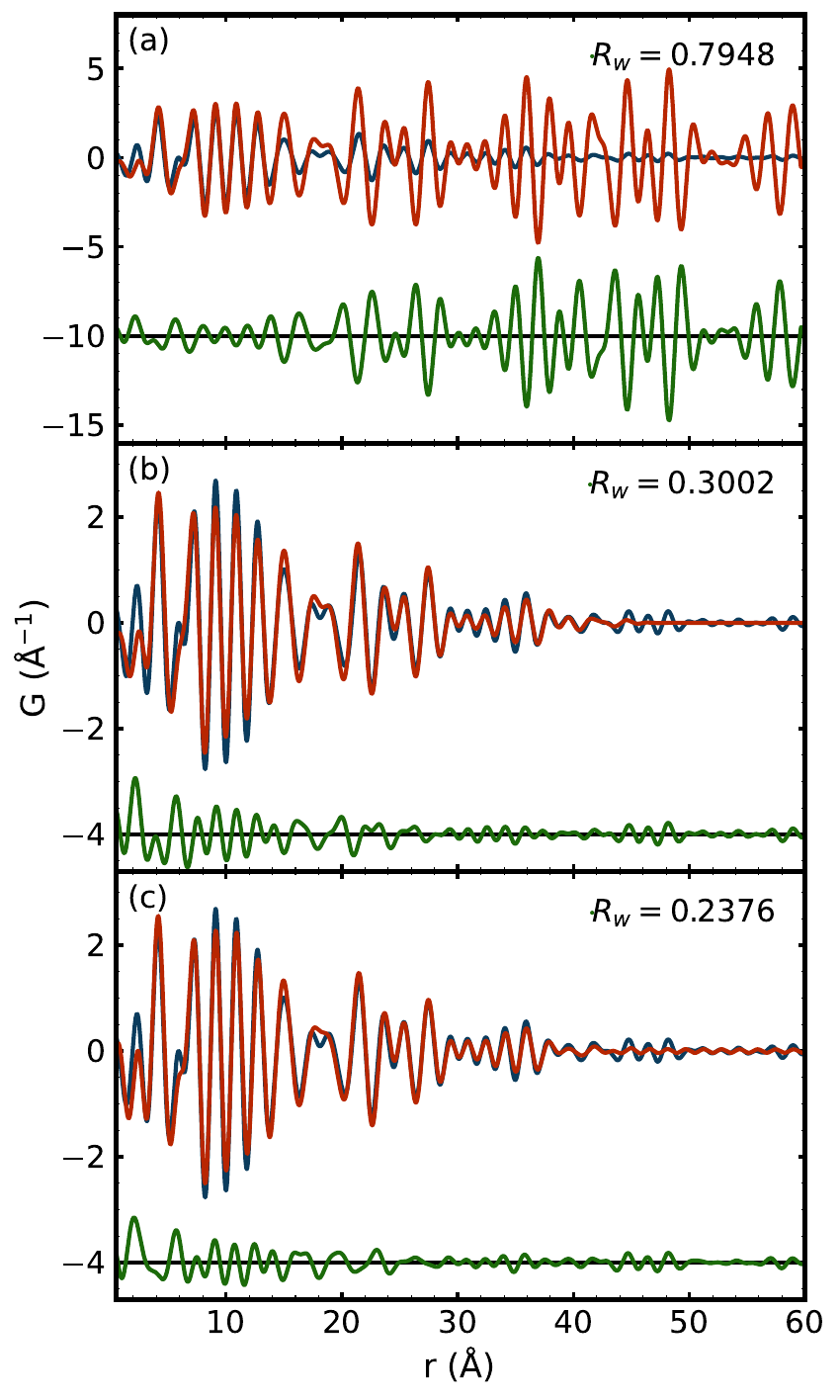}
    \caption{(a) Measured PbS nanocrystal PDF (blue) plotted against a computed PDF of the bulk PbS (red).  (b) The red curve is now the bulk PbS  PDF after applying the spherical nanoparticle \shape morph.  (c) The red curve is a PDF of a spherical PbS nanocrystal computed from the structural model using \pdfgui.
    A difference curve (green) is vertically offset below each plot.}
    \label{fig:shape-morph}
\end{figure}

Here, we have used this approach to estimate the size of PbS nanocrystals with thiostannate (Sn$_2$S$_6$$^{\text{4-}}$) ligands.  These nanocrystals are known to be faceted~\cite{guzelturkUltrafastSymmetryControl25}, but for small diameters, are approximately round. In \fig{shape-morph}(a), we show the PDF measured from the  PbS nanocrystal (blue) against a PDF computed from the bulk structure of PbS~\cite{nodaPbSCrystalStructure87} (red).
We see clear reduction in the signal of the measured data (blue curve) as we go to higher~\ir, indicating that our measured material is definitely nanocrystalline.
We then apply the nanoparticle \shape morph onto the computed bulk PDF. The resulting morphed PDF is shown in red in \fig{shape-morph}(b).  The radius obtained from the morph was $24.51$~\AA. 
For comparison, we used \pdfgui~\cite{farrowPDFfit2PDFgui} to compute a spherical PbS nanocrystal PDF from a structure~\cite{nodaPbSCrystalStructure87} to best fit our measured PDF.  
The resulting refinement resulted in a nanocrystal diameter of $47.72$~\AA\ (radius of $23.86$~\AA) (red curve in \fig{shape-morph}(c)).
The refinement done by \morph (order of seconds) is much faster than the lengthy and human intensive structural refinement done by \pdfgui, with a small relative difference of $1.3\%$ between the two calculated radii.

The \morph-computed and \pdfgui-computed PDFs (red curves in (b) and (c), respectively) also show good agreement up to $r=49$~\AA.  Above this, the \morph-computed PDF flat-lines, while the \pdfgui-computed PDF has small sinusoidal oscillations that account for termination ripples~\cite{egami;b;utbp12}.
This example showcases how \morph can quickly compute nanoparticle and nanocrystal PDFs from bulk PDFs and obtain morphological parameters.

At the time of writing, \morph only includes morphing transformations to account for the effects of a spherical and spheroidal nanoparticle shape on its measured PDF.
If the desired \shape morph is not implemented, please consider requesting it with an issue at the \morph GitHub website~\cite{diffpymorphsoftware25} or by posting a request at the diffpy-users Google group (\url{https://groups.google.com/g/diffpy-users}).
If you are able to implement the morph yourself, please consider posting a pull request at the \morph repository on GitHub.

\subsection{High-Throughput Experiment Use Cases}

\morph has been utilized in high  throughput experimental settings, such as synchrotron beam-times, for a variety of use-cases. 
These include to expose sources of experimental error (e.g. incorrect sample-to-detector distance or misaligned beam center), to identify and remove unwanted signals such as laser induced heating, and to identify optimal processing parameters to compute reliable PDFs from measured \iq data.

A major advantage of using \morph in a high-throughput data setting is its speed and ease of use. 
Rather than performing a lengthy structural refinement, one can obtain morphing parameters in seconds.

\subsubsection{Addressing experimental setup challenges}
\label{sec:exp-setup-mistakes}

When integrating a 2D diffraction image to obtain a 1D scattering function \iq, we require various experimental geometric parameters such as the sample-to-detector distance, location on the detector of the beam-center, tilt angles of the detector away from orthogonality, and wavelength of the incident beam.
These are normally handled by carefully performing calibration measurements. 
In practice, small deviations between calibration and sample datasets can arise due to a multitude of factors such as minor sample or detector repositioning, motor drift, thermal expansion or contraction of experimental components, and other instrumental or environmental effects.
In such cases, \morph can be used to apply small corrective adjustments to the calibration parameters.

\begin{figure}[tbp]
    \centering
    \includegraphics[width=1\linewidth]{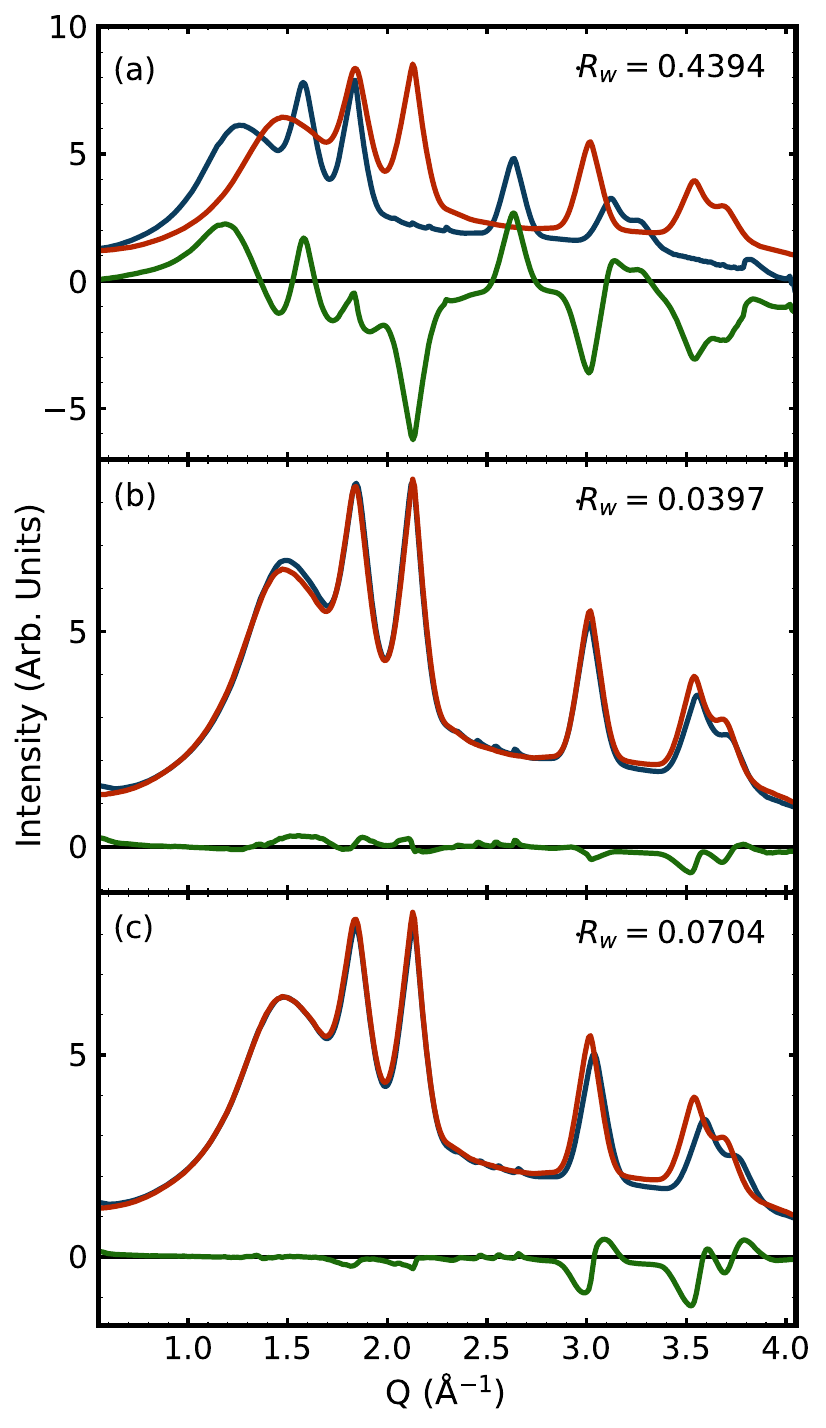}
    \caption{\iqs computed from the same diffraction image using the uncalibrated (blue) and calibrated (red) parameters are plotted in (a).
    In (b) \stretchmorph, \scale, and \shift morphs are applied to the uncalibrated data to maximize agreement with the calibrated data on $Q \in [0.55~\text{\AA},\;4.05~\text{\AA}]$.
    For (c), the same morphs are used, but instead maximizing agreement on the $Q$ region encompassing the first three peaks ($Q \in [0.55~\text{\AA},\;2.35~\text{\AA}]$).
    We include a difference curve (green) below each plot.}
    \label{fig:euc-p2}
\end{figure}

We prepared a sample of PbS nanocrystals with thiostannate (Sn$_2$S$_6$$^{\text{4-}}$) ligands dissolved in an amide solution.
The synthesis of the nanocrystals and ligands follows previous published procedures~\cite{jeongPbSLigandSamplePreparation24}.
A 2D diffraction pattern pattern was obtained from this sample at an XFEL beamline (measured to $\qmax=4.05$~\AA$^{-1}$).
In \fig{euc-p2}(a), we showcase two \iqs computed from this 2D diffraction pattern, but where the integration was carried out using different values for the sample-to-detector distances and beam centers.
The numbers used were rough initial guesses from the geometry as an initial calibration was not available.  This is compared to the final calibrated data (red curve in \fig{euc-p2}(a)). Specifically, the uncalibrated sample-detector distance value used was $D_{\mathrm{unc}} = 64.7$~mm with the calibrated distance found to be $D_{\mathrm{cal}} = 55.7$~mm.  
Additionally, the beam center used for the uncalibrated curve is offset $-2.26$~mm both horizontally and vertically from that found from the calibrated curve.
We can see a significant mismatch between the two curves, but by using the \stretchmorph and \scale morphs along with a ``\shift" morph, in (b), we can obtain very good agreement \fig{euc-p2}(b).  Here, the \shift morph shifts the function to the right by adding a shift constant $\Delta_h$ to the $x$-array to account for the zero-offset.
This helps to account for the misaligned detector center.

Such a morph allows us to make on-the-fly corrections for variability in sample calibration from data-set to data-set.
However, it is also possible to invert the morph parameters to find spectrum-dependent corrections to the calibration values.  

For the correction to the detector distance, we take advantage of the approximation that at
small $Q << \lambda/4\pi$,
\begin{align}
    \epsilon_D &= \frac{D_{\mathrm{cal}} - D_{\mathrm{unc}}}{D_{\mathrm{unc}}}\approx -\eta,\label{eq:delta-D-approx}
\end{align}
where $\eta$ is the \stretchmorph parameter as usual (proof in Supplemental Material, \sect{sup-squeeze}).
To see how well this approximation worked we applied the morphs to maximize agreement in only the lower-$Q$ region (the first three peaks) in \fig{euc-p2}(c).
The \stretchmorph parameter obtained in (c) was $\eta = 0.134$, which gives a correction to the sample-detector distance that is $<4\%$ away from the actual value of of $\epsilon_D = -0.139$.

\begin{figure}[tb]
    \centering
    \includegraphics[width=1\linewidth]{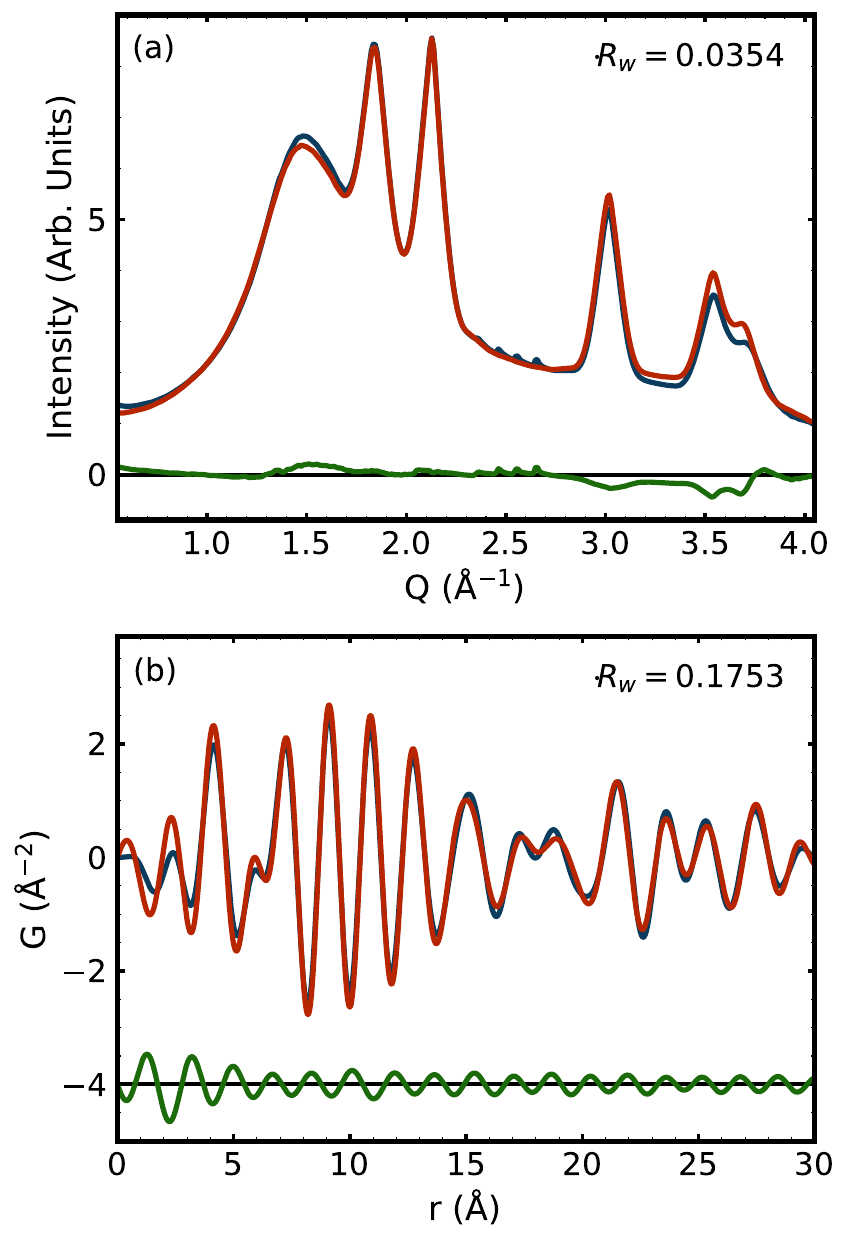}
    \caption{(a) \iqs computed from the uncalibrated but morphed data (blue) and calibrated (red) alignment parameters.
    (b) The PDFs computed from the two \iqs in (a).
    Difference curves (green) are vertically offset below.}
    \label{fig:euc-p2-squeeze}
\end{figure}

\morph includes a ``\squeeze\unskip" morphing transformation as well as the \stretchmorph and \shift that we already introduced.  The \squeeze morph applies a higher-order polynomial to the $x$-array (see \sect{general-x-morphs}).
In \fig{euc-p2-squeeze}(a), we applied a \squeeze using a fifth-order polynomial on the uncalibrated \iq (blue) to attempt to match the properly calibrated \iq (red).
The use of a fifth order polynomial in our \squeeze morph is motivated in the Supplemental Material, \sect{sup-squeeze}.
We can see by comparing \fig{euc-p2}(b,c) and \fig{euc-p2-squeeze}(a) that this quintic polynomial correction can, at the same time, account for the low-\q (near the $1.85$~\AA\ and $2.13$~\AA\ Bragg peaks) and high-$Q > 3.8$~\AA$^{-1}$ regions.
The resulting PDFs are brought well into alignment (\fig{euc-p2-squeeze}(b)).
As such, while the researcher's team is working to identify the new alignment parameters to recompute the \iq correctly, the researcher can still analyze this new data, draw conclusions, and inform the next steps of the experiment using the \morph analysis.


As we discuss in the next section, in principle, for those using a Python program for integration of 2D detector images, such as \pyfai~\cite{KiefferPyFAI25}, \morph can also help obtain the optimal alignment parameters directly through the ``\funcxy\unskip" morph.

\subsubsection{Obtaining data reduction parameters for an optimal PDF}
\label{sec:exp-get-pdf}

So far we have used \morph to carry out rather simple morphing transformations such as \stretchmorph, \shift  and \scale morphs.  Complexity increases with \squeeze that allows for non-linear stretches of the data.
However, in principle the morphing approach can be used for arbitrary transformations to the $x$- and $y$-arrays.  

Here we describe an example where we want to estimate the parameters used in processing raw 1D diffraction dataset all the way to a PDF.  
This processing is done using the \pdfgetxthree software package\cite{juhasPDFgetX3RapidHighly2013} which takes a number of parameters such as how much to scale a background signal, the range of data to use in the Fourier transform, and \emph{rpoly} which controls the flexibility of a polynomial correction done in the process.

\begin{figure}[tbp]
    \centering
    \includegraphics[width=1\linewidth]{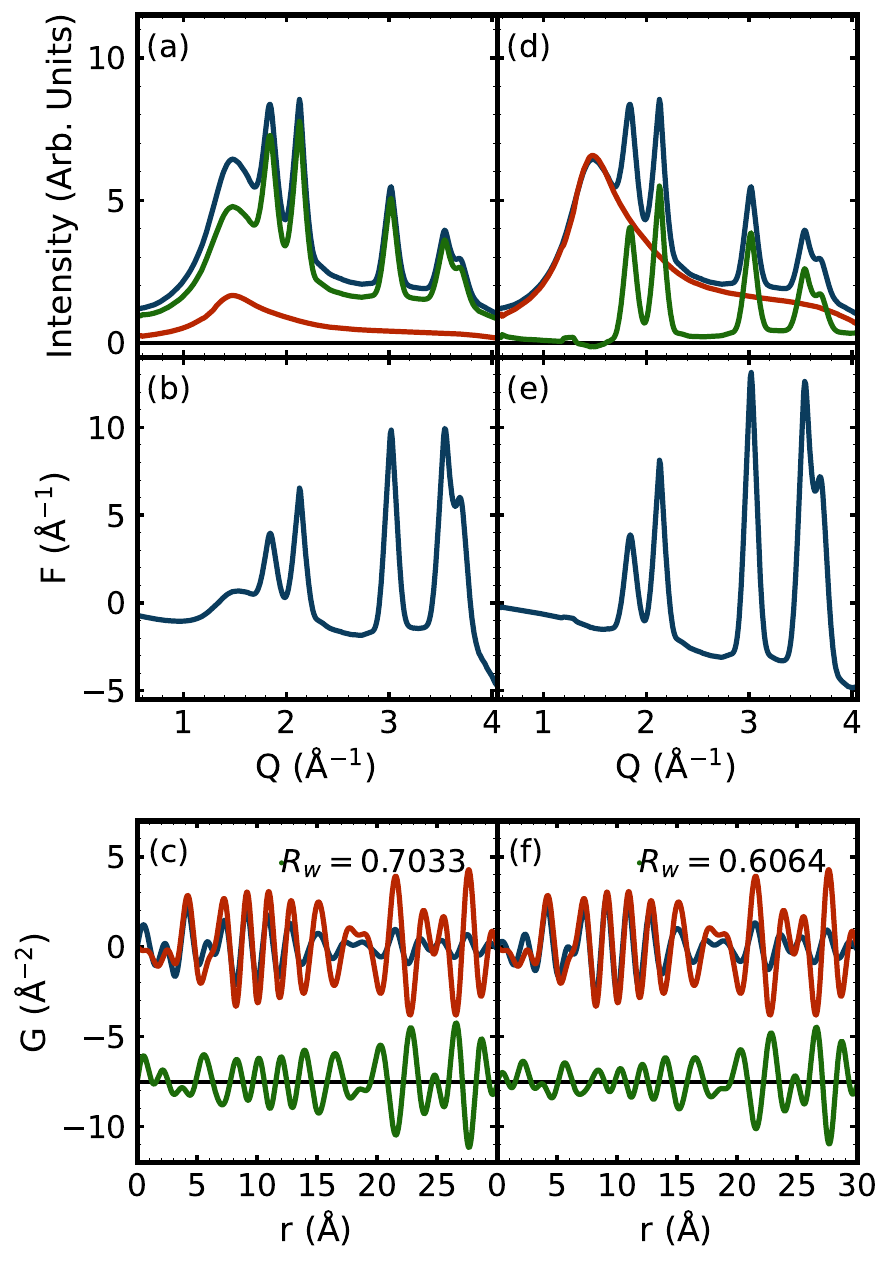}
    \caption{(a) and (d), the 1D scattering intensities of PbS nanoparticles dissolved in amide solution (blue), the raw signal from the empty solvent (red) and (green) the sample signal with the background subtracted. (b) and (e) show the resulting \fq functions obtained by running the green data-set from the panel above through \pdfgetxthree and the blue curves in (c) and (f) the resulting PDFs.  Parameters $rpoly=0.9$ and $bgscale=1$ were used to obtain the curves in (b) and (c) and $bgscale=3.95$ and $rpoly=1.78$ for the curves in (e) and (f). The latter numbers were obtained by \morph. In (c) and (f) the red curve shows the PDF calculated from a  bulk PbS  structure model~\cite{nodaPbSCrystalStructure87}. The green curve is the difference plotted offset below.}
    \label{fig:euc-p3}
\end{figure}

In this example, we have an \iq measured from PbS nanoparticles and a thiostannate salt dissolved in an amide solution (the same sample as \sect{exp-setup-mistakes}).
Our goal is to get a good PDF of the PbS after subtracting the correctly scaled signal from the solvent.
The issue is illustrated in \fig{euc-p3}(a) where the raw signal from the sample (blue curve), scaled background (red curve) and background subtracted signal (green curve) are shown.
After running this through \pdfgetxthree we obtain the \fq shown in \fig{euc-p3}(b).  This has the wrong shape because the background has been under-subtracted and \emph{rpoly} has not been optimized.  

A more ideal background subtraction and \fq is shown in \fig{euc-p3}(d) and (e), respectively.  This was done using the \funcxy morph in \morph and using a PDF calculated from a structural model of PbS as the target to morph on to (red curves in \fig{euc-p3}(c) and (f)).

\pdfgetxthree does the PDF conversion by calling the \texttt{PDFGetter()} functor (which behaves like a function).  This  takes in the input data and a set of parameters such as background scales, \textit{rpoly} and so on and returns out the calculated \fq and \gr.   
Here we used \texttt{PDFGetter()} as the user-supplied morphing function in the \funcxy morph in \morph, allowing it to vary the background scale and \emph{rpoly} parameters with a target function that was a PDF calculated from a structural model for PbS.

This shows how an arbitrarily complicated function can be used for the morphing transformation (here the input and outputs were actually functions in different spaces, reciprocal and real, respectively, and the function involves a polynomial fit and subtraction).  Although the morphing function being applied was complicated, the morph itself was not, in the spirit of the morphing philosophy, with only two well-controlled parameters being refined in such a way that real structural signal could not be lost.

The approach was very fast and allowed thousands of datasets to be rapidly processed each with its own optimized PDF transformation parameters.

The refinement process to obtain optimal PDF calculation parameters is particularly helpful for low \qmax data, especially for parameters like \emph{rpoly}~\cite{juhasPDFgetX3RapidHighly2013}.
A scientist familiar with the PDF method may be able to estimate a good \emph{rpoly} by seeing how it affects the overall shape of the $F(Q)$ function~\cite{juhasPDFgetX3RapidHighly2013}.
However, it is not always possible to measure over a wide enough $Q$-range to make good estimations~\cite{euxfelWorkshopReportQmax24,sapnikXFELHighQmax2025} as we encountered recently when probing the ultra-fast dynamics of a system at a recent x-ray free electron laser (XFEL) experiment. This approach may also be valuable for lab-based sources that may also have low accessible~\qmax.

\begin{figure}[tbp]
    \centering
    \includegraphics[width=1\linewidth]{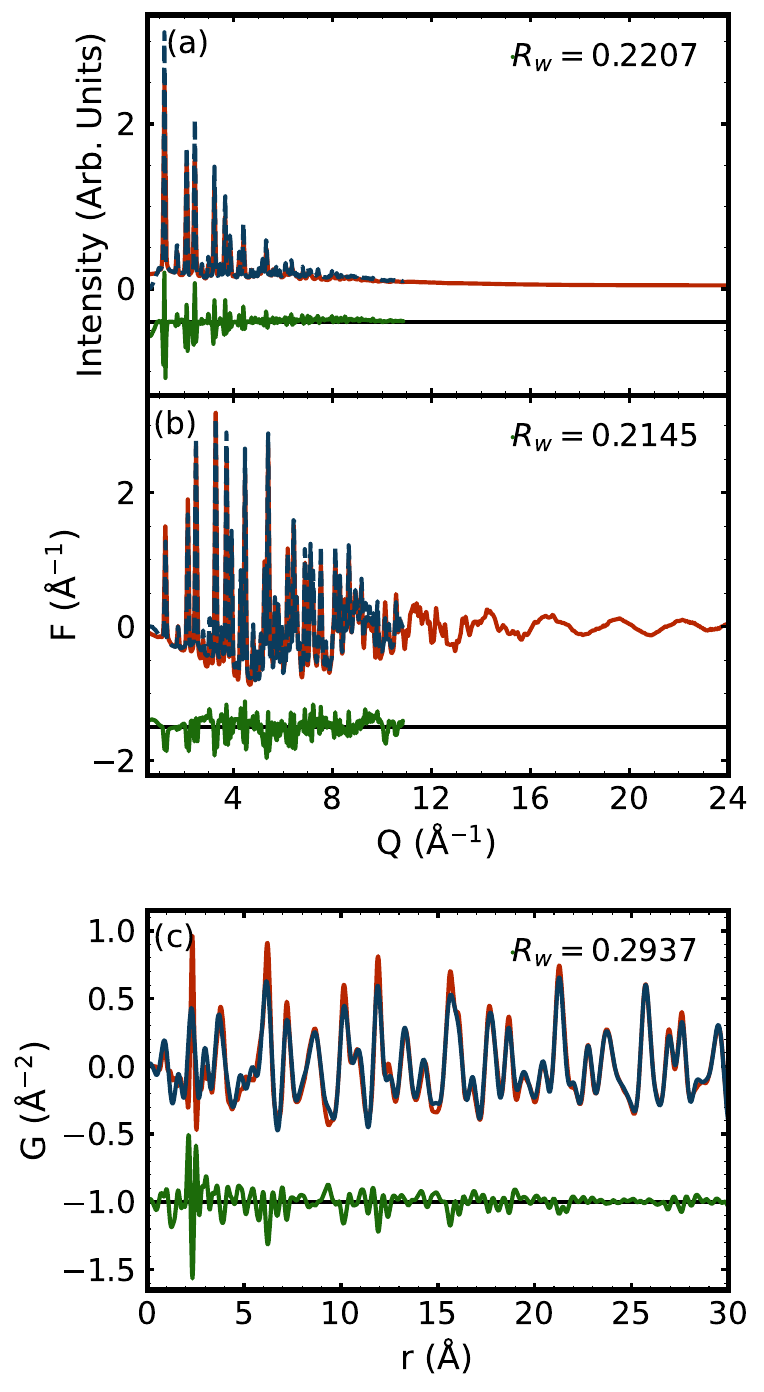}
    \caption{(a) The 1D scattering intensities of NSWS measured at the XFEL over a narrow \q~range (blue) and at the synchrotron over a wide \q~range (red). (b) The $F(Q)$ obtained from the XFEL $I(Q)$ using \pdfgetxthree and morphing (blue), and the target synchrotron $F(Q)$ (red). (c) The PDFs obtained by the morphed XFEL $F(Q)$ (blue) and target synchrotron $F(Q)$ (red).}
    \label{fig:xfel-synchrotron}
\end{figure}

Na$_{\text{2.9}}$Sb$_{\text{0.9}}$W$_{\text{0.1}}$S$_4$ (NSWS) samples were synthesized following previously published procedures~\cite{fuchsNSWSSamplePreparation19}.
In \fig{xfel-synchrotron}(a), we plot \iqs of the NSWS samples measured at at EuXFEL~\cite{deckingEuXFEL07} using the LPD detector at the FXE beamline~\cite{FXE_beamline_overview} (plotted in blue) and measured at the National Synchrotron Light Source II on beamline 28-ID-2 (plotted in red).
The XFEL measurements can be done at much higher time resolution compared to the synchrotron, at the cost of a lower \qmax.  
In (b), we have obtained $F(Q)$ for the synchrotron data (red) by manually tuning the \emph{rpoly} parameter until the large-$Q$ behavior of $F(Q)$ becomes oscillations about zero.
The largest $Q$ ($\qmax=10.85$~\AA) for the XFEL data is too small to carry out the same analysis, making it difficult to tune \emph{rpoly} that way.  
Instead, to obtain the blue curve in \fig{xfel-synchrotron}(b), we again wanted to use \texttt{PDFGetter()} from \pdfgetxthree in the \funcxy morph, but this time give it as a target function the \fq from the synchrotron experiment.  In this case, as well as applying the \texttt{PDFGetter()} corrections, it was necessary to apply a polynomial \squeeze morph due to calibration challenges (see \sect{exp-setup-mistakes}); these were applied together by defining a new Python function that wrapped \texttt{PDFGetter()}. Examples of how to do do this can be found in the \morph documentation.  The resulting PDF, plotted in (c), obtained from the morph-obtained XFEL $F(Q)$ is consistent with that obtained from the synchrotron $F(Q)$, albeit with slightly broader peaks due to the lower~\qmax.

These examples show how morphing provides a way for researchers to rapidly obtain PDFs that are as optimal as possible and process their data in a consistent manner.

\section{Morphing transformations}
\label{sec:morphing-transforms}

In this section we define the current set of morphs available in \morph and develop some theory behind some of the most widely used morph transformations.

In the general setup of morphing, there is a ``target'' signal and a ``morphed" signal.  
The target signal is unchanged, while the morphs are applied to the morphed signal.

The objective of the morphing process is to minimize the residual between the two signals by applying a series of simple ``morphing" transformations. 
This is like fitting, but it is model independent and it is designed simply to mimic physical effects, such as uniform lattice expansion or changes in thermal vibration, though it could be anything.

The signals could be any 1D functions or spectral signals such as a diffraction patterns, atomic pair distribution functions, but could also be Raman, NMR, or something else.

For simplicity we develop the theory with the spectra being PDFs with the understanding that the morphs can be applied to other signals.
Throughout this section, the unmorphed object PDF will be denoted $G(r)$, and the same PDF after morphing will be denoted as $G_m(r;\vec{p})$, where $\vec{p}$  denotes the set of morphing parameters.

\subsubsection{Scaling morph}
Scaling morphs correct the intensity of the morphed PDF by a constant $s$.  Scaling simply captures the fact that a measured signal might be on a different scale due to differences in the incident flux during measurements, instruments used, uncertainties in the data reduction process, or any factor that serves to change the scale of a signal.
The \scale morph can be expressed simply as
\begin{align}
\label{eq:morph_scale_def}
    G_m(r;s) = s \cdot G(r),
\end{align}
where $s$ is the \scale factor.

\subsubsection{Stretching morph}
\label{sec:stretch-morph}
Stretching morphs stretch the PDF linearly along the independent variable axis.
Stretching simulates the effect of a uniform lattice expansion of the structure when $\eta$ is positive, while $-1 < \eta < 0$ simulates a uniform lattice contraction.

The morph is done by linearly interpolating the $r$ grid on to a new stretched $r$-grid, $r'$.  
We chose linear interpolation as most PDF data are stored on dense grids, so errors introduced are small and computational speed is a premium.

In detail, given a \stretchmorph factor $\eta$, the stretched distance, $r'$ is given by
\begin{equation}
r' = \frac{r}{1 + \eta}.
\end{equation}

The resulting \stretchmorph morph is therefore given by
\begin{align}
\label{eq:morph_stretch_def}
    G_m(r;\eta) = G\left(\frac{r}{1 + \eta}\right).
\end{align}

A real uniform lattice expansion with thermal strain $\epsilon$ will also rescale the PDF on the $y$-axis by a factor $(1 + \epsilon)^{-2}$ (see Supplemental Material, \sect{sup-temp}).  Therefore, we recommend using a \scale morph any time a \stretchmorph morph is used to account for intensity changes due to isotropic expansion in the physical system.

\subsubsection{Smearing morph (general)}
\label{sec:smear}
Smearing morphs apply a Gaussian broadening factor to broaden a signal. For example, this will mimic different levels of thermal motion in PDF data (\sect{smear-pdf}).
To accomplish this we carry out a convolution of the signal with a Gaussian function whose width can be varied.

The Gaussian broadening factor $T_\sigma(r)$ has the form
\begin{align}
\label{eq:morph_smear_gaussian_def}
    T_\sigma(r) = \exp \left\{ -\frac{1}{2} \left( \frac{r-r_c}{\sigma} \right)^2 \right\},
\end{align}
where $\sigma$ is the \smear factor, and $r_c = \frac{1}{2}(r_{max} + r_{min})$ is the center of the $r$-grid. 
The choice of $r_c$ is arbitrary, but centering the convolution Gaussian in the middle of the $r$-grid minimizes edge effects in the numerics of the resulting convolution.

The \smear morph then convolves the morphed signal with the broadening function $T_\sigma(r)$ according to
\begin{align}
\label{eq:morph_smear_def}
    R_m(r;\sigma) = \frac{(R \circledast T_\sigma)(r + r_c)}{\sqrt{2\pi\sigma^2}}.
\end{align}
This serves to broaden all peaks in \rr; specifically, the variance of each Gaussian peak is increased by $\sigma^2$ after convolution.  
The $r + r_c$ shifts the smeared curve to ensure its centroid matches that of \rr, and the normalization factor $1/\sqrt{2\pi\sigma^2}$ ensures the two curves have the same integrated magnitudes.  Proof that this operation indeed broadens the peaks of $R(r)$ by $\sigma^2$ while maintaining the centers and integrated magnitudes of each peak can be found in Supplemental Material, \sect{sup-convolution}.

Note that $T_\sigma(r) = T_{-\sigma}(r)$, so a \smear parameter of $+\sigma$ is equivalent to a \smear parameter of $-\sigma$.

\subsubsection{Smearing morph (PDF-specific)}
\label{sec:smear-pdf}

The \smear morph described above can be used to smear any signal.
When that signal is a measured \gr function, it is preferred to convert \gr to the radial distribution function (RDF~\cite{egami;b;utbp12}) before applying the \smear. As such, \morph contains a modified version of the \smear morph, called \smearpdf.
This is because atomic motion acts to broaden peaks in the RDF, \rr, and not the PDF \gr.

The differences are small and an approximately correct result can be obtained by smearing \gr directly. 
However, for more accurate results when dealing with PDFs, a user may use the \smearpdf morph instead of the \smear morph.   At the command line, this is done by using the \texttt{--smear-pdf} option instead of the \texttt{--smear} option.  When using \morph in a Python script, you would use the \texttt{smear\_pdf} parameter instead of \texttt{smear}.  If both \smearpdf and \smear options are used, the program ignores the \smear input and applies the \smearpdf morph only.

For bulk materials, the RDF is related to the PDF by~\cite{farrowBillingeSasPdf09}
\begin{align}
    G(r) &= \frac{R(r)}{r} - 4\pi r \rho_0,
\end{align}
where $\rho_0$ is the atomic density of the material, and $-4\pi\rho_0$ may be referred to as the ``baseline slope".
When \smearpdf is used, the input functions are assumed to be PDFs. The program will first convert it to an RDF, \rr, using either a user-given baseline slope or a best-fit slope (if the user is unable to provide one).
The smear morph then applies a Gaussian broadening factor $T_\sigma(r)$ to \rr.
The morphed $R_m(r)$ is then converted back to a PDF which is then returned as the morphed PDF.

\subsubsection{Shifting Morph}

\label{sec:shift-morph}
Shifting morphs shift the signal horizontally or vertically.  They may be needed to account for constant backgrounds or calibration errors such as a shifted zero-point in a diffraction pattern. Shifting is also useful for data visualization techniques (e.g.\ waterfall plotting).

Given a user-specified horizontal \shift parameter $\Delta_h$ and vertical \shift parameter $\Delta_v$, the \shift morph is expressed simply as
\begin{align}
    G_m(r) &= G(r - \Delta_h) + \Delta_v.
\end{align}

\subsubsection{Nanoparticle shape morph in the PDF}

PDFs measured from a nanoparticle and bulk sample of the same material can look quite different as the particle's shape plays a role in shaping the pair distribution function.  One can approximate the PDF of a nanoparticle $G_{nano}(r)$ from the PDF of a bulk $G_{bulk}(r)$ of the same material by the simple relation~\cite{farrowBillingeSasPdf09}
\begin{align}
    G_{nano}(r) &\approx \gamma_0(r)G_{bulk}(r).\label{eq:nano_bulk_rel}
\end{align}

The \shape and inverse \shape morphs in \morph were designed to compare bulk and nanoparticle PDFs using the nanoparticle characteristic function, $\gamma_0(r; \vec{p})$.  This characteristic function depends on~$r$ and on some parameters $\vec{p}$.

The \shape morphs take input $G(r) = G_{bulk}(r)$ and simply multiply them by the characteristic function, $\gamma_0(r; \vec{p})$,
\begin{align}
    G_m(r;\vec{p}) &= \gamma_0(r; \vec{p}) \cdot G(r).
\end{align}

The inverse \shape morphs take input $G(r) = G_{nano}(r)$ and divide the characteristic function,
\begin{align}
    G_m(r) &= \begin{cases}
        \frac{G(r)}{\gamma_0(r; \vec{p})} & \gamma_0(r; \vec{p}) \neq 0\\
        0 & \gamma_0(r, \vec{p}) = 0
    \end{cases}.
\end{align}

Currently, \morph has nanoparticle characteristic functions for spheres~\cite{kodamaFiniteSizeEffects06} and spheroids~\cite{leiIntrinsicGeometry09}.  The spherical nanoparticle characteristic function, $\gamma_{0,sph}(r; s)$, refines the radius, $s$, of the nanoparticle. The spheroidal nanoparticle characteristic function, $\gamma_{0,srd}(r; s_e, s_p)$, refines the equatorial and polar radii, $s_e$ and $s_p$, respectively.

\subsubsection{General ordinate morph}

In addition to the morphs introduced above, which are designed to account for specific physical effects,  \funcy provides a flexible interface for applying user-defined transformations to the $y$-axis data, enabling comparison of signals influenced by experiment- or system-specific distortions.
The definition of the \funcy morph is,
\begin{align}
G_m(r;f,\vec{p}) = f(r, G(r); \vec{p}),\label{eq:funcy-morph}
\end{align}
where \( f \) is a user-supplied function in the form of a Python function, and $\vec{p}$ represents its parameter dictionary.
The user simply defines a Python function that takes the signal $r$ and $G(r)$ arrays (or equivalently, any $x$ and $y$ arrays) and a  dictionary of parameters then returns the morphed $x$ and $y$ arrays.
This function is passed to the \funcy interface along with the morphed and target functions and the parameters are refined to get a good agreement.

\subsubsection{General grid (abscissa) morphs}
\label{sec:general-x-morphs}

Similarly it is possible to apply user-defined morphs that transform the $x$-axis of the function, generalizing the \stretchmorph morph.  Below, we describe two morphs: \squeeze and \funcx. For reasons discussed at the end of this (sub)subsection, \squeeze should be used when the expected change in the x-axis grid is small, while \funcx could be used otherwise.

\squeeze applies a polynomial transformation to the $x$-axis, where the user can control the order of the polynomial.  This offers a flexible approach for correcting non-linear distortions in signals.
A zero'th order polynomial is actually the \shift morph and a first-order polynomial recovers the behavior of \shift and \stretchmorph morphs applied together.  But the \squeeze interface generalizes this to higher orders.

These morphs have been found useful to account for geometric artifacts introduced by 2D detectors, such as module tilts, curved detection planes, or angle-dependent offsets, as well as structural effects intrinsic to the sample.  Its use was introduced in \sect{exp-setup-mistakes}.

The transformation is defined as
\begin{align}
\Delta r(r; a_0, \cdots, a_n) = a_0 + a_1 r + a_2 r^2 + \cdots + a_n r^n,
\end{align}
where \( \{a_0, a_1, \ldots, a_n\} \) are the polynomial coefficients.  The desired order, $n$, of the polynomial to use is a user defined input.
This is applied as
\begin{align}
r'(r; a_0, \cdots, a_n) = r + \Delta r(r; a_0, \cdots, a_n),\label{eq:squeeze-grid}
\end{align}
resulting in a modified coordinate grid
with the morphed signal evaluated as
\begin{align}
G_m(r') = G(r).\label{eq:grid-morph}
\end{align}
Given a morph and target function, \morph will find the best set of $a_0, \cdots, a_n$ to apply in order to match the target.

To compare to previously-described morphs, if all $a_i = 0$ except $a_0 \neq 0$, \squeeze is equivalent to the horizontal \shift morph with $\Delta_h = a_0$ (\sect{shift-morph}), and if instead all $a_i = 0$ except $a_1 \neq 0$, \squeeze is equivalent to a \stretchmorph morph with $\eta = a_1$ (\sect{stretch-morph}).

The current implementation of the \squeeze morph takes the original function as a two-column array $[[r], [G(r)]]$, where the notation $[\cdot]$ indicates the quantity is an array. Then, \squeeze computes $[r']$ using \eq{grid-morph} to get $[[r'], [G_m(r')]]$.  Finally, \squeeze reinterpolates $G_m(r')$ back onto the original grid $[r]$ and returns the squeezed function interpolated on to the original grid, i.e., $[[r], [G_m(r)]]$. 
The reinterpolation is done for consistency and further processing. 
This is computed using a cubic spline interpolation.
Extrapolation is allowed outside the original domain where needed (and the user will be notified if and where the extrapolation occurs).  Care should be taken by the user if the morph results in a large extrapolation.

The \squeeze morph will warn the user if the polynomial squeeze results in a modified grid, $[r'(r)]$, that is not monotonically increasing.
Non-monotonicity of $r'$ poses a problem in our procedure above, as $G_m(r')$ may no longer be a single-valued function.  This means there is no longer a unique way to interpolate $[[r'], [G_m(r')]]$ onto the grid $[r]$. We resolve this by modifying $G_m(r')$ to be a single-valued function by first identifying all duplicated grid points $r'_{dup}$ in the array $[r'(r)]$;  i.e. $r'_1 = r'_2 = \cdots r'_s = r'_{dup}$ for $s \geq 2$. We then set
\begin{align}
    G_m(r'_{dup}) &= \sum_{i=1}^s G_m(r'_i),
\end{align}
giving $G_m(r')$ a single unique value for each grid point $r'_{dup}$ where it may be multi-valued.
This procedure can result in a $[[r], [G_m(r)]$ that looks vastly different than $[[r], [G(r)]]$ if the function $G_m(r')$ is not smooth or the length of the non-monotonic intervals is large, but seems to work well for small extrapolations that are the most common.

In general, we advise users to avoid a non-monotonic \squeeze. We have found success by (1) reducing the order of the polynomial and (2) first performing safe morphs like \shift and \stretchmorph to get estimates of $a_0$ and $a_1$, and giving those to \squeeze as the initial parameter values.

While \squeeze can only apply polynomials to the grid, the \funcx, in analogy with \funcy, morph allows arbitrary functions. The morphed signal is similarly given by \eq{grid-morph}, though the new grid $r'$ is now given by
\begin{align}
    r'(r; f, \vec{p}) &= f(r, G(r); \vec{p}).\label{eq:funcx-morph}
\end{align}
Here, $f$ is a user-provided Python function, and $\vec{p}$ is a dictionary of function parameters to refine over. The function $f$ should take in both $r$ and $G(r)$ (equivalently, $x$ and $y$) and return the morphed grid $r'(r;\vec{p{}})$ based on the parameters $\vec{p}$.

The implementation of the \funcx morph is slightly different than that of \squeeze. After taking in $[[r], [G(r)]$, we simply return $[[f(r, G(r); \vec{p}])], [G(r)]]$. 
By design, no interpolation on to the original grid is done in this case.  This choice allows arbitrary changes in the independent array, permitting, for example, functions that transform into different spaces, such as the \texttt{PDFGetter()} transform described in \sect{exp-get-pdf}.

The advantage to \funcx is apparent when the morph and target function grids are extremely different as \squeeze could result in strange and uncontrolled behavior. 
However, this means the grid of the morphed function is constantly changing. This makes computing the residual between the morph and target functions difficult, and the calculation of the difference curve can involve varying amounts of interpolation error each iteration of the refinement. As such, we still recommend the \squeeze morph for small changes to the grid as it operates on a single unchanging grid.

\subsubsection{General function morph}
\label{sec:funcxy}

While a user can use the \funcx and \funcy morphs together to change both the grid and data simultaneously, we have defined a \funcxy morph for convenience. This morph is designed to be compatible with PDF calculators such as the \texttt{PDFGetter()} functor in \pdfgetxthree~\cite{juhasPDFgetX3RapidHighly2013}, so one can refine the PDF calculation from a measured intensity function.  This was described in \sect{exp-get-pdf}.

For this morph, the user supplies a single function $\vec{f}$
\begin{align}
    \vec{f}(r, G(r); \vec{p}) &= \left(f_0(r, G(r); \vec{p}), f_1(r, G(r); \vec{p})\right)
\end{align}
that takes in the arrays $r$ and $G(r)$ as well as a dictionary of parameters $\vec{p}$. The function must return two quantities: (a) the desired morph on the grid $f_0$ and, (b) on the function values $f_1$. In Python, this can be done as a tuple $(f_0, f_1)$ or as two comma-separated quantities $f_0, f_1$.

The \funcxy morph is then described by
\begin{align}
    G_m(r'; f_1, \vec{p}) &= f_1(r, G(r); \vec{p}),
\end{align}
where
\begin{align}
    r'(r; f_0, \vec{p}) &= f_0(r, G(r); \vec{p}).
\end{align}
One can see the obvious parallels to Eqs.~\ref{eq:funcy-morph},\ref{eq:funcx-morph}.

\subsubsection{Chaining together multiple morphs}
All morphs (\scale, \stretchmorph, \smear, \smearpdf, \shape, \squeeze, \funcx, \funcy, \funcxy, etc.) can be chained together with other morphs to incorporate multiple corrections.  To do so, a user simply has to enable the morphs they want chained.  

There are some limits to this and the user should use good sense.  In some cases the code will prevent clashes, for example, only one of the \smear and \smearpdf morphs can be used.  If the user enables both on accident only the more specific \smearpdf morph will be done.  Also, since a polynomial \squeeze morph of order zero or higher includes the horizontal \shift morph, the horizontal \shift morph will be disabled when the \squeeze morph is enabled (the vertical \shift morph is not affected).  For the same reason, if the polynomial \squeeze is of order one or higher, the \stretchmorph is disabled.

Users enable morphs by specifying them and initializing the parameter values.  Additionally, there are convenience commands to fix and free previously specified morphs.  Detailed usage can be found in the documentation of the software.

When multiple morphs are specified they are applied a pre-defined order. 
Since we found the optimizer to perform better by applying more complex morphs first, the default order in which enabled morphs are applied is (at the time or writing): \funcxy, \funcx, \funcy, \squeeze, \scale, \stretchmorph, \smear or \smearpdf, \shift (horizontal and vertical are refined simultaneously), then \shape.  This order may change in the future.

\subsubsection{Regression targets}
\label{sec:regression-targets}

For all morphs, the morphing process uses least squares regression to minimize the difference between the morphed and target signals by varying morphing parameters where \rw is used as the objective function.

If the morph and target functions are not on the same grid, the two functions are first interpolated (using linear interpolation) onto a common grid before computing \rw.  The interval of this common grid is the intersection of the two original grid intervals (i.e. if the morphed grid is on the interval $[a, b]$ and the target grid is on the interval $[c, d]$, then the new grid interval is $[\max\{a, c\}, \min\{b, d\}]$).  If the spacing (the distance between consecutive grid points) between the two grids is different, then the common grid uses the larger of the two spacings.  If a grid has uneven spacing (the distance between consecutive grid points vary), then its spacing is defined as the average distance between consecutive grid points.


\section{\morph Software}
\label{sec:morph-software}

\morph is an open-source Python software package that runs on Windows,  MacOS, and Linux operating systems as part of the \href{https://www.diffpy.org/}{\diffpy project}~\cite{diffpyWebsite}.
It can be installed from \href{https://conda-forge.org/}{conda-forge}~\cite{morphCondaForgeRelease25} using conda (recommended) or from \href{https://pypi.org/}{PyPI}~\cite{morphPyPiRelease25} using pip.
The source code is available on GitHub at \url{https://github.com/diffpy/diffpy.morph}.  

The \morph program can be used both using a command line interface (e.g. a bash terminal) or in Python scripts directly. 

The program takes in two input files: the file containing the function to be morphed and the file containing the target function. Readable files contain the data as a multi-column table, where one column contains the x-array and another column has the function values.
This uses the \texttt{loadData()} method in \utils~\cite{diffpyutilssoftware25}.  By default when \texttt{loadData()} encounters two columns it assumes the first column is the $x$- and the second column is the $y$-array but this behavior can be overridden.  
\texttt{loadData()} will skip over header information or can be used to read and return header information. 

The user also provides a set of morphs to apply (e.g. \scale, \stretchmorph, and \smear).
The program then carries out the morphs described above, returning the optimized morphing parameters.  It contains capabilities to plot the morphed function and the reference function with a difference function below, similar to figures in this paper. If requested, the program also returns the morphed function interpolated to the grid of the reference function.

The package also has an API allowing morphs to be called from within Python scripts.

Tutorials for the software and additional code documentation are available in the \morph documentation at \url{https://diffpy.org/diffpy.morph/}.

\section{Conclusion}

We demonstrate morphing as a model-independent tool for comparing 1D spectral data with worked examples that show how useful information may be extracted from diffraction and PDF data.
As \morph uses no model/structural refinement, it is fast, making it ideal for on-the-fly analysis in high-throughput data settings such as a synchrotron experiment.

While initially developed for x-ray PDFs, our examples show that the technique also works for 1D diffraction patterns and neutron diffraction data. Since
\morph is model-independent, it can be used on other 1D functions (e.g. density of states, Raman spectra, flattened image data), though it has not been extensively tested beyond powder diffraction patterns and PDFs.

\section{Acknowledgements}

Algorithm and software development, paper writing, as well as Billinge-group involvement in the experiments and analysis, was supported by
U.S. Department of Energy, Office of Science, Office of Basic Energy Sciences (DOE-BES)
under contract No. DE-SC0024141.
Initial development (supporting PJ and CF) of \morph was done under the Distributed Analysis of Neutron Scattering Data (DANSE) project funded by the US National Science Foundation (NSF) under DMR-0520547. Further developments by CHL were supported by the Next Generation Synthesis Center (GENESIS), an Energy Frontier Research Center funded by the U.S. Department of Energy, Office of Science, Basic Energy Sciences under Award Number DE-SC0019212.

X-ray PDF measurements of the Iridium Telluride samples were conducted on beamline 28-ID-2 of the National Synchrotron Light Source II, a U.S.  Department of Energy (DOE) Office of Science User Facility operated for the DOE Office of Science by Brookhaven National Laboratory under Contract No.  DE-SC0012704.  This contract also supported CP's synthesis of the IrTe$_2$ and Ir$_{\text{0.8}}$Rh$_{\text{0.2}}$Te$_2$ samples. Crystal growth of the Ir$_{\text{0.95}}$Pt$_{\text{0.05}}$Te$_2$ samples by HDZ was supported by the National Science Foundation Grant No. DMR-2003117.
ESB acknowledges funding provided by the Institute of Physics Belgrade, through the grant by the Ministry of Science, Technological Development and Innovation of the Republic of Serbia.

X-ray PDF measurements of the PbS nanocrystals were taken at the SACLA beamline~2 at SPring-8 faclility in Hyogo, Japan.  NSG acknowledges support for the data acquisition by US-DOE-BES Contract No. DE-SC0019375 and a David and Lucile Packard Foundation Fellowship for Science and Engineering.
VRKW's involvement in the experiment was supported by NSF Graduate Research Fellowship No. DGE1106400.
AJ and DVT acknowledge support from the U.S. Department of Energy, Office of Science, Basic Energy Sciences, Materials Sciences and Engineering Division, under grant DE-SC0025256.
AJ was also partially supported by the Kwanjeong Educational Foundation.
We would like to thank Yuya Kubota and Jungmin Kang, Tadashi Togashi, Taito Osaka, and Kensuke Tono for help with the data acquisition. 

The NSWS experiments were carried out at the European XFEL in Schenefeld, Gemany at beamline FXE under proposal \#8015.  The authors would like to thank Paul Evans, Peter Zalden, and Christopher Milne for their help with these experiments, and Sanjit Ghose for help with collecting synchrotron data on these samples at bealmine 28-ID-2 at NSLS-II (funding noted above). LN would like to thank the Natural Sciences and Engineering Research Council of Canada (NSERC) for funding through a platform Discovery Grant and a Canada Research Chair.

The Neutron PDF measurements of YSZ used resources at beamline BL-1B (NOMAD) at the Spallation Neutron Source, a DOE Office of Science User Facility operated by the Oak Ridge National Laboratory. Special thanks is given to the SNS staff, Joerg Neuefeind, Jue Liu, Emily Van Aucken, and Dante Quirinale for aid in data collection and reduction as well as experiment design and planning. Thanks is also given to Margarita Petrova from the Page group for aiding in data collection, sample preparation, and data reduction.

\setcounter{section}{0}
\setcounter{figure}{0}
\setcounter{table}{0}
\setcounter{equation}{0}

\renewcommand{\thesection}{S\Roman{section}}
\renewcommand{\thefigure}{S\arabic{figure}}
\renewcommand{\thetable}{S\arabic{table}}
\renewcommand{\theequation}{S\arabic{equation}}

\section*{Supplemental Material}
\subsection{Peak shifting in absence of a phase transition}

\label{sec:figs}


In \sect{identify-phase-transition}, we showed two PDFs of Ir$_{\text{0.8}}$Rh$_{\text{0.2}}$Te$_2$ measured at $250$~K and $300$~K and claimed the peaks of the higher-temperature PDF are shifted (to the right) of the lower-temperature PDF peaks.

If this is the dominant change, the difference between a function $f(x)$ and a slightly right-shifted function $f(x + \Delta x)$ should be correlated with the derivative $f'(x)$ since
\begin{align}
    \Delta x f'(x) &\approx f(x + \Delta x) - f(x).\label{derivative-shift-relation}
\end{align}
to first order in $\Delta x > 0$.

\begin{figure}[tbp]
    \centering
    \includegraphics[width=1\columnwidth]{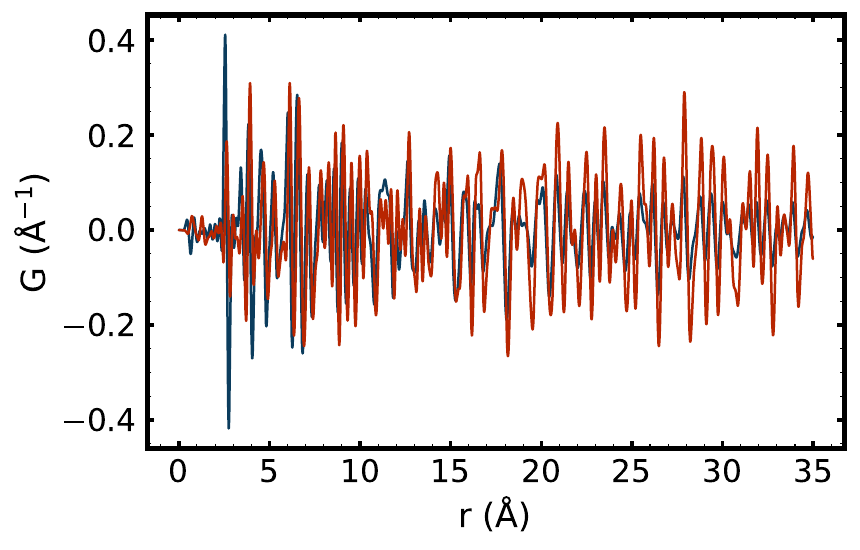}
    \caption{The difference curve (red) between the PDFs of Ir$_{\text{0.8}}$Rh$_{\text{0.2}}$Te$_2$ measured at $250$~K minus the PDF measured at $300$~K is plotted against the derivative (blue) of the $250$~K PDF.  The derivative has been rescaled by the grid spacing $\Delta r$.  The two curves are visually correlated, especially at high-\ir. }
    \label{sup-fig:grad-diff-250K-300K}
\end{figure}

Therefore, if our claim that the peaks of the $300$~K PDF $g_{300}(r)$ are shifted from the peaks of the $250$~K PDF $g_{250}(r)$ is valid, we expect to see correlation between $\Delta g(r) = g_{300}(r) - g_{250}(r)$ and $g'_{250}(r)$.  In Fig.~\ref{sup-fig:grad-diff-250K-300K}, we have plotted $\Delta g(r)$ (in red) and $\Delta r \cdot g'_{250}(r)$ (in blue), where $\Delta r$ is the grid spacing for $g_{250}(r)$.
The gradient was computed using forward differences and rescaled by $\Delta r$ for comparison. The two curves are visually correlated in the high~\ir region supporting the hypothesis that the difference curve originates from a shift of the signal to higher-\ir.

\begin{figure}[tbp]
    \centering
\includegraphics[width=1\columnwidth]{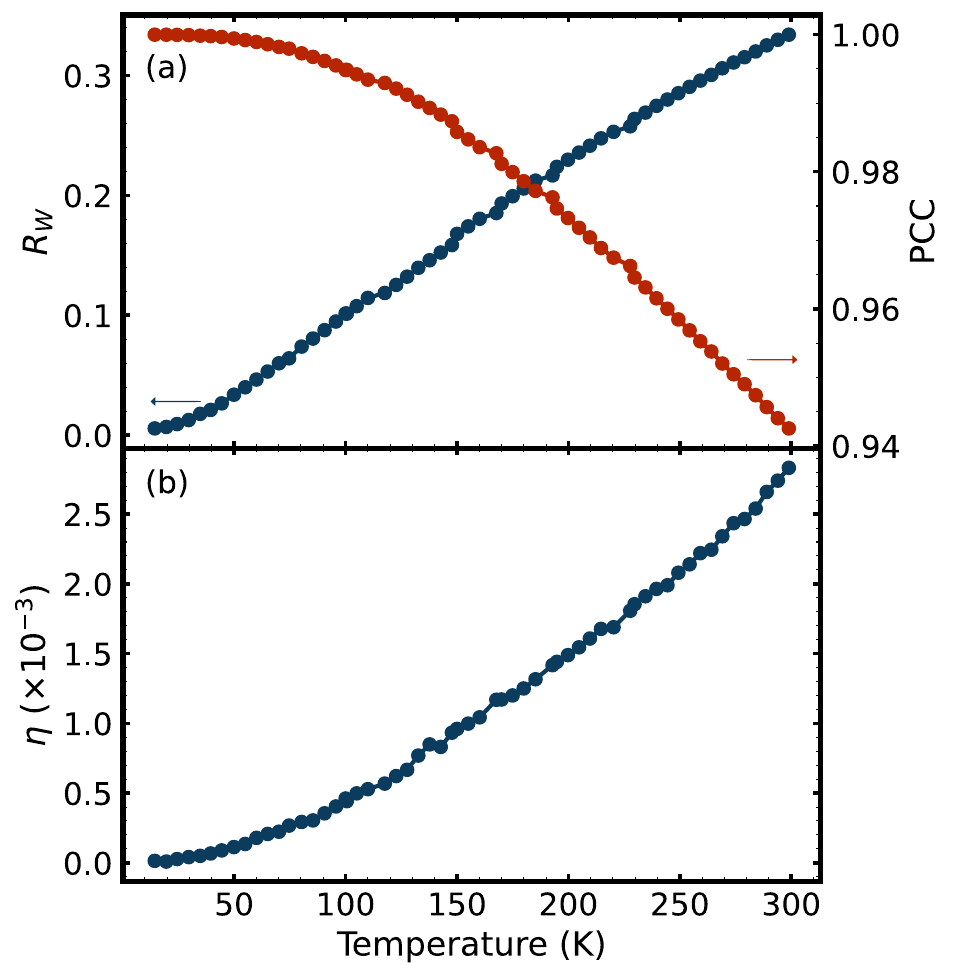}
    \caption{(a) \rw (blue) and PCC (red) from morphing the $9$~K PDF onto PDFs at each temperature point for Ir$_{\text{0.8}}$Rh$_{\text{0.2}}$Te$_2$ measured with x-rays at temperatures ranging from $9$~K to $299$~K. (b) The resulting \stretchmorph parameter $\eta$ used in each morph. The curves are all relatively smooth when compared to \fig{IrTe2-All}, indicating no interesting structural change occurs across this temperature range.}
    \label{sup-fig:Ir0.8Rh0.2Te2-All}
\end{figure}

\begin{figure}[tbp]
    \centering
\includegraphics[width=1\columnwidth]{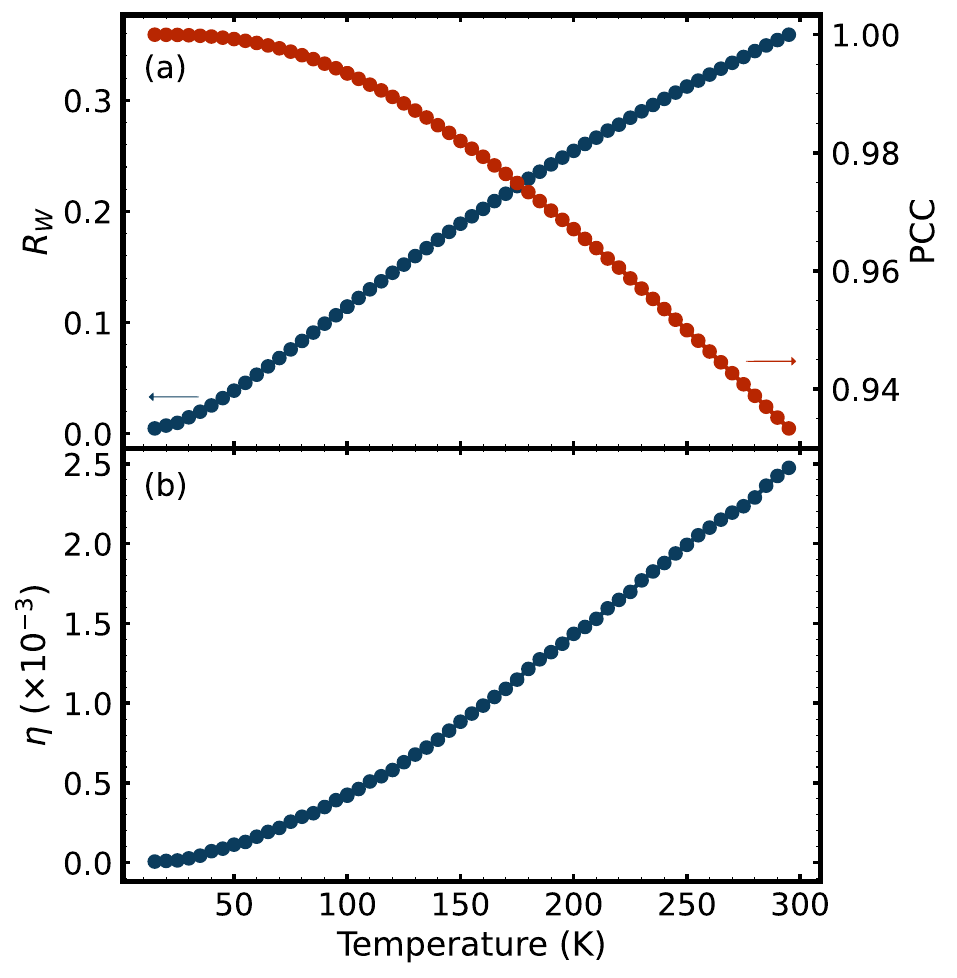}
    \caption{(a) \rw (blue) and PCC (red) from morphing the $10$~K PDF onto PDFs at each temperature point for Ir$_{\text{0.95}}$Pt$_{\text{0.05}}$Te$_2$ measured with x-rays at temperatures ranging from $10$~K to $295$~K. (b) The resulting \stretchmorph parameter $\eta$ used in each morph.  Like in Fig.~\ref{sup-fig:Ir0.8Rh0.2Te2-All}, we see no interesting structural change in this temperature range.}
    \label{sup-fig:Ir0.95Pt0.05Te2-All}
\end{figure}

In Fig.~\ref{fig:IrTe2-All} we showed what plots of \rw, PCC, and $\eta$, the stretch parameter, look like in the case where a sample is going through a phase transition. In that case, the sample was IrTe$_2$.  Here we show the same plots for cases where no phase transformation is occuring. Figs.~\ref{sup-fig:Ir0.8Rh0.2Te2-All} shows the case for 
Ir$_{\text{0.8}}$Rh$_{\text{0.2}}$Te$_2$ and \ref{sup-fig:Ir0.95Pt0.05Te2-All} the case for Ir$_{\text{0.95}}$Pt$_{\text{0.05}}$Te$_2$.
In both plots, we see a smooth change in the \stretchmorph parameter, indicating no structural change.
This is supported by (a) of these figures, where the \rw and PCC for each morph is plotted, also as functions of temperature.  They too smoothly vary as a function of temperature, with the \rw rising and PCC falling.

Because there is no structural phase transition Ir$_{\text{0.8}}$Rh$_{\text{0.2}}$Te$_2$ and Ir$_{\text{0.95}}$Pt$_{\text{0.05}}$Te$_2$ from $10$~K to $300$~K, we expect a majority of the difference curve signal between PDFs of these samples measured at these temperatures to be from heating effects.  These can be corrected for and removed by \morph, as shown in \fig{10-300-Morph}.

\subsection{Derivation of the effects on the PDF of a uniform lattice expansion}
\label{sec:sup-temp}

To begin, we define the radial distribution function (RDF).  For atom $i$, its partial radial distribution $R_i(r)$ is defined such that $R_i(r)dr$ gives the number of atoms that are a distance between $r$ and $r+dr$ away from $i$.  The definition of $R_i(r)$ excludes self-counting, so $R_i(0) = 0$.

The total RDF, $R(r)$, is the stoichiometric average of the partial RDFs of each atom in the material, or
\begin{align}
    R(r) &= \frac{1}{\#\;atoms}\sum_{i=1}^{\#\;atoms} R_i(r).
\end{align}
Thus, $\int_a^b R(r)dr$ gives the number of atomic pairs (per atom) with a separation distance between $a$ and $b$.

The RDF is related to the PDF $G(r)$ by~\cite{farrowBillingeSasPdf09}
\begin{align}
    G(r) &= \frac{R(r)}{r} - 4\pi r\rho_0\gamma_0(r),
\end{align}
where $\rho_0$ is the atomic number density, and $\gamma_0(r)$ is the nanoparticle characteristic function (defined in Supplemental Material, \sect{sup-shape}).

When a material expands isotropically with strain $\epsilon$, all distances between pairs of atoms increase by a factor of $1 + \epsilon > 0$ (compression occurs for $-1 < \epsilon < 0$).  Thus, the number of atomic pairs with separation distances between $a$ and $b$ before expansion occurs equals the number of atomic pairs with separation distances between $(1+\epsilon) a$ and $(1+\epsilon) b$ after.  The total number of atoms in the material is unchanged by the expansion. Defining $R(r)$ to be the RDF pre-expansion and $R'(r)$ to be that post-expansion,
\begin{align}
    \label{eq:RDF_RDF-prime}
    \int_a^b R(r)dr &= \int_{(1+\epsilon) a}^{(1+\epsilon) b} R'(r)dr.
\end{align}
Applying a change of variables $r \mapsto r/(1+\epsilon)$ to the left hand side of \eq{RDF_RDF-prime} gives
\begin{align}
    \int_a^b R(r)dr &= \int_{(1+\epsilon) a}^{(1+\epsilon) b} \frac{R(r/(1+\epsilon))}{1+\epsilon}dr.
    \label{eq:RDF_stretched}
\end{align}
Since the relations in \eq{RDF_RDF-prime} and \eq{RDF_stretched} hold for all $0 \leq a \leq b$,
\begin{align}
    R'(r) &= \frac{R(r/(1+\epsilon))}{1+\epsilon}\label{eq:R-stretched-R-relation}
\end{align}
for all $r \geq 0$.

The corresponding pre-expansion PDF is
\begin{align}
    G(r) = \frac{R(r)}{r} - 4\pi r \rho_0\gamma_0(r),
\end{align}
and the post-expansion PDF is
\begin{align}
    G'(r) &= \frac{R'(r)}{r} - 4\pi r \rho_0\gamma_0(r)\\
    &= \frac{R(r/(1+\epsilon))}{(1+\epsilon) r} - 4\pi r\rho'_0\gamma'_0(r).
\end{align}
The $\rho_0, \gamma_0(r)$ are the pre-expansion atomic density and nanoparticle form function, and $\rho'_0, \gamma'_0(r)$ the same quantities post-expansion.  Due to the expansion, the volume of the material increases by $(1+\epsilon)^3$, while the total number of atoms remains constant, so
\begin{align}
    \rho'_0 &= \frac{1}{(1+\epsilon)^3}\rho_0.
\end{align}
The post-expansion nanoparticle characteristic function scales as
\begin{align}
    \label{eq:shape_factor_scale}
    \gamma'_0(r) &= \gamma_0\left(\frac{r}{1 + \epsilon}\right).
\end{align}
(See Supplemental Material, \sect{sup-shape} for proof.)

Finally, we conclude that the PDF after expansion is
\begin{align}
    \;\;\;\;G'(r) &= \frac{R(r/(1+\epsilon))}{(1+\epsilon)r} - 4\pi r\frac{\rho_0\gamma_0(r/(1+\epsilon))}{(1+\epsilon)^3}\\
    &= \frac{G(r/(1+\epsilon))}{(1+\epsilon)^2}.\label{eq:Gr-stretch-conclusion}
\end{align}

Therefore, applying a \stretchmorph parameter $\eta = \epsilon$ and a \scale parameter $s = (1 + \epsilon)^{-2}$ to \gr models isotropic expansion with strain $\epsilon$.

It is important to note that an isotropic expansion (with strain $\epsilon)$ generally denotes that time-averaged atomic distances are multiplied by a factor $1 + \epsilon$.  These time-averaged atomic distances generally correspond to peaks in the PDF.  Therefore, we can still conclude from our derivation of \eq{Gr-stretch-conclusion} that peaks in $G'(r)$ are located at the same $r$-positions as peaks in $G(r/(1+\epsilon))$.

We note that this treatment strictly holds only for a uniform expansion.  This is unlikely to be a good approximation for structures with highly anisotropic structures or when there exist hierarchical bonding strengths; for example, molecular solids where a large lattice expansion will not be reflected in large changes in intra-molecular PDF.

It is not necessarily the case, however, that the relation \eq{Gr-stretch-conclusion} holds for all parts of the PDF.  Since atoms are not stationary, the RDF is a time-dependent function $R(r, t)$. Thus, so too is the PDF $G(r, t)$.  The PDF $G(r)$ we obtain from a material experimentally can be thought of as the time-averaged PDF of that material over the data collection period.  The conditions causing a uniform lattice expansion (e.g. increasing temperature) can also change how atomic positions are distributed over time and affect the PDF in a manner distinct from \eq{Gr-stretch-conclusion}.  Other morphs are required to account for these changes.  (For example, in \sect{sup-convolution}, we show how the smearing morph is used to correct for one such effect.)

With these caveats said, in our experience, the stretch morph can still be a useful approximate first step even when the conditions discussed above are not perfectly met.

\subsection{Uniform Lattice Expansion Effects on Reciprocal Space Functions}
\label{sec:sup-temp-iq}

We showed above how a uniform lattice expansion with strain $\epsilon$ stretches the grid of the PDF by a factor $1 + \epsilon$. Below we show that the same expansion instead stretches the reciprocal space ($Q$-space) function grid by a factor $(1 + \epsilon)^{-1}$.

In particular, we derive the relations Eqs.~\ref{eq:IcQ-stretch}, \ref{eq:IdQ-stretch}, \ref{eq:SQ-stretch}, and \ref{eq:FQ-stretch} in order to show that isotropic expansion of a material by strain $\epsilon$ stretches the grid of the reciprocal space functions \iq, \sq, and \fq by $(1 + \epsilon)^{-1}$.

While the change in \sq, \fq, and 1D neutron scattering intensities due to lattice expansion can be fully explained by the aforementioned grid stretch, this is not strictly the case for the 1D x-ray and electron scattering intensities.  However, we have observed that in practice, for small $|\epsilon| < 0.01$, it is still a very accurate approximation with little error.  This error, we will show, grows linearly with $|\epsilon|$.

\subsubsection{Effect on scattering intensities, \iq}
\label{sec:sup-iq-stretch}

In a system with $N$ atoms, the full 1D coherent scattering intensity function (of an isotropic sample) is given by by~\cite{farrowBillingeSasPdf09}
\begin{align}
    I_c(Q) &= \sum_{i,j=1}^N b_j^*b_i\frac{\sin(Qr_{ij})}{Qr_{ij}},\label{eq:IQ-bi}
\end{align}
where $r_{ij}$ is the distance between atoms $i$ and $j$. Removing the self-scattering gives the 1D discrete scattering intensity
\begin{align}
    I_d(Q) &= I_c(Q) - \sum_{i=1}^N b_j^*b_i.
\end{align}
In the case of neutron scattering, the $b_i$ denote the coherent neutron scattering lengths that are independent of~\q. Therefore, upon an isotropic expansion with strain $\epsilon$, the interatomic distances $r_{ij}$ become $(1 + \epsilon)r_{ij}$. The coherent scattering intensity post-$\epsilon$-strain is then
\begin{align}
    I_{c,\;n^0}^\epsilon(Q) &= \sum_{i,j=1}^N b_j^*b_i\frac{\sin(Q(1+\epsilon)r_{ij})}{Q(1+\epsilon)r_{ij}}\\
    &= I_{c,\;n^0}((1+\epsilon)Q),\label{eq:IcQ-stretch}
\end{align}
Thus, if a \stretchmorph morph is applied on $I_{c,\;n^0}(Q)$ to match $I_{c,\;n^0}^\epsilon(Q)$ as target, the \stretchmorph parameter $\eta_{iq}$ will be
\begin{align}
    \eta_{iq} &= \frac{1}{1 + \epsilon} - 1\\
    &\approx -\epsilon.\label{eq:stretch-iq-icq-approximation}
\end{align}
The same relation holds for the discrete scattering intensity:
\begin{align}
    I_{d,\;n^0}^\epsilon(Q) &= I_{d,\;n^0}((1+\epsilon)Q).\label{eq:IdQ-stretch}
\end{align}

In the case of x-ray (and electron) scattering, we replace the $b_i$ in \eq{IQ-bi} with \q-dependent atomic form factor functions $f_i(Q)$. Here, the scattering intensity function post-expansion is given by
\begin{align}
    I_{c,\;xray/e^-}^\epsilon(Q;\epsilon) &= \sum_{i,j=1}^N f_j(Q)^*f_i(Q)\frac{\sin((1+\epsilon)Qr_{ij})}{(1+\epsilon)Qr_{ij}}\label{eq:iq-stretched-physical}\\
    \overset{\text{In General}}&{\neq} I_{c,\;xray/e^-}((1+\epsilon)Q)\\
    &= \sum_{i,j=1}^N f_j((1+\epsilon)Q)^*f_i((1+\epsilon)Q)\nonumber\\
    &\phantom{=} \times \frac{\sin((1+\epsilon)Qr_{ij})}{(1+\epsilon)Qr_{ij}}.
\end{align}
Our simple stretch no longer works as the $f_i(Q)$ are unchanged by the isotropic expansion. 

In practice, this error has very little effect on our morphing results.  We will showcase this on a set of simulated data and then on real experimental data.

For the simulated data, we consider an Ir$-$Te bond of length $r_{Ir,Te} = 3$~\AA~\cite{legerIrTe2BondLengths00}.
Its contribution to the x-ray scattering intensity $I_{c,\;xray}(Q)$ is
\begin{align}
    \mathcal{C}_\mathrm{Ir,Te}(Q) &= f_\mathrm{Ir}(Q)^*f_\mathrm{Te}(Q)\frac{\sin(Qr_\mathrm{Ir,Te})}{Qr_\mathrm{Ir,Te}}.
\end{align}
To simulate how a strain $\epsilon$ acts on the \iq, we can apply a \stretchmorph morph with \stretchmorph parameter $\eta = \epsilon$. The \stretchmorph morph simulated contribution from this bond is
\begin{align}
    \mathcal{C}_\mathrm{Ir,Te}((1+\epsilon)Q) &= f_\mathrm{Ir}((1+\epsilon)Q)^*f_\mathrm{Te}((1+\epsilon)Q)\nonumber\\
    &\phantom{=} \times \frac{\sin((1+\epsilon)Qr_\mathrm{Ir,Te})}{(1+\epsilon)Qr_\mathrm{Ir,Te}}.
\end{align}
However, in reality, an $\epsilon$ strain on a material will produce a scattering intensity following $I^\epsilon_{c,\;xray}(Q)$ (given by \eq{iq-stretched-physical}), with the contribution by the Ir$-$Te bond as
\begin{align}
    \mathcal{C}_\mathrm{Ir,Te}^\epsilon(Q) &= f_\mathrm{Ir}(Q)^*f_{Te}(Q)\frac{\sin((1+\epsilon)Qr_\mathrm{Ir,Te})}{(1+\epsilon)Qr_\mathrm{Ir,Te}}\\
    &\neq \mathcal{C}_\mathrm{Ir,Te}((1+\epsilon)Q).
\end{align}
We will show, however, that the \stretchmorph morph still well-approximates the actual effects of straining (i.e. $\mathcal{C}_\mathrm{Ir,Te}((1+\epsilon)Q) \approx \mathcal{C}_\mathrm{Ir,Te}^\epsilon(Q)$).

We can compute $\mathcal{C}^\epsilon_{Ir,Te}(Q)$ and $\mathcal{C}_{Ir,Te}((1+\epsilon)Q)$ by approximating Gaussians,
\begin{align}
    f_i(Q) &\approx c_i + \sum_{k=1}^4 a_{i,k}e^{-b_{i,k}\left(\frac{Q}{4\pi}\right)^2},\label{eq:four-gaussian-approx}
\end{align}
with coefficients $a_{i,k}, b_{i,k}, c_i \in \mathbb{R}$ for Ir and Te taken from the literature~\cite{brownAtomicFormFactors06}.

\begin{figure}[tbp]
    \centering
    \includegraphics[width=1\columnwidth]{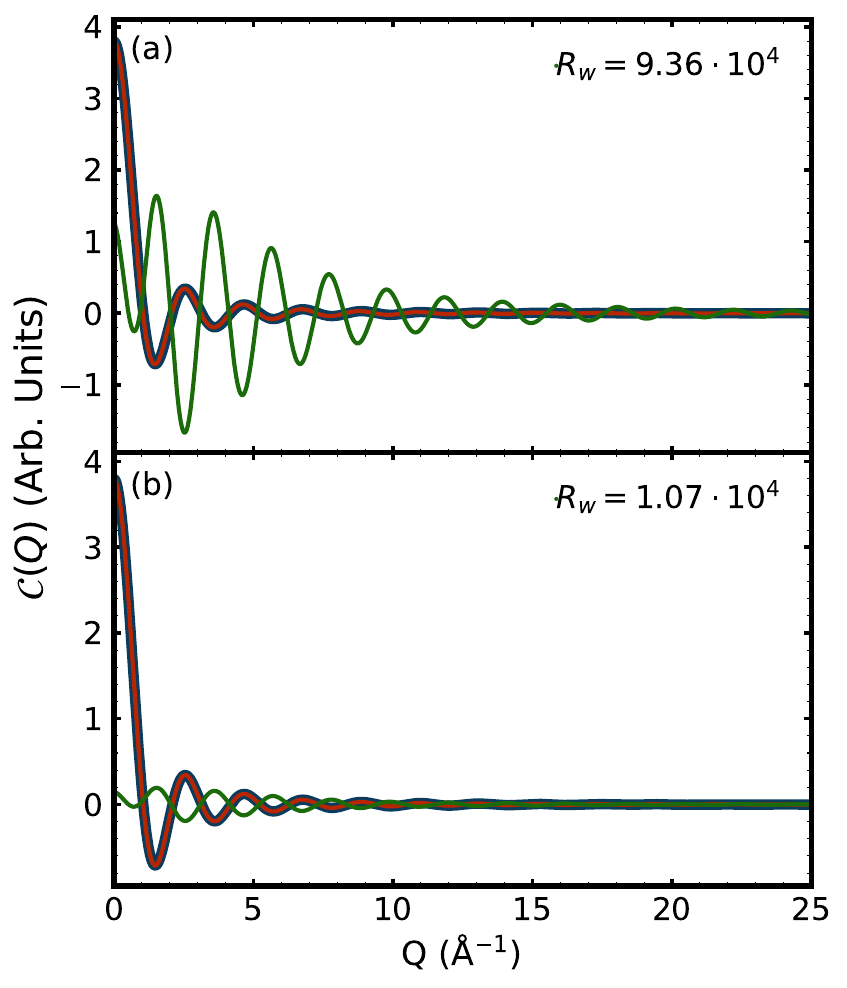}
    \caption{(a) $\mathcal{C}_{Ir,Te}(Q)$ (blue) and $\mathcal{C}^\epsilon_{Ir,Te}(Q)$ (red) computed for $\epsilon = 0.01$. A \scale and \stretchmorph morph has been applied to $\mathcal{C}_{Ir,Te}(Q)$.  The difference curve (green) is scaled $\times 1000$.  (b) The same plot as (a), but for $\epsilon = 0.001$.  All \rw values are computed on the grid $Q \in [0~\text{\AA},250~\text{\AA}]$ with grid spacing $dQ = 0.01$.}
    \label{sup-fig:Ir-Te-bond-IQ-contribution}
\end{figure}

In Fig.~\ref{sup-fig:Ir-Te-bond-IQ-contribution}(a), we compute and plot $\mathcal{C}^\epsilon_\mathrm{Ir,Te}(Q)$ (in red) for $\epsilon = 0.01$.  We have morphed $\mathcal{C}_\mathrm{Ir,Te}(Q)$ to match  $\mathcal{C}^\epsilon_\mathrm{Ir,Te}(Q)$ using the morphing parameters in \tabl{stretch-error-empirical}.
\begin{table}[tbp]
\caption{Test of the effects of the \q~dependence of the x-ray form factor on the accuracy of the \morph parameters, $\eta$ (\stretchmorph) and $s$ (\scale). If the approximation is good we expect $s$ to be close to unity and $\eta$ to be close to $-\epsilon$.  We also expect \rw to be small.  These conditions are satisfied for reasonable strains of $0.01$ and $0.001$.  We also find that the error scales with~$\epsilon$.} 
    \centering
    \begin{tabular}{|c|c|c|c|c|c|c|}
    \hline
    & \multicolumn{2}{|c|}{\textbf{Ir-Te}} & \multicolumn{2}{|c|}{\textbf{Ir-Ir}} & \multicolumn{2}{|c|}{\textbf{Te-Te}}\\
    \hline
    $\boldsymbol{\epsilon}$ & $\mathbf{0.01}$ & $\mathbf{0.001}$ & $\mathbf{0.01}$ & $\mathbf{0.001}$ & $\mathbf{0.01}$ & $\mathbf{0.001}$\\
    \hline
    $\mathbf{10^4(s-1)}$ & $3.34$ & $0.385$ & $2.64$ & $0.319$ & $3.94$ & $0.443$\\
    \hline
    $\mathbf{10^4}(\boldsymbol{\eta}+\boldsymbol{\epsilon})$ & $1.80$ & $0.0613$ & $1.52$ & $0.0367$ & $2.14$ & $0.0932$\\
    \hline
    $\mathbf{10^4\rw}$ & $9.36$ & $1.07$ & $9.04$ & $1.04$ & $9.84$ & $1.11$\\
    \hline
    \end{tabular}
    \label{table:stretch-error-empirical}
\end{table}
The \rw of the morph is extremely small at $\sim 10^{-3}$.  The \scale parameter $s \approx 1$ and \stretchmorph parameter $\eta \approx -\epsilon$, meaning the $\eta$ parameter we extract from morphing \iqs will accurately model the strain $\epsilon$ according to \eq{stretch-iq-icq-approximation}.

Changing to $\epsilon = 0.001$ ($\epsilon$ is reduced by an order of magnitude), we see in Fig.~\ref{sup-fig:Ir-Te-bond-IQ-contribution}(b) the \rw is also reduced by about an order of magnitude.  We also see in \tabl{stretch-error-empirical} the new $1-s$ and $\eta + \epsilon$ are both an order of magnitude closer to zero compared to the $\epsilon = 0.01$ case, indicating that we can even more accurately extract the strain from the \stretchmorph parameter.

Repeating these calculations for a simulated Ir$-$Ir and Te$-$Te bond (both with length $3$~\AA) in \tabl{stretch-error-empirical}, we see again that the \stretchmorph morphed contribution $\mathcal{C}_{i,j}((1+\epsilon)Q)$ still accurately models the actual contribution $\mathcal{C}^\epsilon_\mathrm{Ir,Te}((1+\epsilon)Q)$, with extremely small $\rw$ discrepancies and $\eta \approx -\epsilon$.

We also observe the same trend as Fig.~\ref{sup-fig:Ir-Te-bond-IQ-contribution} where \rw, $1-s$, and $\eta + \epsilon$ for $\epsilon = 0.001$ are an order of magnitude smaller compared to the parameters when $\epsilon = 0.01$.  This is not a coincidence, as we will show in \sect{iq-error-bound}.

Finally, we will demonstrate on real experimental data that the errors from applying Eqs.~\ref{eq:IcQ-stretch},\ref{eq:IdQ-stretch} are small, and that they are still good models for x-ray \iq data.
In \sect{sup-temp}, we showed that a strain~$\epsilon$ corresponds to a \stretchmorph parameter $\eta_{gr} = \epsilon$ for a morph between PDFs.  If  Eqs.~\ref{eq:IcQ-stretch},\ref{eq:IdQ-stretch} apply, the \stretchmorph parameter between corresponding \iqs $\eta_{iq}$ will follow
\begin{align}
    1 + \eta_{gr} &= \frac{1}{1 + \eta_{iq}}.\label{eq:stretch-gr-iq-relation}
\end{align}

\begin{figure}[tbp]
    \centering
    \includegraphics[width=1\columnwidth]{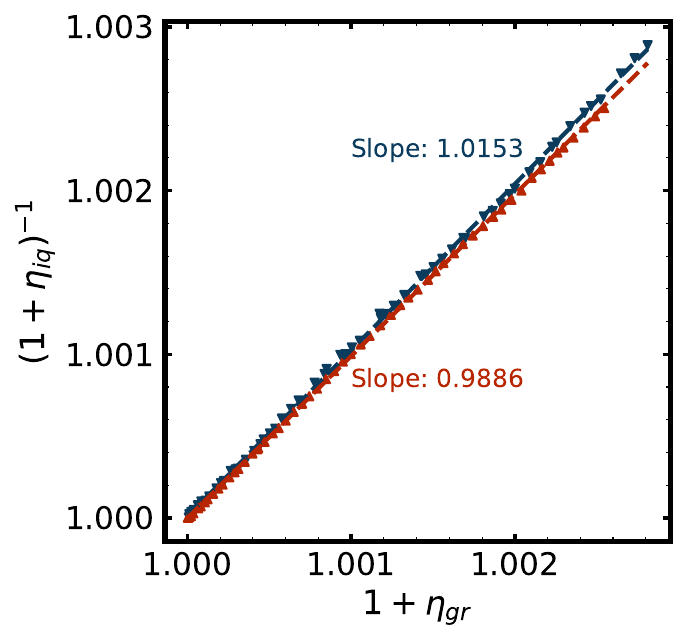}
    \caption{
    The $G(r)$ stretches, $1 + \eta_{gr}$, and reciprocal of the $I(Q)$ stretches, $(1 + \eta_{iq})^{-1}$, for the Ir$_{\text{0.8}}$Rh$_{\text{0.2}}$Te$_2$ sample (blue) and the Ir$_{\text{0.95}}$Pt$_{\text{0.05}}$Te$_2$ sample (red).
    The correlation coefficients are $0.99980$ for Ir$_{\text{0.8}}$Rh$_{\text{0.2}}$Te$_2$ and $0.99993$ for Ir$_{\text{0.95}}$Pt$_{\text{0.05}}$Te$_2$.}
    \label{sup-fig:Gr-Iq-stretches}
\end{figure}

We measured \iqs of samples of Ir$_{\text{0.8}}$Rh$_{\text{0.2}}$Te$_2$ and Ir$_{\text{0.95}}$Pt$_{\text{0.05}}$Te$_2$ in the temperature range $\sim 10$~K to $\sim 300$~K. From each \iq, a PDF was computed.  To obtain the PDF \stretchmorph parameters $\eta_{gr}$, we morphed low-temperature ($9$~K for the Rh-substituted sample and $10$~K for the Pt-substituted sample) PDFs against higher-temperature (up to $295$~K for the Rh-substituted sample and $299$~K for the Pt-substituted sample).  We performed the same process on the 1D diffraction patterns extracted at the same temperatures to obtain $\eta_{iq}$.  In Fig.~\ref{sup-fig:Gr-Iq-stretches}, we plot $1 + \eta_{gr}$ vs. $(1 + \eta_{iq})^{-1}$ for the two samples.
The slopes of the plots above are within $1.14\%$ from the value of $1$ predicted by \eq{stretch-gr-iq-relation}.  As such, we can use Eqs.~\ref{eq:IcQ-stretch},\ref{eq:IdQ-stretch} to model the effect of a strain $\epsilon$ on the \iqs, showing \morph can be used just as effectively on \iq data as PDF data.

\subsubsection{Analytically bounding the \iq error}
\label{sec:iq-error-bound}

In \tabl{stretch-error-empirical}, it can be seen that the error (the \rws) from applying Eq.~\ref{eq:IcQ-stretch} seems to decrease linearly with $|\epsilon|$.  This can be shown analytically.

We start by quantifying the error at each $Q$ as
\begin{align}
    \text{err}(Q) &\equiv \abs{I_{c,\;strain,\;xray}(Q; \epsilon) - I_c((1+\epsilon)Q)}.\label{eq:IQ-stretch-error}
\end{align}
We will show for x-ray scattering (within a $Q$-range of $0~\text{\AA}\leq Q \leq 25~\text{\AA}$), 
\begin{align}
    \text{err}(Q) &\lesssim \mathcal{K}\abs{\epsilon}\label{eq:error-upper-bound}
\end{align}
for some constant $\mathcal{K}$ independent of $Q$.
In words, we will show that the error is (approximately) linearly upper-bounded by the magnitude of the strain.

By applying Eqs.~\ref{eq:IQ-bi},\ref{eq:four-gaussian-approx}, we see that the error in \eq{IQ-stretch-error} is
\begin{widetext}
\begin{align}
    \text{err}(Q) &= \abs{\left(2\sum_{i,j=1}^N \sum_{k=1}^4 C_{i,j,k}h(Q; B_{j,k}, \epsilon) + \sum_{i,j=1}^N \sum_{k,\ell=1}^4 A_{i,j,k,\ell}h(Q; B_{i,j,k,\ell}, \epsilon)\right)\frac{\sin((1+\epsilon)Qr_{i,j})}{(1+\epsilon)Qr_{i,j}}}
\end{align}
\end{widetext}
where
\begin{align}
    h(Q; B, \epsilon) &= e^{-BQ^2} - e^{-B((1+\epsilon)Q)^2},\\
    A_{i,j,k,\ell} &= a_{ik}a_{j\ell},\\
    B_{i,j} &= \frac{b_{i,j}}{16\pi^2},\\
    B_{i,j,k,\ell} &= \frac{b_{i,k} + b_{j,\ell}}{16\pi^2},\\
    C_{i,j,k} &= c_ia_{jk}.
\end{align}

For $\epsilon > 0$, $0 < h(Q;B,\epsilon)$, with maximum attained at
\begin{align}
    Q^* &= \sqrt{\frac{2\ln(1+\epsilon)}{(2\epsilon + \epsilon^2)B}}.
\end{align}
For $\epsilon < 0$, $h(Q; B, \epsilon) < 0$, with minimum attained at $Q^*$.  For both cases,
\begin{align}
    h(Q^*;B,\epsilon) &= \exp\left\{-\frac{2\ln(1+\epsilon)}{2\epsilon+\epsilon^2}\right\}\left(\frac{2\epsilon + \epsilon^2}{(1 + \epsilon)^2}\right)\\
    &\approx \frac{2\epsilon}{e},
\end{align}
where this approximation holds for small $\epsilon < 1$. For $\epsilon = 0.01$, the approximation holds up to a relative error of $0.5\%$.
Then, as $|\text{sinc}(x)| \leq 1$, we can very loosely upper-bound the error
\begin{align}
    \text{err}(Q) &\leq 2\sum_{i,j=1}^N \sum_{k=1}^4 \abs{C_{i,j,k}h(Q; B_{j,k}, \epsilon)}\nonumber\\
    &\phantom{=} + \sum_{i,j=1}^N \sum_{k,\ell=1}^4 \abs{A_{i,j,k,\ell}h(Q; B_{i,j,k,\ell}, \epsilon)}\\
    &\lesssim \frac{2|\epsilon|}{e}\left(2\sum_{i,j=1}^N\sum_{k=1}^4 |C_{i,j,k}| + \sum_{i,j=1}^N\sum_{k,\ell=1}^4 |A_{i,j,k,\ell}|\right).
\end{align}
As such, we have demonstrated our desired relation \eq{error-upper-bound}.

A similar derivation can be done to show the error from applying \eq{IdQ-stretch},
\begin{align}
    \abs{I_{d, e, xray}(Q) - I_d((1 + \epsilon)Q)},
\end{align}
is also (approximately) linear in $|\epsilon|$.


\subsubsection{Effect on scattering structure functions \sq and \fq}

In the process of computing the PDF \gr from \iq, we compute two intermediate reciprocal-space functions \sq and \fq. We show below how these functions are defined and how an isotropic expansion with strain $\epsilon$ also stretches their grid by $(1 + \epsilon)^{-1}$. The equations below use the \q-independent coherent neutron scattering length $b_i$, but all derivations follow equivalently if one chooses to use the $Q$-dependent form factors $f_i(Q)$ instead.

The total scattering structure function, \sq, is the discrete scattering intensity normalized by the average scattering factor~\cite{farrowBillingeSasPdf09}:
\begin{align}
    S(Q) &= \frac{NI_d(Q)}{\sum_{i,j=1}^N b_j^*b_i} + 1.\label{eq:def-SQ}
\end{align}
For a finite number of scatterers, $N, I_d(Q)$ are finite and the scattering intensities are non-zero, ensuring \sq is finite as long as the denominator of \eq{def-SQ} is non-zero. We will assume this is the case, as only very rare combinations of materials will cause the scattering length products to perfectly cancel.

The reduced total scattering structure function \fq is then obtained from \sq through
\begin{align}
    F(Q) &= Q(S(Q) - 1).\label{eq:SQ-to-FQ}
\end{align}
Since \sq and thus $S(0)$ is finite, $F(0) = 0$.

We finally obtain the PDF $G(r)$ through the relation~\cite{farrowBillingeSasPdf09}
\begin{align}
    G(r) &= \frac{2}{\pi}\int_0^\infty F(Q)\sin(Qr)dQ.\label{eq:FQ-to-GR}
\end{align}
To be pedantic, since we include the small angle scattering $Q$-values in the integral above, the result obtained is actually
\begin{align}
    \frac{R(r)}{r} &= \frac{2}{\pi}\int_0^\infty F(Q)\sin(Qr)dQ,\label{eq:FQ-to-RR}
\end{align}
where $R(r)$ is the radial distribution function explored in \sect{sup-temp}. Our derivations below will use the more commonly used (albeit inaccurate) expression \eq{FQ-to-GR}, but all results follow equivalently by substituting \eq{FQ-to-GR} with \eq{FQ-to-RR} and \eq{sec-IQ-stretch} with \eq{R-stretched-R-relation}.

Now, let us consider a PDF $G(r)$ of a material and PDF $G_e(r)$ of the same material having undergone uniform expansion with strain $\epsilon$.
\begin{align}
    G_e(r) &= \frac{G(r/(1+\epsilon))}{(1+\epsilon)^2},\label{eq:sec-IQ-stretch}
\end{align}
Let $S(Q), F(Q)$ be the functions used in the computation of $G(r)$ and $S_e(Q), F_e(Q)$ used in computing $G_e(r)$. To avoid dealing with $Q$-dependent form factors, we will work backward to obtain expressions for the functions post-uniform-expansion in terms of their original counterparts.

Combining \eq{FQ-to-GR} and \eq{sec-IQ-stretch} gives
\begin{align}
    G_e(r) &= \frac{2}{\pi}\int_0^\infty \frac{F(Q)}{1 + \epsilon}\sin\left(\frac{Qr}{1 + \epsilon}\right)\frac{dQ}{1 + \epsilon}\\
    &= \int_0^\infty \frac{F((1+\epsilon)Q')}{1 + \epsilon}\sin(Q'r)dQ',\label{eq:expanded-FQ}
\end{align}
where we have applied a change of variables $Q' = Q/(1+\epsilon)$ on the last step.
From \eq{FQ-to-GR},
\begin{align}
    G_e(r) &= \frac{2}{\pi}\int_0^\infty F_e(Q)\sin(Qr)dQ,\label{eq:expanded-FQ-to-GR}
\end{align}
and subtracting \eq{expanded-FQ} gives
\begin{align}
    0 &= \int_0^\infty \left(F_e(Q) - \frac{F((1+\epsilon)Q)}{1+\epsilon}\right)\sin(Qr)dQ.
\end{align}
Since $F_e(0) = 0$ and $F(0) = 0$,
\begin{align}
    F_e(0) - \frac{F(0)}{1 + \epsilon} &= 0,
\end{align}
so by the inverse sine Fourier transform of an odd function~\cite{boasFourierSine2006},
\begin{align}
    F_e(Q) - \frac{F((1+\epsilon)Q)}{1+\epsilon} &\propto \int_0^\infty 0\sin(Qr)dr\\
    &= 0.
\end{align}
Therefore
\begin{align}
    F_e(Q) &= \frac{F((1+\epsilon)Q)}{1+\epsilon}.\label{eq:FQ-stretch}
\end{align}

Then, from \eq{SQ-to-FQ},
\begin{align}
    Q(S_e(Q) - 1) &= Q(S((1+\epsilon)Q) - 1).
\end{align}
This means for all $Q > 0$,
\begin{align}
    S_e(Q) &= S((1+\epsilon)Q).\label{eq:SQ-stretch}
\end{align}
For continuous $S_e(Q), S(Q)$, this means $S_e(0) = S(0)$ as well (and thus \eq{SQ-stretch} also holds at $Q = 0$).

\subsection{Modeling Thermal Smearing on the PDF}
\label{sec:sup-convolution}

Peaks in the radial distribution function (see Supplemental Material, \sect{sup-temp} for definition) obtained from measured PDFs have approximately Gaussian shapes due to Debye-Waller effects.  The variance $\sigma_p^2$ of each peak $p$ is given by the atomic displacement parameters (ADP) of the atoms contributing to~$p$.
The ADP can depend on dynamic (temperature-dependent) and static factors~\cite{dinnebierBillingeOverviewPrinciples18}.  Models, such as the Debye model, generally separate the two.  Per this model, the mean square ADP, denoted $\overline{u^2}$, is given by
\begin{align}
    \label{eq:ADP_separation}
    \overline{u^2} &= A_i(T) + A_{i,static},
\end{align}
where $A_i(T)$ increases with temperature.  Specifically, the Debye model treats the temperature-dependent component as~\cite{dinnebierBillingeOverviewPrinciples18}
\begin{align}
    \label{eq:debye}
    A_i(T) &= \frac{3h^2T^2}{4\pi^2M_ik_B\Theta_D^3} \int_0^{\Theta_D/T} \frac{x}{e^x-1}dx\nonumber\\
    &\phantom{=}+ \frac{3h^2}{16\pi^2M_ik_B\Theta_D}
\end{align}
where $M$ is the mass of the atom $i$, $\Theta_D$ is the Debye temperature of the (crystal) material, $h$ is Planck's constant, and $k_B$ is Boltzmann's constant.
Since
\begin{align}
    f(T) = T^2\int_0^{\Theta_D/T} \frac{x}{e^x-1}dx
\end{align}
is a strictly increasing function of $T \in \mathbb{R}^+$ for $\Theta_D > 0$, increasing $T$ causes $A_i(T)$ to increase by some positive quantity $\sigma^2_i > 0$.  Increases in the ADP translate to broadening of each Guassian peak in the RDF.

Motivated by the Debye model, the \smear morph broadens all Gaussian peaks in a radial distribution function by an increased variance of $\sigma^2$.  As $\sigma^2 \geq 0$, only thermal broadening due to increases in temperature can be captured, so it recommended to use \morph to morph a lower-temperature sample to match a target higher-temperature sample.

It is of note that the Debye model indicates that the temperature-dependent component of the ADP depends on the mass of the atom (see \eq{debye}).  As such, increasing temperature will broaden each peak in the RDF differently.  \morph only captures the ``average" broadening, but still works well to reduce features in the difference due to thermal broadening.  A user can also tell \morph to evaluate the smearing on an $r$ interval capturing a single peak of interest.

The exact operation done by \morph is given in \eq{morph_smear_def}.  Below, we prove that this broadens all peaks in $R(r)$ by an additional variance $\sigma^2$.

Assume our radial distribution function can be decomposed into a sum of Gaussians
\begin{align}
    \label{eq:RDF_gaussian_sum}
    R(r) &= \sum_k I_kN(r; \mu_k, \sigma_k^2),
\end{align}
where each $I_k \in \mathbb{R}^+$ is the amplitude multiplied to a normalized Gaussian $N(r; \mu_k, \sigma_k^2)$ following
\begin{align}
    N(r; \mu_k, \sigma_k^2) &= \frac{1}{\sqrt{2\pi\sigma_k^2}}\exp\left\{-\frac{1}{2}\left(\frac{r - \mu_k}{\sigma_k}\right)^2\right\}
\end{align}
with mean $\mu_k$ and variance $\sigma_k^2$.  Convolving the Gaussian broadening factor $T_\sigma(r)$ from \eq{morph_smear_gaussian_def} with a single Gaussian $N_k(r) \equiv N(r; \mu_k, \sigma_k^2)$ gives
\begin{align}
    (N_k \circledast T_\sigma)(r) &= \int_{-\infty}^\infty N_k(r') \cdot T_\sigma(r - r')dr'\\
    &= \int_{-\infty}^\infty \frac{1}{\sqrt{2\pi\sigma_k^2}}\exp\left\{-\frac{1}{2}\left(\frac{r' - \mu_k}{\sigma_k}\right)^2\right\}\nonumber\\
    &\phantom{=} \times \exp\left\{-\frac{1}{2}\left(\frac{(r - r') - r_c}{\sigma}\right)^2\right\}dr'\\
    &= \frac{1}{\sqrt{2\pi\sigma_k^2}} \int_{-\infty}^\infty \exp\left\{-\frac{f(r')}{2\beta} + \alpha\right\}dr'\label{eq:single_integral},
\end{align}
where we have defined
\begin{align}
    \alpha &\equiv \frac{(r - (\mu_k + r_c))^2}{2(\sigma_k^2 + \sigma^2)},\\
    \beta &\equiv \frac{\sigma^2\sigma_k^2}{\sigma_k^2 + \sigma^2},\\
    f(r') &\equiv \left(r' - \frac{\sigma^2\mu_k + \sigma^2_k(r_c - r)}{\sigma_k^2 + \sigma^2}\right)^2.
\end{align}
Some rearrangement lets us evaluate the integral
\begin{align}
    (N_k \circledast T_\sigma)(r) &= \frac{\exp\left\{-\alpha\right\}}{\sqrt{2\pi\sigma_k^2}}\int_{-\infty}^\infty \exp\left\{-\frac{f(r')}{2\beta}\right\}dr'\\
    &= \sqrt{\frac{\beta}{\sigma_k^2}}\exp\left\{-\frac{1}{2}\frac{(r - (\mu_k + r_c))^2}{\sigma_k^2 + \sigma^2}\right\}\\
    &= \sqrt{2\pi\sigma^2} N(r; \mu_k + r_c, \sigma_k^2 + \sigma^2).
\end{align}
As the convolution operator is linear, convolving $R(r)$ from \eq{RDF_gaussian_sum} with $T_\sigma(r)$ gives
\begin{align}
    (R \circledast T_\sigma)(r) &= \sum_k I_k \sqrt{2\pi\sigma^2} N(r;\mu_k + r_c, \sigma_k^2 + \sigma^2).
\end{align}
Therefore,
\begin{align}
    \frac{(R \circledast T_\sigma)(r + r_c)}{\sqrt{2\pi \sigma^2}} &= \sum_k I_k N(r;\mu_k, \sigma_k^2 + \sigma^2),
\end{align}
which is $R(r)$, but with each Gaussian peak broadened by an additional variance $\sigma^2$.

\subsection{Nanoparticle Characteristic Function}
\label{sec:sup-shape}

The shape of a nanoparticle plays a role in shaping the PDF through the nanoparticle characteristic function $\gamma_0(r)$ (see \eq{nano_bulk_rel}).  In this section, we detail how the form factor is computed and prove the equality in \eq{shape_factor_scale}.

To begin, define the shape function $s(\vec{r})$, which takes value $1$ for $\vec{r}$ within the nanoparticle and $0$ for $\vec{r}$ outside. The unaveraged particle form factor $\gamma_0(\vec{r})$ is the autocorrelation of this shape function averaged over the nanoparticle volume $V$, computed as~\cite{farrowBillingeSasPdf09}
\begin{align}
    \gamma_0(\vec{r}) &= \frac{1}{V} \iiint s(\vec{r}')s(\vec{r}' + \vec{r})d\vec{r}'.\label{eq:nff}
\end{align}
We refer to the angle-averaged version of this quantity, denoted $\gamma_0(r)$, as the nanoparticle characteristic function, which is given by~\cite{farrowBillingeSasPdf09}
\begin{align}
    \gamma_0(r) &= \frac{\int d\phi \int d\theta \sin(\theta)r^2\gamma_0(\vec{r})}{\int d\phi \int d\theta r^2\sin(\theta)}.\label{eq:nff-aa}
\end{align}

There is an important relationship between the nanoparticle characteristic function of a material before and after uniform expansion, used in Supplemental Material, \sect{sup-temp}. Given a material has undergone isotropic lattice expansion with thermal strain $-1 < \epsilon$, the new form factor $\gamma'_0(r)$ is a stretched version of the pre-expansion form factor: $\gamma'_0(r) = \gamma_0(r/(1+\epsilon))$. To show this, we first note that expansion maps points $\vec{r}$ in the original material to $\vec{r}' = (1+\epsilon)\vec{r}$.  Thus, the shape function post-expansion is related to that pre-expansion by
\begin{align}
    s'((1+\epsilon)\vec{r}) = s(\vec{r}).
\end{align}
The volume of the nanoparticle is increased by a scale factor $(1+\epsilon)^3$ due to expansion. Following \eq{nff}, the post-expansion nanoparticle characteristic function follows
\begin{align}
    \gamma'_0(\vec{r}) &= \frac{1}{\kappa^3V} \iiint s'(\vec{r}')s'(\vec{r}' + \vec{r})d\vec{r}',
\end{align}
where
\begin{align}
    \kappa &\equiv 1+\epsilon
\end{align}
is defined to condense the following equations.  Applying a change of variables $\vec{r}' \mapsto \vec{r}'/\kappa$ gives:
\begin{align}
    \gamma'_0(\vec{r}) &= \frac{1}{\kappa^3V} \iiint s'(\kappa\vec{r}')s'(\kappa\vec{r}' + \vec{r})\kappa^3d\vec{r}'\\
    &= \frac{1}{V} \iiint s(\vec{r}')s(\vec{r}' + \vec{r}/\kappa)d\vec{r}'\\
    &= \gamma_0(\vec{r}/\kappa).
\end{align}
Finally, taking the angle-averaged integrals from \eq{nff-aa} gives the desired relation in \eq{shape_factor_scale}.

\subsection{A theoretical motivation for the polynomial grid morph}
\label{sec:sup-squeeze}


A common setup for a diffraction experiment uses a flat detector and a (x-ray/neutron/electron) source beam perpendicular to the detector. Scattered rays travel off the sample to the detector some distance away. Let's say that we measured a sample-to-detector distance of $D'$ distinct from the actual detector distance $D$. We will show that when the relative error is small ($|D' - D|/D < 0.1$), this results in a polynomial change to the grid of the measured 1D scattering intensity \iq. This motivates the morphs in \sect{general-x-morphs}.

For elastic scattering in the setup above, a point on the detector that is a distance $d$ from the detector center corresponds to a $Q$ of~\cite{dinnebier01GeneralIntroPowderDiffraction2008,egami;b;utbp12}
\begin{align}
    Q &= \frac{4\pi\sin\left(\frac{1}{2}\arctan(d/D)\right)}{\lambda},
\end{align}
but our computed $Q'$ using the incorrect $D'$ is
\begin{align}
    Q' &= \frac{4\pi\sin\left(\frac{1}{2}\arctan(d/D')\right)}{\lambda}.
\end{align}
There is no simple polynomial relation between $Q$ and $Q'$.

However, letting
\begin{align}
    \epsilon_D &= \frac{D' - D}{D}
\end{align}
be small ($|\epsilon_D| < 0.1$), we can Taylor approximate to first order
\begin{align}
    Q' &\approx Q - \frac{4\pi}{\lambda}\frac{(d/D)\cos\left(\frac{1}{2}\arctan(d/D)\right)}{2((d/D)^2 + 1)}\epsilon_D.\label{eq:D-taylor}
\end{align}
Then, by using
\begin{align}
    \frac{d}{D} &= \tan\left(2\arcsin\left(\frac{\lambda}{4\pi}Q\right)\right)\label{eq:d-D-ratio}
\end{align}
and some trigonometric identities (further below in this section), this simplifies to a polynomial action on $Q$ per
\begin{align}
    Q' &\approx Q - \epsilon_D\left(2\left(\frac{\lambda}{4\pi}\right)^4Q^5 - 3\left(\frac{\lambda}{4\pi}\right)^2Q^3 + Q\right).\label{eq:D-poly}
\end{align}
For $|\epsilon_D| < 0.1$, the relative error of the approximation is $\sim 1\%$ for small $Q$, and this error decays to $0$ for larger $Q$.
For $-40/41 < \epsilon_D < 1/2$, $Q'$ in \eq{D-poly} is a monotonically increasing function of $Q$ on the observable range $0 \leq Q \leq 4\pi/\lambda$.

If we assume $Q << \lambda/4\pi$, we can further approximate
\begin{align}
    Q' &\approx (1-\epsilon_D)Q.
\end{align}
This shows that the \stretchmorph morph can account for this misalignment in low $Q$ using \stretchmorph parameter $\eta = -\epsilon_D$.

Here at the end, we detail the full derivation of the polynomial approximation in \eq{D-poly} from the Taylor approximation. For convenience, define
\begin{align}
    q &\equiv \frac{\lambda}{4\pi}Q.
\end{align}
We start by using the double angle formula on \eq{d-D-ratio} to get
\begin{align}
    \frac{d}{D} &= \frac{2\tan(\arcsin(q))}{1 - \tan^2(\arcsin(q))}\\
    &= \frac{2q\sqrt{1 - q^2}}{1 - 2q^2},\label{eq:d-D-ratio-q}
\end{align}
where we have used
\begin{align}
    \tan(\arcsin(x)) &= \frac{x}{\sqrt{1 - x^2}}.
\end{align}
Then, we use the half angle formula on \eq{D-taylor} to get
\begin{align}
    \cos\left(\frac{1}{2}\arctan(d/D)\right) &= \sqrt{\frac{1 + \cos(\arctan(d/D))}{2}}\\
    &= \sqrt{\frac{1 + \frac{1}{\sqrt{1 + (d/D)^2}}}{2}},\label{eq:D-intermediary}
\end{align}
where we have used
\begin{align}
    \cos(\arctan(x)) &= \frac{1}{\sqrt{1 + x^2}}.
\end{align}
Plugging in \eq{d-D-ratio-q} into \eq{D-intermediary} and simplifying gives
\begin{align}
    \cos\left(\frac{1}{2}\arctan(d/D)\right) &= \sqrt{1 - q^2}.\label{eq:D-cos-arctan-relation}
\end{align}
We can now use Eqs.~\ref{eq:d-D-ratio-q},\ref{eq:D-cos-arctan-relation} to obtain a polynomial relation
\begin{align}
    \frac{(d/D)\cos\left(\frac{1}{2}\arctan(d/D)\right)}{2((d/D)^2 + 1)} &= q(1 - q^2)(1 - 2q^2).
\end{align}
Applying this to \eq{D-taylor} gives the desired \eq{D-poly}.

\bibliography{ay_pdfmorph,billinge-group-bib,bg-pdf-standards}

\begin{thebibliography}{53}%
\makeatletter
\providecommand \@ifxundefined [1]{%
 \@ifx{#1\undefined}
}%
\providecommand \@ifnum [1]{%
 \ifnum #1\expandafter \@firstoftwo
 \else \expandafter \@secondoftwo
 \fi
}%
\providecommand \@ifx [1]{%
 \ifx #1\expandafter \@firstoftwo
 \else \expandafter \@secondoftwo
 \fi
}%
\providecommand \natexlab [1]{#1}%
\providecommand \enquote  [1]{``#1''}%
\providecommand \bibnamefont  [1]{#1}%
\providecommand \bibfnamefont [1]{#1}%
\providecommand \citenamefont [1]{#1}%
\providecommand \href@noop [0]{\@secondoftwo}%
\providecommand \href [0]{\begingroup \@sanitize@url \@href}%
\providecommand \@href[1]{\@@startlink{#1}\@@href}%
\providecommand \@@href[1]{\endgroup#1\@@endlink}%
\providecommand \@sanitize@url [0]{\catcode `\\12\catcode `\$12\catcode `\&12\catcode `\#12\catcode `\^12\catcode `\_12\catcode `\%12\relax}%
\providecommand \@@startlink[1]{}%
\providecommand \@@endlink[0]{}%
\providecommand \url  [0]{\begingroup\@sanitize@url \@url }%
\providecommand \@url [1]{\endgroup\@href {#1}{\urlprefix }}%
\providecommand \urlprefix  [0]{URL }%
\providecommand \Eprint [0]{\href }%
\providecommand \doibase [0]{http://dx.doi.org/}%
\providecommand \selectlanguage [0]{\@gobble}%
\providecommand \bibinfo  [0]{\@secondoftwo}%
\providecommand \bibfield  [0]{\@secondoftwo}%
\providecommand \translation [1]{[#1]}%
\providecommand \BibitemOpen [0]{}%
\providecommand \bibitemStop [0]{}%
\providecommand \bibitemNoStop [0]{.\EOS\space}%
\providecommand \EOS [0]{\spacefactor3000\relax}%
\providecommand \BibitemShut  [1]{\csname bibitem#1\endcsname}%
\let\auto@bib@innerbib\@empty
\bibitem [{\citenamefont {Dinnebier}\ and\ \citenamefont {Billinge}(2008)}]{dinnebier01GeneralIntroPowderDiffraction2008}%
  \BibitemOpen
  \bibfield  {author} {\bibinfo {author} {\bibfnamefont {R.~E.}\ \bibnamefont {Dinnebier}}\ and\ \bibinfo {author} {\bibfnamefont {S.~J.~L.}\ \bibnamefont {Billinge}},\ }in\ \href {\doibase 10.1039/9781847558237-00001} {\emph {\bibinfo {booktitle} {Powder {{Diffraction}}}}},\ \bibinfo {editor} {edited by\ \bibinfo {editor} {\bibfnamefont {R.~E.}\ \bibnamefont {Dinnebier}}\ and\ \bibinfo {editor} {\bibfnamefont {S.~J.~L.}\ \bibnamefont {Billinge}}}\ (\bibinfo  {publisher} {Royal Society of Chemistry},\ \bibinfo {address} {Cambridge},\ \bibinfo {year} {2008})\ pp.\ \bibinfo {pages} {1--19}\BibitemShut {NoStop}%
\bibitem [{\citenamefont {Egami}\ and\ \citenamefont {Billinge}(2012)}]{egami;b;utbp12}%
  \BibitemOpen
  \bibfield  {author} {\bibinfo {author} {\bibfnamefont {T.}~\bibnamefont {Egami}}\ and\ \bibinfo {author} {\bibfnamefont {S.~J.~L.}\ \bibnamefont {Billinge}},\ }\href@noop {} {\emph {\bibinfo {title} {Underneath the {{Bragg Peaks}}: {{Structural Analysis}} of {{Complex Materials}}}}},\ \bibinfo {edition} {2nd}\ ed.,\ \bibinfo {series} {Pergamon Materials Series}\ No.~\bibinfo {number} {16}\ (\bibinfo  {publisher} {Elsevier},\ \bibinfo {address} {Amsterdam},\ \bibinfo {year} {2012})\BibitemShut {NoStop}%
\bibitem [{\citenamefont {Billinge}\ and\ \citenamefont {Jensen}(2024)}]{billi;b;apdfap24}%
  \BibitemOpen
  \bibfield  {author} {\bibinfo {author} {\bibfnamefont {S.~J.~L.}\ \bibnamefont {Billinge}}\ and\ \bibinfo {author} {\bibfnamefont {K.~M.~{\O}.}\ \bibnamefont {Jensen}},\ }\href@noop {} {\emph {\bibinfo {title} {Atomic {{Pair Distribution Function Analysis}} - {{A Primer}}}}},\ International {{Union}} of {{Crystallography Texts}} on {{Crystallography}}\ (\bibinfo  {publisher} {Oxford University Press},\ \bibinfo {address} {Oxford, England},\ \bibinfo {year} {2024})\BibitemShut {NoStop}%
\bibitem [{\citenamefont {Pearson}\ and\ \citenamefont {Galton}(1895)}]{pearsonVIINoteRegression1895}%
  \BibitemOpen
  \bibfield  {author} {\bibinfo {author} {\bibfnamefont {K.}~\bibnamefont {Pearson}}\ and\ \bibinfo {author} {\bibfnamefont {F.}~\bibnamefont {Galton}},\ }\href {\doibase 10.1098/rspl.1895.0041} {\bibfield  {journal} {\bibinfo  {journal} {Proceedings of the Royal Society of London}\ }\textbf {\bibinfo {volume} {58}},\ \bibinfo {pages} {240} (\bibinfo {year} {1895})}\BibitemShut {NoStop}%
\bibitem [{\citenamefont {\phantom{}Regents of the University~of California}(1988)}]{bsd3clause88}%
  \BibitemOpen
  \bibfield  {author} {\bibinfo {author} {\bibnamefont {\phantom{}Regents of the University~of California}},\ }\href {https://opensource.org/license/bsd-3-clause} {\enquote {\bibinfo {title} {The 3-clause bsd license},}\ } (\bibinfo {year} {1988})\BibitemShut {NoStop}%
\bibitem [{\citenamefont {Yang}\ \emph {et~al.}(2025{\natexlab{a}})\citenamefont {Yang}, \citenamefont {Farrow}, \citenamefont {Liu}, \citenamefont {Kitsu},\ and\ \citenamefont {Billinge}}]{morphPyPiRelease25}%
  \BibitemOpen
  \bibfield  {author} {\bibinfo {author} {\bibfnamefont {A.}~\bibnamefont {Yang}}, \bibinfo {author} {\bibfnamefont {C.~L.}\ \bibnamefont {Farrow}}, \bibinfo {author} {\bibfnamefont {C.-H.}\ \bibnamefont {Liu}}, \bibinfo {author} {\bibfnamefont {L.}~\bibnamefont {Kitsu}}, \ and\ \bibinfo {author} {\bibfnamefont {S.~J.~L.}\ \bibnamefont {Billinge}},\ }\href {https://pypi.org/project/diffpy.morph/} {\enquote {\bibinfo {title} {diffpy.morph (pypi package)},}\ } (\bibinfo {year} {2025}{\natexlab{a}}),\ \Eprint {http://arxiv.org/abs/https://pypi.org/project/diffpy.morph} {https://pypi.org/project/diffpy.morph} \BibitemShut {NoStop}%
\bibitem [{\citenamefont {conda-forge community}(2025)}]{morphCondaForgeRelease25}%
  \BibitemOpen
  \bibfield  {author} {\bibinfo {author} {\bibnamefont {conda-forge community}},\ }\href {https://anaconda.org/channels/conda-forge/packages/diffpy.morph/overview} {\enquote {\bibinfo {title} {diffpy.morph (conda-forge package)},}\ } (\bibinfo {year} {2025}),\ \Eprint {http://arxiv.org/abs/https://anaconda.org/channels/conda-forge/packages/diffpy.morph/overview} {https://anaconda.org/channels/conda-forge/packages/diffpy.morph/overview} \BibitemShut {NoStop}%
\bibitem [{\citenamefont {Lazarevi\ifmmode~\acute{c}\else \'{c}\fi{}}\ \emph {et~al.}(2014{\natexlab{a}})\citenamefont {Lazarevi\ifmmode~\acute{c}\else \'{c}\fi{}}, \citenamefont {Bozin}, \citenamefont {\ifmmode \check{S}\else \v{S}\fi{}\ifmmode \acute{c}\else \'{c}\fi{}epanovi\ifmmode~\acute{c}\else \'{c}\fi{}}, \citenamefont {Opa\ifmmode \check{c}\else \v{c}\fi{}i\ifmmode~\acute{c}\else \'{c}\fi{}}, \citenamefont {Lei}, \citenamefont {Petrovic},\ and\ \citenamefont {Popovi\ifmmode~\acute{c}\else \'{c}\fi{}}}]{pretrovicIrTe2SamplePreparation14}%
  \BibitemOpen
  \bibfield  {author} {\bibinfo {author} {\bibfnamefont {N.}~\bibnamefont {Lazarevi\ifmmode~\acute{c}\else \'{c}\fi{}}}, \bibinfo {author} {\bibfnamefont {E.~S.}\ \bibnamefont {Bozin}}, \bibinfo {author} {\bibfnamefont {M.}~\bibnamefont {\ifmmode \check{S}\else \v{S}\fi{}\ifmmode \acute{c}\else \'{c}\fi{}epanovi\ifmmode~\acute{c}\else \'{c}\fi{}}}, \bibinfo {author} {\bibfnamefont {M.}~\bibnamefont {Opa\ifmmode \check{c}\else \v{c}\fi{}i\ifmmode~\acute{c}\else \'{c}\fi{}}}, \bibinfo {author} {\bibfnamefont {H.}~\bibnamefont {Lei}}, \bibinfo {author} {\bibfnamefont {C.}~\bibnamefont {Petrovic}}, \ and\ \bibinfo {author} {\bibfnamefont {Z.~V.}\ \bibnamefont {Popovi\ifmmode~\acute{c}\else \'{c}\fi{}}},\ }\href {\doibase 10.1103/PhysRevB.89.224301} {\bibfield  {journal} {\bibinfo  {journal} {Phys. Rev. B}\ }\textbf {\bibinfo {volume} {89}},\ \bibinfo {pages} {224301} (\bibinfo {year} {2014}{\natexlab{a}})}\BibitemShut {NoStop}%
\bibitem [{\citenamefont {Liu}\ \emph {et~al.}(2018)\citenamefont {Liu}, \citenamefont {Lei}, \citenamefont {Wang}, \citenamefont {Abeykoon}, \citenamefont {Warren}, \citenamefont {Bozin},\ and\ \citenamefont {Petrovic}}]{petrovicIrRhTe2SamplePreparation18}%
  \BibitemOpen
  \bibfield  {author} {\bibinfo {author} {\bibfnamefont {Y.}~\bibnamefont {Liu}}, \bibinfo {author} {\bibfnamefont {H.}~\bibnamefont {Lei}}, \bibinfo {author} {\bibfnamefont {K.}~\bibnamefont {Wang}}, \bibinfo {author} {\bibfnamefont {M.}~\bibnamefont {Abeykoon}}, \bibinfo {author} {\bibfnamefont {J.~B.}\ \bibnamefont {Warren}}, \bibinfo {author} {\bibfnamefont {E.}~\bibnamefont {Bozin}}, \ and\ \bibinfo {author} {\bibfnamefont {C.}~\bibnamefont {Petrovic}},\ }\href {\doibase 10.1103/PhysRevB.98.094519} {\bibfield  {journal} {\bibinfo  {journal} {Phys. Rev. B}\ }\textbf {\bibinfo {volume} {98}},\ \bibinfo {pages} {094519} (\bibinfo {year} {2018})}\BibitemShut {NoStop}%
\bibitem [{\citenamefont {Yang}\ \emph {et~al.}(2012)\citenamefont {Yang}, \citenamefont {Choi}, \citenamefont {Oh}, \citenamefont {Hogan}, \citenamefont {Horibe}, \citenamefont {Kim}, \citenamefont {Min},\ and\ \citenamefont {Cheong}}]{yangIrRhTe2SamplePreparation12}%
  \BibitemOpen
  \bibfield  {author} {\bibinfo {author} {\bibfnamefont {J.~J.}\ \bibnamefont {Yang}}, \bibinfo {author} {\bibfnamefont {Y.~J.}\ \bibnamefont {Choi}}, \bibinfo {author} {\bibfnamefont {Y.~S.}\ \bibnamefont {Oh}}, \bibinfo {author} {\bibfnamefont {A.}~\bibnamefont {Hogan}}, \bibinfo {author} {\bibfnamefont {Y.}~\bibnamefont {Horibe}}, \bibinfo {author} {\bibfnamefont {K.}~\bibnamefont {Kim}}, \bibinfo {author} {\bibfnamefont {B.~I.}\ \bibnamefont {Min}}, \ and\ \bibinfo {author} {\bibfnamefont {S.-W.}\ \bibnamefont {Cheong}},\ }\href {\doibase 10.1103/PhysRevLett.108.116402} {\bibfield  {journal} {\bibinfo  {journal} {Phys. Rev. Lett.}\ }\textbf {\bibinfo {volume} {108}},\ \bibinfo {pages} {116402} (\bibinfo {year} {2012})}\BibitemShut {NoStop}%
\bibitem [{\citenamefont {Pyon}\ \emph {et~al.}(2012)\citenamefont {Pyon}, \citenamefont {Kudo},\ and\ \citenamefont {Nohara}}]{pyonIrPtTe2SamplePreparation12}%
  \BibitemOpen
  \bibfield  {author} {\bibinfo {author} {\bibfnamefont {S.}~\bibnamefont {Pyon}}, \bibinfo {author} {\bibfnamefont {K.}~\bibnamefont {Kudo}}, \ and\ \bibinfo {author} {\bibfnamefont {M.}~\bibnamefont {Nohara}},\ }\href {\doibase 10.1143/JPSJ.81.053701} {\bibfield  {journal} {\bibinfo  {journal} {Journal of the Physical Society of Japan}\ }\textbf {\bibinfo {volume} {81}},\ \bibinfo {pages} {053701} (\bibinfo {year} {2012})}\BibitemShut {NoStop}%
\bibitem [{\citenamefont {Yu}\ \emph {et~al.}(2018{\natexlab{a}})\citenamefont {Yu}, \citenamefont {Banerjee}, \citenamefont {Lei}, \citenamefont {Sinclair}, \citenamefont {Abeykoon}, \citenamefont {Zhou}, \citenamefont {Petrovic}, \citenamefont {Guguchia},\ and\ \citenamefont {Bozin}}]{banerjeeBozinLocalFlucctuatingDimers18}%
  \BibitemOpen
  \bibfield  {author} {\bibinfo {author} {\bibfnamefont {R.}~\bibnamefont {Yu}}, \bibinfo {author} {\bibfnamefont {S.}~\bibnamefont {Banerjee}}, \bibinfo {author} {\bibfnamefont {H.~C.}\ \bibnamefont {Lei}}, \bibinfo {author} {\bibfnamefont {R.}~\bibnamefont {Sinclair}}, \bibinfo {author} {\bibfnamefont {M.}~\bibnamefont {Abeykoon}}, \bibinfo {author} {\bibfnamefont {H.~D.}\ \bibnamefont {Zhou}}, \bibinfo {author} {\bibfnamefont {C.}~\bibnamefont {Petrovic}}, \bibinfo {author} {\bibfnamefont {Z.}~\bibnamefont {Guguchia}}, \ and\ \bibinfo {author} {\bibfnamefont {E.~S.}\ \bibnamefont {Bozin}},\ }\href {\doibase 10.1103/PhysRevB.97.174515} {\bibfield  {journal} {\bibinfo  {journal} {Phys. Rev. B}\ }\textbf {\bibinfo {volume} {97}},\ \bibinfo {pages} {174515} (\bibinfo {year} {2018}{\natexlab{a}})}\BibitemShut {NoStop}%
\bibitem [{\citenamefont {Jeong}\ \emph {et~al.}(2003)\citenamefont {Jeong}, \citenamefont {Heffner}, \citenamefont {Graf},\ and\ \citenamefont {Billinge}}]{jeongCorrelatedAtomicMotion03}%
  \BibitemOpen
  \bibfield  {author} {\bibinfo {author} {\bibfnamefont {I.-K.}\ \bibnamefont {Jeong}}, \bibinfo {author} {\bibfnamefont {R.~H.}\ \bibnamefont {Heffner}}, \bibinfo {author} {\bibfnamefont {M.~J.}\ \bibnamefont {Graf}}, \ and\ \bibinfo {author} {\bibfnamefont {S.~J.~L.}\ \bibnamefont {Billinge}},\ }\href {\doibase 10.1103/PhysRevB.67.104301} {\bibfield  {journal} {\bibinfo  {journal} {Phys. Rev. B}\ }\textbf {\bibinfo {volume} {67}},\ \bibinfo {pages} {104301} (\bibinfo {year} {2003})}\BibitemShut {NoStop}%
\bibitem [{\citenamefont {Yu}\ \emph {et~al.}(2018{\natexlab{b}})\citenamefont {Yu}, \citenamefont {Banerjee}, \citenamefont {Lei}, \citenamefont {Abeykoon}, \citenamefont {Petrovic}, \citenamefont {Guguchia},\ and\ \citenamefont {Bozin}}]{banerjeeBozinDSTransition18}%
  \BibitemOpen
  \bibfield  {author} {\bibinfo {author} {\bibfnamefont {R.}~\bibnamefont {Yu}}, \bibinfo {author} {\bibfnamefont {S.}~\bibnamefont {Banerjee}}, \bibinfo {author} {\bibfnamefont {H.}~\bibnamefont {Lei}}, \bibinfo {author} {\bibfnamefont {M.}~\bibnamefont {Abeykoon}}, \bibinfo {author} {\bibfnamefont {C.}~\bibnamefont {Petrovic}}, \bibinfo {author} {\bibfnamefont {Z.}~\bibnamefont {Guguchia}}, \ and\ \bibinfo {author} {\bibfnamefont {E.~S.}\ \bibnamefont {Bozin}},\ }\href {\doibase 10.1103/PhysRevB.98.134506} {\bibfield  {journal} {\bibinfo  {journal} {Phys. Rev. B}\ }\textbf {\bibinfo {volume} {98}},\ \bibinfo {pages} {134506} (\bibinfo {year} {2018}{\natexlab{b}})}\BibitemShut {NoStop}%
\bibitem [{\citenamefont {Farrow}\ \emph {et~al.}(2007)\citenamefont {Farrow}, \citenamefont {Juh\'as}, \citenamefont {Liu}, \citenamefont {Bryndin}, \citenamefont {Božin}, \citenamefont {Bloch}, \citenamefont {Proffen},\ and\ \citenamefont {Billinge}}]{farrowPDFfit2PDFgui}%
  \BibitemOpen
  \bibfield  {author} {\bibinfo {author} {\bibfnamefont {C.~L.}\ \bibnamefont {Farrow}}, \bibinfo {author} {\bibfnamefont {P.}~\bibnamefont {Juh\'as}}, \bibinfo {author} {\bibfnamefont {J.~W.}\ \bibnamefont {Liu}}, \bibinfo {author} {\bibfnamefont {D.}~\bibnamefont {Bryndin}}, \bibinfo {author} {\bibfnamefont {E.~S.}\ \bibnamefont {Božin}}, \bibinfo {author} {\bibfnamefont {J.}~\bibnamefont {Bloch}}, \bibinfo {author} {\bibfnamefont {T.}~\bibnamefont {Proffen}}, \ and\ \bibinfo {author} {\bibfnamefont {S.~J.~L.}\ \bibnamefont {Billinge}},\ }\href {\doibase 10.1088/0953-8984/19/33/335219} {\bibfield  {journal} {\bibinfo  {journal} {Journal of Physics: Condensed Matter}\ }\textbf {\bibinfo {volume} {19}},\ \bibinfo {pages} {335219} (\bibinfo {year} {2007})}\BibitemShut {NoStop}%
\bibitem [{\citenamefont {Frandsen}\ \emph {et~al.}(2023)\citenamefont {Frandsen}, \citenamefont {Yang},\ and\ \citenamefont {Billinge}}]{pdfttpdata23}%
  \BibitemOpen
  \bibfield  {author} {\bibinfo {author} {\bibfnamefont {B.}~\bibnamefont {Frandsen}}, \bibinfo {author} {\bibfnamefont {L.}~\bibnamefont {Yang}}, \ and\ \bibinfo {author} {\bibfnamefont {S.~J.~L.}\ \bibnamefont {Billinge}},\ }\href@noop {} {\enquote {\bibinfo {title} {{pdfttp\_data}},}\ } (\bibinfo {year} {2023}),\ \Eprint {http://arxiv.org/abs/{https://github.com/Billingegroup/pdfttp\_data}} {{https://github.com/Billingegroup/pdfttp\_data}} \BibitemShut {NoStop}%
\bibitem [{\citenamefont {Grüneisen}(1912)}]{gruneisenTheoryMonatomic1912}%
  \BibitemOpen
  \bibfield  {author} {\bibinfo {author} {\bibfnamefont {E.}~\bibnamefont {Grüneisen}},\ }\href {\doibase https://doi.org/10.1002/andp.19123441202} {\bibfield  {journal} {\bibinfo  {journal} {Annalen der Physik}\ }\textbf {\bibinfo {volume} {344}},\ \bibinfo {pages} {257} (\bibinfo {year} {1912})}\BibitemShut {NoStop}%
\bibitem [{\citenamefont {Chang}(1967)}]{changBulkModulusTemperatureDependence1967}%
  \BibitemOpen
  \bibfield  {author} {\bibinfo {author} {\bibfnamefont {Y.}~\bibnamefont {Chang}},\ }\href {\doibase https://doi.org/10.1016/0022-3697(67)90101-1} {\bibfield  {journal} {\bibinfo  {journal} {Journal of Physics and Chemistry of Solids}\ }\textbf {\bibinfo {volume} {28}},\ \bibinfo {pages} {697} (\bibinfo {year} {1967})}\BibitemShut {NoStop}%
\bibitem [{\citenamefont {Schroeder}(2021)}]{schroederIntroduction21}%
  \BibitemOpen
  \bibfield  {author} {\bibinfo {author} {\bibfnamefont {D.}~\bibnamefont {Schroeder}},\ }\href@noop {} {\emph {\bibinfo {title} {An Introduction to Thermal Physics}}}\ (\bibinfo  {publisher} {Oxford University Press},\ \bibinfo {year} {2021})\BibitemShut {NoStop}%
\bibitem [{\citenamefont {Anderson}(1966)}]{andersonDerivationWachtman1966}%
  \BibitemOpen
  \bibfield  {author} {\bibinfo {author} {\bibfnamefont {O.~L.}\ \bibnamefont {Anderson}},\ }\href {\doibase 10.1103/PhysRev.144.553} {\bibfield  {journal} {\bibinfo  {journal} {Phys. Rev.}\ }\textbf {\bibinfo {volume} {144}},\ \bibinfo {pages} {553} (\bibinfo {year} {1966})}\BibitemShut {NoStop}%
\bibitem [{\citenamefont {Lazarevi\ifmmode~\acute{c}\else \'{c}\fi{}}\ \emph {et~al.}(2014{\natexlab{b}})\citenamefont {Lazarevi\ifmmode~\acute{c}\else \'{c}\fi{}}, \citenamefont {Bozin}, \citenamefont {\ifmmode \check{S}\else \v{S}\fi{}\ifmmode \acute{c}\else \'{c}\fi{}epanovi\ifmmode~\acute{c}\else \'{c}\fi{}}, \citenamefont {Opa\ifmmode \check{c}\else \v{c}\fi{}i\ifmmode~\acute{c}\else \'{c}\fi{}}, \citenamefont {Lei}, \citenamefont {Petrovic},\ and\ \citenamefont {Popovi\ifmmode~\acute{c}\else \'{c}\fi{}}}]{lazarevicProbingIrTePRC14}%
  \BibitemOpen
  \bibfield  {author} {\bibinfo {author} {\bibfnamefont {N.}~\bibnamefont {Lazarevi\ifmmode~\acute{c}\else \'{c}\fi{}}}, \bibinfo {author} {\bibfnamefont {E.~S.}\ \bibnamefont {Bozin}}, \bibinfo {author} {\bibfnamefont {M.}~\bibnamefont {\ifmmode \check{S}\else \v{S}\fi{}\ifmmode \acute{c}\else \'{c}\fi{}epanovi\ifmmode~\acute{c}\else \'{c}\fi{}}}, \bibinfo {author} {\bibfnamefont {M.}~\bibnamefont {Opa\ifmmode \check{c}\else \v{c}\fi{}i\ifmmode~\acute{c}\else \'{c}\fi{}}}, \bibinfo {author} {\bibfnamefont {H.}~\bibnamefont {Lei}}, \bibinfo {author} {\bibfnamefont {C.}~\bibnamefont {Petrovic}}, \ and\ \bibinfo {author} {\bibfnamefont {Z.~V.}\ \bibnamefont {Popovi\ifmmode~\acute{c}\else \'{c}\fi{}}},\ }\href {\doibase 10.1103/PhysRevB.89.224301} {\bibfield  {journal} {\bibinfo  {journal} {Phys. Rev. B}\ }\textbf {\bibinfo {volume} {89}},\ \bibinfo {pages} {224301} (\bibinfo {year} {2014}{\natexlab{b}})}\BibitemShut {NoStop}%
\bibitem [{\citenamefont {Ushakov}\ \emph {et~al.}(2015)\citenamefont {Ushakov}, \citenamefont {Navrotsky}, \citenamefont {Weber},\ and\ \citenamefont {Neuefeind}}]{ushakovYSZHighTemperatureThermalExpansion15}%
  \BibitemOpen
  \bibfield  {author} {\bibinfo {author} {\bibfnamefont {S.~V.}\ \bibnamefont {Ushakov}}, \bibinfo {author} {\bibfnamefont {A.}~\bibnamefont {Navrotsky}}, \bibinfo {author} {\bibfnamefont {R.~J.~K.}\ \bibnamefont {Weber}}, \ and\ \bibinfo {author} {\bibfnamefont {J.~C.}\ \bibnamefont {Neuefeind}},\ }\href {https://doi.org/10.1111/jace.13767} {\bibfield  {journal} {\bibinfo  {journal} {{Journal of the American Ceramic Society}}\ }\textbf {\bibinfo {volume} {98}},\ \bibinfo {pages} {3381} (\bibinfo {year} {2015})}\BibitemShut {NoStop}%
\bibitem [{\citenamefont {Aldebert}(1984)}]{aldebertPyrometryLimitationsHighTemperature84}%
  \BibitemOpen
  \bibfield  {author} {\bibinfo {author} {\bibfnamefont {P.}~\bibnamefont {Aldebert}},\ }\href {\doibase 10.1051/rphysap:01984001909064900} {\bibfield  {journal} {\bibinfo  {journal} {{Revue de Physique Appliqu{\'e}e}}\ }\textbf {\bibinfo {volume} {19}},\ \bibinfo {pages} {649} (\bibinfo {year} {1984})}\BibitemShut {NoStop}%
\bibitem [{\citenamefont {Ahmad}\ \emph {et~al.}(2015)\citenamefont {Ahmad}, \citenamefont {Van~Campen}, \citenamefont {Fields}, \citenamefont {Yu}, \citenamefont {Pool}, \citenamefont {Parilla}, \citenamefont {Ginley}, \citenamefont {Van~Hest},\ and\ \citenamefont {Toney}}]{imteyazRapidThermalProcessingChamber15}%
  \BibitemOpen
  \bibfield  {author} {\bibinfo {author} {\bibfnamefont {M.~I.}\ \bibnamefont {Ahmad}}, \bibinfo {author} {\bibfnamefont {D.~G.}\ \bibnamefont {Van~Campen}}, \bibinfo {author} {\bibfnamefont {J.~D.}\ \bibnamefont {Fields}}, \bibinfo {author} {\bibfnamefont {J.}~\bibnamefont {Yu}}, \bibinfo {author} {\bibfnamefont {V.~L.}\ \bibnamefont {Pool}}, \bibinfo {author} {\bibfnamefont {P.~A.}\ \bibnamefont {Parilla}}, \bibinfo {author} {\bibfnamefont {D.~S.}\ \bibnamefont {Ginley}}, \bibinfo {author} {\bibfnamefont {M.~F. A.~M.}\ \bibnamefont {Van~Hest}}, \ and\ \bibinfo {author} {\bibfnamefont {M.~F.}\ \bibnamefont {Toney}},\ }\href {\doibase 10.1063/1.4904848} {\bibfield  {journal} {\bibinfo  {journal} {Review of Scientific Instruments}\ }\textbf {\bibinfo {volume} {86}},\ \bibinfo {pages} {013902} (\bibinfo {year} {2015})}\BibitemShut {NoStop}%
\bibitem [{\citenamefont {{Witz, Grégoire and Shklover, Valery and Steurer, Walter and Bachegowda, Sharath and Bossmann, Hans-Peter}}(2007)}]{witzPhaseDiagram07}%
  \BibitemOpen
  \bibfield  {author} {\bibinfo {author} {\bibnamefont {{Witz, Grégoire and Shklover, Valery and Steurer, Walter and Bachegowda, Sharath and Bossmann, Hans-Peter}}},\ }\href {https://doi.org/10.1111/j.1551-2916.2007.01785.x} {\bibfield  {journal} {\bibinfo  {journal} {{Journal of the American Ceramic Society}}\ }\textbf {\bibinfo {volume} {90}},\ \bibinfo {pages} {2935} (\bibinfo {year} {2007})}\BibitemShut {NoStop}%
\bibitem [{\citenamefont {Hayashi}\ \emph {et~al.}(2005)\citenamefont {Hayashi}, \citenamefont {Saitou}, \citenamefont {Maruyama}, \citenamefont {Inaba}, \citenamefont {Kawamura},\ and\ \citenamefont {Mori}}]{hayashiYSZThermalExpansionCoefficient05}%
  \BibitemOpen
  \bibfield  {author} {\bibinfo {author} {\bibfnamefont {H.}~\bibnamefont {Hayashi}}, \bibinfo {author} {\bibfnamefont {T.}~\bibnamefont {Saitou}}, \bibinfo {author} {\bibfnamefont {N.}~\bibnamefont {Maruyama}}, \bibinfo {author} {\bibfnamefont {H.}~\bibnamefont {Inaba}}, \bibinfo {author} {\bibfnamefont {K.}~\bibnamefont {Kawamura}}, \ and\ \bibinfo {author} {\bibfnamefont {M.}~\bibnamefont {Mori}},\ }\href {\doibase https://doi.org/10.1016/j.ssi.2004.08.021} {\bibfield  {journal} {\bibinfo  {journal} {Solid State Ionics}\ }\textbf {\bibinfo {volume} {176}},\ \bibinfo {pages} {613} (\bibinfo {year} {2005})}\BibitemShut {NoStop}%
\bibitem [{\citenamefont {Krogstad}\ \emph {et~al.}(2015)\citenamefont {Krogstad}, \citenamefont {Gao}, \citenamefont {Bai}, \citenamefont {Wang}, \citenamefont {Lipkin},\ and\ \citenamefont {Levi}}]{krogstadTetragonalYSZHighTemperatureThermalExpansion15}%
  \BibitemOpen
  \bibfield  {author} {\bibinfo {author} {\bibfnamefont {J.~A.}\ \bibnamefont {Krogstad}}, \bibinfo {author} {\bibfnamefont {Y.}~\bibnamefont {Gao}}, \bibinfo {author} {\bibfnamefont {J.}~\bibnamefont {Bai}}, \bibinfo {author} {\bibfnamefont {J.}~\bibnamefont {Wang}}, \bibinfo {author} {\bibfnamefont {D.~M.}\ \bibnamefont {Lipkin}}, \ and\ \bibinfo {author} {\bibfnamefont {C.~G.}\ \bibnamefont {Levi}},\ }\href {https://doi.org/10.1111/jace.13249} {\bibfield  {journal} {\bibinfo  {journal} {{Journal of the American Ceramic Society}}\ }\textbf {\bibinfo {volume} {98}},\ \bibinfo {pages} {247} (\bibinfo {year} {2015})}\BibitemShut {NoStop}%
\bibitem [{\citenamefont {Goyal}\ and\ \citenamefont {Singh}(2020)}]{goyalSizeProperties20}%
  \BibitemOpen
  \bibfield  {author} {\bibinfo {author} {\bibfnamefont {M.}~\bibnamefont {Goyal}}\ and\ \bibinfo {author} {\bibfnamefont {M.}~\bibnamefont {Singh}},\ }\href {\doibase 10.1007/s00339-020-3327-9} {\bibfield  {journal} {\bibinfo  {journal} {Applied Physics A}\ }\textbf {\bibinfo {volume} {126}},\ \bibinfo {pages} {176} (\bibinfo {year} {2020})}\BibitemShut {NoStop}%
\bibitem [{\citenamefont {Tolbert}\ and\ \citenamefont {Alivisatos}(1994)}]{tolbertCdSeSizeProperties94}%
  \BibitemOpen
  \bibfield  {author} {\bibinfo {author} {\bibfnamefont {S.~H.}\ \bibnamefont {Tolbert}}\ and\ \bibinfo {author} {\bibfnamefont {A.~P.}\ \bibnamefont {Alivisatos}},\ }\href {\doibase 10.1126/science.265.5170.373} {\bibfield  {journal} {\bibinfo  {journal} {Science}\ }\textbf {\bibinfo {volume} {265}},\ \bibinfo {pages} {373} (\bibinfo {year} {1994})}\BibitemShut {NoStop}%
\bibitem [{\citenamefont {Farrow}\ and\ \citenamefont {Billinge}(2009)}]{farrowBillingeSasPdf09}%
  \BibitemOpen
  \bibfield  {author} {\bibinfo {author} {\bibfnamefont {C.~L.}\ \bibnamefont {Farrow}}\ and\ \bibinfo {author} {\bibfnamefont {S.~J.~L.}\ \bibnamefont {Billinge}},\ }\href {\doibase 10.1107/S0108767309009714} {\bibfield  {journal} {\bibinfo  {journal} {Acta Crystallographica Section A}\ }\textbf {\bibinfo {volume} {65}},\ \bibinfo {pages} {232} (\bibinfo {year} {2009})}\BibitemShut {NoStop}%
\bibitem [{\citenamefont {Banerjee}\ \emph {et~al.}(2019)\citenamefont {Banerjee}, \citenamefont {Zangiabadi}, \citenamefont {{Mahdavi-Shakib}}, \citenamefont {Husremovic}, \citenamefont {Frederick}, \citenamefont {Barmak}, \citenamefont {Austin},\ and\ \citenamefont {Billinge}}]{banerjeeQuantitativeStructuralCharacterization2019c}%
  \BibitemOpen
  \bibfield  {author} {\bibinfo {author} {\bibfnamefont {S.}~\bibnamefont {Banerjee}}, \bibinfo {author} {\bibfnamefont {A.}~\bibnamefont {Zangiabadi}}, \bibinfo {author} {\bibfnamefont {A.}~\bibnamefont {{Mahdavi-Shakib}}}, \bibinfo {author} {\bibfnamefont {S.}~\bibnamefont {Husremovic}}, \bibinfo {author} {\bibfnamefont {B.~G.}\ \bibnamefont {Frederick}}, \bibinfo {author} {\bibfnamefont {K.}~\bibnamefont {Barmak}}, \bibinfo {author} {\bibfnamefont {R.~N.}\ \bibnamefont {Austin}}, \ and\ \bibinfo {author} {\bibfnamefont {S.~J.~L.}\ \bibnamefont {Billinge}},\ }\href {\doibase 10.1021/acsanm.9b01246} {\bibfield  {journal} {\bibinfo  {journal} {ACS Appl. Nano Mater.}\ }\textbf {\bibinfo {volume} {2}},\ \bibinfo {pages} {6268} (\bibinfo {year} {2019})}\BibitemShut {NoStop}%
\bibitem [{\citenamefont {Banerjee}\ \emph {et~al.}(2020)\citenamefont {Banerjee}, \citenamefont {Liu}, \citenamefont {Jensen}, \citenamefont {Juh{\'a}s}, \citenamefont {Lee}, \citenamefont {Tofanelli}, \citenamefont {Ackerson}, \citenamefont {Murray},\ and\ \citenamefont {Billinge}}]{banerjeeClusterminingApproachDetermining2020a}%
  \BibitemOpen
  \bibfield  {author} {\bibinfo {author} {\bibfnamefont {S.}~\bibnamefont {Banerjee}}, \bibinfo {author} {\bibfnamefont {C.-H.}\ \bibnamefont {Liu}}, \bibinfo {author} {\bibfnamefont {K.~M.~{\O}.}\ \bibnamefont {Jensen}}, \bibinfo {author} {\bibfnamefont {P.}~\bibnamefont {Juh{\'a}s}}, \bibinfo {author} {\bibfnamefont {J.~D.}\ \bibnamefont {Lee}}, \bibinfo {author} {\bibfnamefont {M.}~\bibnamefont {Tofanelli}}, \bibinfo {author} {\bibfnamefont {C.~J.}\ \bibnamefont {Ackerson}}, \bibinfo {author} {\bibfnamefont {C.~B.}\ \bibnamefont {Murray}}, \ and\ \bibinfo {author} {\bibfnamefont {S.~J.~L.}\ \bibnamefont {Billinge}},\ }\href {\doibase 10.1107/S2053273319013214} {\bibfield  {journal} {\bibinfo  {journal} {Acta Crystallogr A Found Adv}\ }\textbf {\bibinfo {volume} {76}},\ \bibinfo {pages} {24} (\bibinfo {year} {2020})}\BibitemShut {NoStop}%
\bibitem [{\citenamefont {Tao}\ \emph {et~al.}(2024)\citenamefont {Tao}, \citenamefont {Billet}, \citenamefont {De~Roo},\ and\ \citenamefont {Billinge}}]{taoRapidModellingLocalStructure24}%
  \BibitemOpen
  \bibfield  {author} {\bibinfo {author} {\bibfnamefont {S.}~\bibnamefont {Tao}}, \bibinfo {author} {\bibfnamefont {J.}~\bibnamefont {Billet}}, \bibinfo {author} {\bibfnamefont {J.}~\bibnamefont {De~Roo}}, \ and\ \bibinfo {author} {\bibfnamefont {S.~J.~L.}\ \bibnamefont {Billinge}},\ }\href {\doibase 10.1021/acs.chemmater.3c03002} {\bibfield  {journal} {\bibinfo  {journal} {Chemistry of Materials}\ }\textbf {\bibinfo {volume} {36}},\ \bibinfo {pages} {10912} (\bibinfo {year} {2024})}\BibitemShut {NoStop}%
\bibitem [{\citenamefont {Masadeh}\ \emph {et~al.}(2007)\citenamefont {Masadeh}, \citenamefont {Bo\ifmmode~\check{z}\else \v{z}\fi{}in}, \citenamefont {Farrow}, \citenamefont {Paglia}, \citenamefont {Juh\'as}, \citenamefont {Billinge}, \citenamefont {Karkamkar},\ and\ \citenamefont {Kanatzidis}}]{masadehCdSeNano07}%
  \BibitemOpen
  \bibfield  {author} {\bibinfo {author} {\bibfnamefont {A.~S.}\ \bibnamefont {Masadeh}}, \bibinfo {author} {\bibfnamefont {E.~S.}\ \bibnamefont {Bo\ifmmode~\check{z}\else \v{z}\fi{}in}}, \bibinfo {author} {\bibfnamefont {C.~L.}\ \bibnamefont {Farrow}}, \bibinfo {author} {\bibfnamefont {G.}~\bibnamefont {Paglia}}, \bibinfo {author} {\bibfnamefont {P.}~\bibnamefont {Juh\'as}}, \bibinfo {author} {\bibfnamefont {S.~J.~L.}\ \bibnamefont {Billinge}}, \bibinfo {author} {\bibfnamefont {A.}~\bibnamefont {Karkamkar}}, \ and\ \bibinfo {author} {\bibfnamefont {M.~G.}\ \bibnamefont {Kanatzidis}},\ }\href {\doibase 10.1103/PhysRevB.76.115413} {\bibfield  {journal} {\bibinfo  {journal} {Phys. Rev. B}\ }\textbf {\bibinfo {volume} {76}},\ \bibinfo {pages} {115413} (\bibinfo {year} {2007})}\BibitemShut {NoStop}%
\bibitem [{\citenamefont {Guzelturk}\ \emph {et~al.}(2025)\citenamefont {Guzelturk}, \citenamefont {Portner}, \citenamefont {Ondry}, \citenamefont {Ghanbarzadeh}, \citenamefont {Tarantola}, \citenamefont {Jeong}, \citenamefont {Field}, \citenamefont {Chandler}, \citenamefont {Wieman}, \citenamefont {Hopper}, \citenamefont {Watkins}, \citenamefont {Yu}, \citenamefont {Cheng}, \citenamefont {Lin}, \citenamefont {Luo}, \citenamefont {Kramer}, \citenamefont {Shen}, \citenamefont {Reid}, \citenamefont {Borkiewicz}, \citenamefont {Ruett}, \citenamefont {Zhang}, \citenamefont {Lindenberg}, \citenamefont {Ma}, \citenamefont {Schaller}, \citenamefont {Talapin},\ and\ \citenamefont {Cotts}}]{guzelturkUltrafastSymmetryControl25}%
  \BibitemOpen
  \bibfield  {author} {\bibinfo {author} {\bibfnamefont {B.}~\bibnamefont {Guzelturk}}, \bibinfo {author} {\bibfnamefont {J.}~\bibnamefont {Portner}}, \bibinfo {author} {\bibfnamefont {J.}~\bibnamefont {Ondry}}, \bibinfo {author} {\bibfnamefont {S.}~\bibnamefont {Ghanbarzadeh}}, \bibinfo {author} {\bibfnamefont {M.}~\bibnamefont {Tarantola}}, \bibinfo {author} {\bibfnamefont {A.}~\bibnamefont {Jeong}}, \bibinfo {author} {\bibfnamefont {T.}~\bibnamefont {Field}}, \bibinfo {author} {\bibfnamefont {A.~M.}\ \bibnamefont {Chandler}}, \bibinfo {author} {\bibfnamefont {E.}~\bibnamefont {Wieman}}, \bibinfo {author} {\bibfnamefont {T.~R.}\ \bibnamefont {Hopper}}, \bibinfo {author} {\bibfnamefont {N.~E.}\ \bibnamefont {Watkins}}, \bibinfo {author} {\bibfnamefont {J.}~\bibnamefont {Yu}}, \bibinfo {author} {\bibfnamefont {X.}~\bibnamefont {Cheng}}, \bibinfo {author} {\bibfnamefont {M.-F.}\ \bibnamefont {Lin}}, \bibinfo {author} {\bibfnamefont {D.}~\bibnamefont {Luo}}, \bibinfo {author} {\bibfnamefont {P.~L.}\
  \bibnamefont {Kramer}}, \bibinfo {author} {\bibfnamefont {X.}~\bibnamefont {Shen}}, \bibinfo {author} {\bibfnamefont {A.~H.}\ \bibnamefont {Reid}}, \bibinfo {author} {\bibfnamefont {O.}~\bibnamefont {Borkiewicz}}, \bibinfo {author} {\bibfnamefont {U.}~\bibnamefont {Ruett}}, \bibinfo {author} {\bibfnamefont {X.}~\bibnamefont {Zhang}}, \bibinfo {author} {\bibfnamefont {A.~M.}\ \bibnamefont {Lindenberg}}, \bibinfo {author} {\bibfnamefont {J.}~\bibnamefont {Ma}}, \bibinfo {author} {\bibfnamefont {R.~D.}\ \bibnamefont {Schaller}}, \bibinfo {author} {\bibfnamefont {D.~V.}\ \bibnamefont {Talapin}}, \ and\ \bibinfo {author} {\bibfnamefont {B.~L.}\ \bibnamefont {Cotts}},\ }\href {\doibase https://doi.org/10.1002/adma.202414196} {\bibfield  {journal} {\bibinfo  {journal} {Advanced Materials}\ }\textbf {\bibinfo {volume} {37}},\ \bibinfo {pages} {2414196} (\bibinfo {year} {2025})}\BibitemShut {NoStop}%
\bibitem [{\citenamefont {Noda}\ \emph {et~al.}(1987)\citenamefont {Noda}, \citenamefont {Masumoto}, \citenamefont {Ohba}, \citenamefont {Saito}, \citenamefont {Toriumi}, \citenamefont {Iwata},\ and\ \citenamefont {Shibuya}}]{nodaPbSCrystalStructure87}%
  \BibitemOpen
  \bibfield  {author} {\bibinfo {author} {\bibfnamefont {Y.}~\bibnamefont {Noda}}, \bibinfo {author} {\bibfnamefont {K.}~\bibnamefont {Masumoto}}, \bibinfo {author} {\bibfnamefont {S.}~\bibnamefont {Ohba}}, \bibinfo {author} {\bibfnamefont {Y.}~\bibnamefont {Saito}}, \bibinfo {author} {\bibfnamefont {K.}~\bibnamefont {Toriumi}}, \bibinfo {author} {\bibfnamefont {Y.}~\bibnamefont {Iwata}}, \ and\ \bibinfo {author} {\bibfnamefont {I.}~\bibnamefont {Shibuya}},\ }\href {\doibase 10.1107/S0108270187091509} {\bibfield  {journal} {\bibinfo  {journal} {Acta Crystallographica Section C}\ }\textbf {\bibinfo {volume} {43}},\ \bibinfo {pages} {1443} (\bibinfo {year} {1987})}\BibitemShut {NoStop}%
\bibitem [{\citenamefont {Yang}\ \emph {et~al.}(2025{\natexlab{b}})\citenamefont {Yang}, \citenamefont {Farrow}, \citenamefont {Liu}, \citenamefont {Kitsu},\ and\ \citenamefont {Billinge}}]{diffpymorphsoftware25}%
  \BibitemOpen
  \bibfield  {author} {\bibinfo {author} {\bibfnamefont {A.}~\bibnamefont {Yang}}, \bibinfo {author} {\bibfnamefont {C.~L.}\ \bibnamefont {Farrow}}, \bibinfo {author} {\bibfnamefont {C.-H.}\ \bibnamefont {Liu}}, \bibinfo {author} {\bibfnamefont {L.}~\bibnamefont {Kitsu}}, \ and\ \bibinfo {author} {\bibfnamefont {S.~J.~L.}\ \bibnamefont {Billinge}},\ }\href@noop {} {\enquote {\bibinfo {title} {{diffpy.morph}},}\ } (\bibinfo {year} {2025}{\natexlab{b}}),\ \Eprint {http://arxiv.org/abs/{https://github.com/diffpy/diffpy.morph}} {{https://github.com/diffpy/diffpy.morph}} \BibitemShut {NoStop}%
\bibitem [{\citenamefont {Jeong}\ \emph {et~al.}(2024)\citenamefont {Jeong}, \citenamefont {Portner}, \citenamefont {Tanner}, \citenamefont {C.}, \citenamefont {Zhou}, \citenamefont {Mi}, \citenamefont {Tazoui}, \citenamefont {Lee}, \citenamefont {Wall}, \citenamefont {Ginsberg},\ and\ \citenamefont {Talapin}}]{jeongPbSLigandSamplePreparation24}%
  \BibitemOpen
  \bibfield  {author} {\bibinfo {author} {\bibfnamefont {A.}~\bibnamefont {Jeong}}, \bibinfo {author} {\bibfnamefont {J.}~\bibnamefont {Portner}}, \bibinfo {author} {\bibfnamefont {C.~P.~N.}\ \bibnamefont {Tanner}}, \bibinfo {author} {\bibfnamefont {O.~J.}\ \bibnamefont {C.}}, \bibinfo {author} {\bibfnamefont {C.}~\bibnamefont {Zhou}}, \bibinfo {author} {\bibfnamefont {Z.}~\bibnamefont {Mi}}, \bibinfo {author} {\bibfnamefont {Y.~A.}\ \bibnamefont {Tazoui}}, \bibinfo {author} {\bibfnamefont {B.}~\bibnamefont {Lee}}, \bibinfo {author} {\bibfnamefont {V.~R.~K.}\ \bibnamefont {Wall}}, \bibinfo {author} {\bibfnamefont {N.~S.}\ \bibnamefont {Ginsberg}}, \ and\ \bibinfo {author} {\bibfnamefont {D.~V.}\ \bibnamefont {Talapin}},\ }\href {\doibase 10.1021/acsnano.4c06033} {\bibfield  {journal} {\bibinfo  {journal} {ACS Nano}\ }\textbf {\bibinfo {volume} {18}},\ \bibinfo {pages} {33864} (\bibinfo {year} {2024})}\BibitemShut {NoStop}%
\bibitem [{\citenamefont {Kieffer}\ \emph {et~al.}(2025)\citenamefont {Kieffer}, \citenamefont {Orlans}, \citenamefont {Coquelle}, \citenamefont {Debionne}, \citenamefont {Basu}, \citenamefont {Homs}, \citenamefont {Santoni},\ and\ \citenamefont {De~Sanctis}}]{KiefferPyFAI25}%
  \BibitemOpen
  \bibfield  {author} {\bibinfo {author} {\bibfnamefont {J.}~\bibnamefont {Kieffer}}, \bibinfo {author} {\bibfnamefont {J.}~\bibnamefont {Orlans}}, \bibinfo {author} {\bibfnamefont {N.}~\bibnamefont {Coquelle}}, \bibinfo {author} {\bibfnamefont {S.}~\bibnamefont {Debionne}}, \bibinfo {author} {\bibfnamefont {S.}~\bibnamefont {Basu}}, \bibinfo {author} {\bibfnamefont {A.}~\bibnamefont {Homs}}, \bibinfo {author} {\bibfnamefont {G.}~\bibnamefont {Santoni}}, \ and\ \bibinfo {author} {\bibfnamefont {D.}~\bibnamefont {De~Sanctis}},\ }\href {\doibase 10.1107/S1600576724011038} {\bibfield  {journal} {\bibinfo  {journal} {Journal of Applied Crystallography}\ }\textbf {\bibinfo {volume} {58}},\ \bibinfo {pages} {138} (\bibinfo {year} {2025})}\BibitemShut {NoStop}%
\bibitem [{\citenamefont {Juhas}\ \emph {et~al.}(2013)\citenamefont {Juhas}, \citenamefont {Davis}, \citenamefont {Farrow},\ and\ \citenamefont {Billinge}}]{juhasPDFgetX3RapidHighly2013}%
  \BibitemOpen
  \bibfield  {author} {\bibinfo {author} {\bibfnamefont {P.}~\bibnamefont {Juhas}}, \bibinfo {author} {\bibfnamefont {T.}~\bibnamefont {Davis}}, \bibinfo {author} {\bibfnamefont {C.~L.}\ \bibnamefont {Farrow}}, \ and\ \bibinfo {author} {\bibfnamefont {S.~J.~L.}\ \bibnamefont {Billinge}},\ }\href {\doibase 10.1107/S0021889813005190} {\bibfield  {journal} {\bibinfo  {journal} {J Appl Crystallogr}\ }\textbf {\bibinfo {volume} {46}},\ \bibinfo {pages} {560} (\bibinfo {year} {2013})},\ \Eprint {http://arxiv.org/abs/1211.7126} {arXiv:1211.7126 [cond-mat]} \BibitemShut {NoStop}%
\bibitem [{\citenamefont {Bertolotti}\ and\ \citenamefont {Robinson}(2024)}]{euxfelWorkshopReportQmax24}%
  \BibitemOpen
  \bibfield  {author} {\bibinfo {author} {\bibfnamefont {F.}~\bibnamefont {Bertolotti}}\ and\ \bibinfo {author} {\bibfnamefont {I.}~\bibnamefont {Robinson}},\ }\href@noop {} {\enquote {\bibinfo {title} {Scientific opportunities with very hard xfel radiation},}\ }\bibinfo {howpublished} {XFEL.EU WR-2024-002} (\bibinfo {year} {2024})\BibitemShut {NoStop}%
\bibitem [{\citenamefont {Sapnik}\ \emph {et~al.}(2025)\citenamefont {Sapnik}, \citenamefont {Chater}, \citenamefont {Keeble}, \citenamefont {Evans}, \citenamefont {Bertolotti}, \citenamefont {Guagliardi}, \citenamefont {St{\o}ckler}, \citenamefont {Harbourne}, \citenamefont {Borup}, \citenamefont {Silberg}, \citenamefont {Descamps}, \citenamefont {Prescher}, \citenamefont {Klee}, \citenamefont {Phelipeau}, \citenamefont {Ullah}, \citenamefont {Medina}, \citenamefont {Bird}, \citenamefont {Kaznelson}, \citenamefont {Lynn}, \citenamefont {Goodwin}, \citenamefont {Iversen}, \citenamefont {Crepisson}, \citenamefont {Bozin}, \citenamefont {Jensen}, \citenamefont {McBride}, \citenamefont {Neder}, \citenamefont {Robinson}, \citenamefont {Wark}, \citenamefont {Andrzejewski}, \citenamefont {Boesenberg}, \citenamefont {Brambrink}, \citenamefont {Camarda}, \citenamefont {Cerantola}, \citenamefont {Goede}, \citenamefont {H{\"{o}}ppner}, \citenamefont {Humphries}, \citenamefont {Konopkova}, \citenamefont {Kujala},
  \citenamefont {Michelat}, \citenamefont {Nakatsutsumi}, \citenamefont {Pelka}, \citenamefont {Preston}, \citenamefont {Randolph}, \citenamefont {Roeper}, \citenamefont {Schmidt}, \citenamefont {Strohm}, \citenamefont {Tang}, \citenamefont {Talkovski}, \citenamefont {Zastrau}, \citenamefont {Appel},\ and\ \citenamefont {Keen}}]{sapnikXFELHighQmax2025}%
  \BibitemOpen
  \bibfield  {author} {\bibinfo {author} {\bibfnamefont {A.~F.}\ \bibnamefont {Sapnik}}, \bibinfo {author} {\bibfnamefont {P.~A.}\ \bibnamefont {Chater}}, \bibinfo {author} {\bibfnamefont {D.~S.}\ \bibnamefont {Keeble}}, \bibinfo {author} {\bibfnamefont {J.~S.~O.}\ \bibnamefont {Evans}}, \bibinfo {author} {\bibfnamefont {F.}~\bibnamefont {Bertolotti}}, \bibinfo {author} {\bibfnamefont {A.}~\bibnamefont {Guagliardi}}, \bibinfo {author} {\bibfnamefont {L.~J.}\ \bibnamefont {St{\o}ckler}}, \bibinfo {author} {\bibfnamefont {E.~A.}\ \bibnamefont {Harbourne}}, \bibinfo {author} {\bibfnamefont {A.~B.}\ \bibnamefont {Borup}}, \bibinfo {author} {\bibfnamefont {R.~S.}\ \bibnamefont {Silberg}}, \bibinfo {author} {\bibfnamefont {A.}~\bibnamefont {Descamps}}, \bibinfo {author} {\bibfnamefont {C.}~\bibnamefont {Prescher}}, \bibinfo {author} {\bibfnamefont {B.~D.}\ \bibnamefont {Klee}}, \bibinfo {author} {\bibfnamefont {A.}~\bibnamefont {Phelipeau}}, \bibinfo {author} {\bibfnamefont {I.}~\bibnamefont {Ullah}}, \bibinfo
  {author} {\bibfnamefont {K.~G.}\ \bibnamefont {Medina}}, \bibinfo {author} {\bibfnamefont {T.~A.}\ \bibnamefont {Bird}}, \bibinfo {author} {\bibfnamefont {V.}~\bibnamefont {Kaznelson}}, \bibinfo {author} {\bibfnamefont {W.}~\bibnamefont {Lynn}}, \bibinfo {author} {\bibfnamefont {A.~L.}\ \bibnamefont {Goodwin}}, \bibinfo {author} {\bibfnamefont {B.~B.}\ \bibnamefont {Iversen}}, \bibinfo {author} {\bibfnamefont {C.}~\bibnamefont {Crepisson}}, \bibinfo {author} {\bibfnamefont {E.~S.}\ \bibnamefont {Bozin}}, \bibinfo {author} {\bibfnamefont {K.~M.~{\O}.}\ \bibnamefont {Jensen}}, \bibinfo {author} {\bibfnamefont {E.~E.}\ \bibnamefont {McBride}}, \bibinfo {author} {\bibfnamefont {R.~B.}\ \bibnamefont {Neder}}, \bibinfo {author} {\bibfnamefont {I.}~\bibnamefont {Robinson}}, \bibinfo {author} {\bibfnamefont {J.~S.}\ \bibnamefont {Wark}}, \bibinfo {author} {\bibfnamefont {M.}~\bibnamefont {Andrzejewski}}, \bibinfo {author} {\bibfnamefont {U.}~\bibnamefont {Boesenberg}}, \bibinfo {author} {\bibfnamefont
  {E.}~\bibnamefont {Brambrink}}, \bibinfo {author} {\bibfnamefont {C.}~\bibnamefont {Camarda}}, \bibinfo {author} {\bibfnamefont {V.}~\bibnamefont {Cerantola}}, \bibinfo {author} {\bibfnamefont {S.}~\bibnamefont {Goede}}, \bibinfo {author} {\bibfnamefont {H.}~\bibnamefont {H{\"{o}}ppner}}, \bibinfo {author} {\bibfnamefont {O.~S.}\ \bibnamefont {Humphries}}, \bibinfo {author} {\bibfnamefont {Z.}~\bibnamefont {Konopkova}}, \bibinfo {author} {\bibfnamefont {N.}~\bibnamefont {Kujala}}, \bibinfo {author} {\bibfnamefont {T.}~\bibnamefont {Michelat}}, \bibinfo {author} {\bibfnamefont {M.}~\bibnamefont {Nakatsutsumi}}, \bibinfo {author} {\bibfnamefont {A.}~\bibnamefont {Pelka}}, \bibinfo {author} {\bibfnamefont {T.~R.}\ \bibnamefont {Preston}}, \bibinfo {author} {\bibfnamefont {L.}~\bibnamefont {Randolph}}, \bibinfo {author} {\bibfnamefont {M.}~\bibnamefont {Roeper}}, \bibinfo {author} {\bibfnamefont {A.}~\bibnamefont {Schmidt}}, \bibinfo {author} {\bibfnamefont {C.}~\bibnamefont {Strohm}}, \bibinfo {author}
  {\bibfnamefont {M.}~\bibnamefont {Tang}}, \bibinfo {author} {\bibfnamefont {P.}~\bibnamefont {Talkovski}}, \bibinfo {author} {\bibfnamefont {U.}~\bibnamefont {Zastrau}}, \bibinfo {author} {\bibfnamefont {K.}~\bibnamefont {Appel}}, \ and\ \bibinfo {author} {\bibfnamefont {D.~A.}\ \bibnamefont {Keen}},\ }\href {\doibase 10.1107/S205225252500538X} {\bibfield  {journal} {\bibinfo  {journal} {IUCrJ}\ }\textbf {\bibinfo {volume} {12}},\ \bibinfo {pages} {531} (\bibinfo {year} {2025})}\BibitemShut {NoStop}%
\bibitem [{\citenamefont {Fuchs}\ \emph {et~al.}(2020)\citenamefont {Fuchs}, \citenamefont {Culver}, \citenamefont {Till},\ and\ \citenamefont {Zeier}}]{fuchsNSWSSamplePreparation19}%
  \BibitemOpen
  \bibfield  {author} {\bibinfo {author} {\bibfnamefont {T.}~\bibnamefont {Fuchs}}, \bibinfo {author} {\bibfnamefont {S.~P.}\ \bibnamefont {Culver}}, \bibinfo {author} {\bibfnamefont {P.}~\bibnamefont {Till}}, \ and\ \bibinfo {author} {\bibfnamefont {W.~G.}\ \bibnamefont {Zeier}},\ }\href {\doibase 10.1021/acsenergylett.9b02537} {\bibfield  {journal} {\bibinfo  {journal} {ACS Energy Letters}\ }\textbf {\bibinfo {volume} {5}},\ \bibinfo {pages} {146} (\bibinfo {year} {2020})}\BibitemShut {NoStop}%
\bibitem [{\citenamefont {Decking}(2007)}]{deckingEuXFEL07}%
  \BibitemOpen
  \bibfield  {author} {\bibinfo {author} {\bibfnamefont {W.}~\bibnamefont {Decking}},\ }in\ \href@noop {} {\emph {\bibinfo {booktitle} {Brilliant Light in Life and Material Sciences}}},\ \bibinfo {editor} {edited by\ \bibinfo {editor} {\bibfnamefont {V.}~\bibnamefont {Tsakanov}}\ and\ \bibinfo {editor} {\bibfnamefont {H.}~\bibnamefont {Wiedemann}}}\ (\bibinfo  {publisher} {Springer Netherlands},\ \bibinfo {address} {Dordrecht},\ \bibinfo {year} {2007})\ pp.\ \bibinfo {pages} {21--30}\BibitemShut {NoStop}%
\bibitem [{\citenamefont {Galler}\ \emph {et~al.}(2019)\citenamefont {Galler}, \citenamefont {Gawelda}, \citenamefont {Biednov}, \citenamefont {Bomer}, \citenamefont {Britz}, \citenamefont {Brockhauser}, \citenamefont {Choi}, \citenamefont {Diez}, \citenamefont {Frankenberger}, \citenamefont {French}, \citenamefont {Görries}, \citenamefont {Hart}, \citenamefont {Hauf}, \citenamefont {Khakhulin}, \citenamefont {Knoll}, \citenamefont {Korsch}, \citenamefont {Kubicek}, \citenamefont {Kuster}, \citenamefont {Lang}, \citenamefont {Alves~Lima}, \citenamefont {Otte}, \citenamefont {Schulz}, \citenamefont {Zalden},\ and\ \citenamefont {Bressler}}]{FXE_beamline_overview}%
  \BibitemOpen
  \bibfield  {author} {\bibinfo {author} {\bibfnamefont {A.}~\bibnamefont {Galler}}, \bibinfo {author} {\bibfnamefont {W.}~\bibnamefont {Gawelda}}, \bibinfo {author} {\bibfnamefont {M.}~\bibnamefont {Biednov}}, \bibinfo {author} {\bibfnamefont {C.}~\bibnamefont {Bomer}}, \bibinfo {author} {\bibfnamefont {A.}~\bibnamefont {Britz}}, \bibinfo {author} {\bibfnamefont {S.}~\bibnamefont {Brockhauser}}, \bibinfo {author} {\bibfnamefont {T.-K.}\ \bibnamefont {Choi}}, \bibinfo {author} {\bibfnamefont {M.}~\bibnamefont {Diez}}, \bibinfo {author} {\bibfnamefont {P.}~\bibnamefont {Frankenberger}}, \bibinfo {author} {\bibfnamefont {M.}~\bibnamefont {French}}, \bibinfo {author} {\bibfnamefont {D.}~\bibnamefont {Görries}}, \bibinfo {author} {\bibfnamefont {M.}~\bibnamefont {Hart}}, \bibinfo {author} {\bibfnamefont {S.}~\bibnamefont {Hauf}}, \bibinfo {author} {\bibfnamefont {D.}~\bibnamefont {Khakhulin}}, \bibinfo {author} {\bibfnamefont {M.}~\bibnamefont {Knoll}}, \bibinfo {author} {\bibfnamefont {T.}~\bibnamefont
  {Korsch}}, \bibinfo {author} {\bibfnamefont {K.}~\bibnamefont {Kubicek}}, \bibinfo {author} {\bibfnamefont {M.}~\bibnamefont {Kuster}}, \bibinfo {author} {\bibfnamefont {P.}~\bibnamefont {Lang}}, \bibinfo {author} {\bibfnamefont {F.}~\bibnamefont {Alves~Lima}}, \bibinfo {author} {\bibfnamefont {F.}~\bibnamefont {Otte}}, \bibinfo {author} {\bibfnamefont {S.}~\bibnamefont {Schulz}}, \bibinfo {author} {\bibfnamefont {P.}~\bibnamefont {Zalden}}, \ and\ \bibinfo {author} {\bibfnamefont {C.}~\bibnamefont {Bressler}},\ }\href {\doibase 10.1107/S1600577519006647} {\bibfield  {journal} {\bibinfo  {journal} {Journal of Synchrotron Radiation}\ }\textbf {\bibinfo {volume} {26}},\ \bibinfo {pages} {1432} (\bibinfo {year} {2019})}\BibitemShut {NoStop}%
\bibitem [{\citenamefont {Kodama}\ \emph {et~al.}(2006)\citenamefont {Kodama}, \citenamefont {Iikubo}, \citenamefont {Taguchi},\ and\ \citenamefont {Shamoto}}]{kodamaFiniteSizeEffects06}%
  \BibitemOpen
  \bibfield  {author} {\bibinfo {author} {\bibfnamefont {K.}~\bibnamefont {Kodama}}, \bibinfo {author} {\bibfnamefont {S.}~\bibnamefont {Iikubo}}, \bibinfo {author} {\bibfnamefont {T.}~\bibnamefont {Taguchi}}, \ and\ \bibinfo {author} {\bibfnamefont {S.-i.}\ \bibnamefont {Shamoto}},\ }\href {\doibase 10.1107/S0108767306034635} {\bibfield  {journal} {\bibinfo  {journal} {Acta Crystallographica Section A}\ }\textbf {\bibinfo {volume} {62}},\ \bibinfo {pages} {444} (\bibinfo {year} {2006})}\BibitemShut {NoStop}%
\bibitem [{\citenamefont {Lei}\ \emph {et~al.}(2009)\citenamefont {Lei}, \citenamefont {de~Graff}, \citenamefont {Thorpe}, \citenamefont {Wells},\ and\ \citenamefont {Sartbaeva}}]{leiIntrinsicGeometry09}%
  \BibitemOpen
  \bibfield  {author} {\bibinfo {author} {\bibfnamefont {M.}~\bibnamefont {Lei}}, \bibinfo {author} {\bibfnamefont {A.~M.~R.}\ \bibnamefont {de~Graff}}, \bibinfo {author} {\bibfnamefont {M.~F.}\ \bibnamefont {Thorpe}}, \bibinfo {author} {\bibfnamefont {S.~A.}\ \bibnamefont {Wells}}, \ and\ \bibinfo {author} {\bibfnamefont {A.}~\bibnamefont {Sartbaeva}},\ }\href {\doibase 10.1103/PhysRevB.80.024118} {\bibfield  {journal} {\bibinfo  {journal} {Phys. Rev. B}\ }\textbf {\bibinfo {volume} {80}},\ \bibinfo {pages} {024118} (\bibinfo {year} {2009})}\BibitemShut {NoStop}%
\bibitem [{\citenamefont {\phantom{}DiffPy team}(2025)}]{diffpyWebsite}%
  \BibitemOpen
  \bibfield  {author} {\bibinfo {author} {\bibnamefont {\phantom{}DiffPy team}},\ }\href {https://www.diffpy.org/} {\enquote {\bibinfo {title} {Diffpy - atomic structure analysis in python},}\ } (\bibinfo {year} {2025}),\ \Eprint {http://arxiv.org/abs/https://www.diffpy.org/} {https://www.diffpy.org/} \BibitemShut {NoStop}%
\bibitem [{\citenamefont {Juh\'as}\ \emph {et~al.}(2025)\citenamefont {Juh\'as}, \citenamefont {Farrow}, \citenamefont {Yang}, \citenamefont {Knox}, \citenamefont {Yang},\ and\ \citenamefont {Billinge}}]{diffpyutilssoftware25}%
  \BibitemOpen
  \bibfield  {author} {\bibinfo {author} {\bibfnamefont {P.}~\bibnamefont {Juh\'as}}, \bibinfo {author} {\bibfnamefont {C.~L.}\ \bibnamefont {Farrow}}, \bibinfo {author} {\bibfnamefont {X.}~\bibnamefont {Yang}}, \bibinfo {author} {\bibfnamefont {K.~R.}\ \bibnamefont {Knox}}, \bibinfo {author} {\bibfnamefont {A.}~\bibnamefont {Yang}}, \ and\ \bibinfo {author} {\bibfnamefont {S.~J.~L.}\ \bibnamefont {Billinge}},\ }\href@noop {} {\enquote {\bibinfo {title} {{diffpy.utils}},}\ } (\bibinfo {year} {2025}),\ \Eprint {http://arxiv.org/abs/{https://github.com/diffpy/diffpy.utils}} {{https://github.com/diffpy/diffpy.utils}} \BibitemShut {NoStop}%
\bibitem [{\citenamefont {Léger}\ \emph {et~al.}(2000)\citenamefont {Léger}, \citenamefont {Pereira}, \citenamefont {Haines}, \citenamefont {Jobic},\ and\ \citenamefont {Brec}}]{legerIrTe2BondLengths00}%
  \BibitemOpen
  \bibfield  {author} {\bibinfo {author} {\bibfnamefont {J.}~\bibnamefont {Léger}}, \bibinfo {author} {\bibfnamefont {A.}~\bibnamefont {Pereira}}, \bibinfo {author} {\bibfnamefont {J.}~\bibnamefont {Haines}}, \bibinfo {author} {\bibfnamefont {S.}~\bibnamefont {Jobic}}, \ and\ \bibinfo {author} {\bibfnamefont {R.}~\bibnamefont {Brec}},\ }\href {\doibase https://doi.org/10.1016/S0022-3697(99)00230-9} {\bibfield  {journal} {\bibinfo  {journal} {Journal of Physics and Chemistry of Solids}\ }\textbf {\bibinfo {volume} {61}},\ \bibinfo {pages} {27} (\bibinfo {year} {2000})}\BibitemShut {NoStop}%
\bibitem [{\citenamefont {Brown}\ \emph {et~al.}(2006)\citenamefont {Brown}, \citenamefont {Fox}, \citenamefont {Maslen}, \citenamefont {O'Keefe},\ and\ \citenamefont {Willis}}]{brownAtomicFormFactors06}%
  \BibitemOpen
  \bibfield  {author} {\bibinfo {author} {\bibfnamefont {P.~J.}\ \bibnamefont {Brown}}, \bibinfo {author} {\bibfnamefont {A.~G.}\ \bibnamefont {Fox}}, \bibinfo {author} {\bibfnamefont {E.~N.}\ \bibnamefont {Maslen}}, \bibinfo {author} {\bibfnamefont {M.~A.}\ \bibnamefont {O'Keefe}}, \ and\ \bibinfo {author} {\bibfnamefont {B.~T.~M.}\ \bibnamefont {Willis}},\ }\enquote {\bibinfo {title} {Intensity of diffracted intensities},}\ in\ \href {\doibase 10.1107/97809553602060000600} {\emph {\bibinfo {booktitle} {International Tables for Crystallography Volume C: Mathematical, physical and chemical tables}}},\ \bibinfo {editor} {edited by\ \bibinfo {editor} {\bibfnamefont {E.}~\bibnamefont {Prince}}}\ (\bibinfo  {publisher} {Springer Netherlands},\ \bibinfo {address} {Dordrecht},\ \bibinfo {year} {2006})\ pp.\ \bibinfo {pages} {554--595}\BibitemShut {NoStop}%
\bibitem [{\citenamefont {Boas}(2006)}]{boasFourierSine2006}%
  \BibitemOpen
  \bibfield  {author} {\bibinfo {author} {\bibfnamefont {M.~L.}\ \bibnamefont {Boas}},\ }\href@noop {} {\emph {\bibinfo {title} {{Mathematical methods in the physical sciences}}}}\ (\bibinfo  {publisher} {Wiley},\ \bibinfo {address} {Hoboken, NJ},\ \bibinfo {year} {2006})\BibitemShut {NoStop}%
\bibitem [{\citenamefont {Dinnebier}\ and\ \citenamefont {Billinge}(2018)}]{dinnebierBillingeOverviewPrinciples18}%
  \BibitemOpen
  \bibfield  {author} {\bibinfo {author} {\bibfnamefont {R.~E.}\ \bibnamefont {Dinnebier}}\ and\ \bibinfo {author} {\bibfnamefont {S.~J.~L.}\ \bibnamefont {Billinge}},\ }\href {\doibase 10.1107/97809553602060000935} {\bibfield  {journal} {\bibinfo  {journal} {International Tables for Crystallography}\ }\textbf {\bibinfo {volume} {H}},\ \bibinfo {pages} {2} (\bibinfo {year} {2018})}\BibitemShut {NoStop}%
\end{thebibliography}%
\bibliographystyle{apsrev4-1}



%



%
%


\end{document}